\documentclass[openacc]{rsproca_new}%%%%where rsproca is the template name
\usepackage{mathtools}
\usepackage[utf8]{inputenc}
\usepackage{url}
\pdfoutput=1

\newcommand{\sn}[1]{\S~\ref{sec:#1}}

\newcommand\hg[4]{{}_2F_1\biggl(\!\!\begin{array}{c}{#1},{#2}\\{#3}%
\end{array}\!\!\Big|\,{#4}\biggr)}
\newcommand{\cor}[1]{\left\langle #1 \right\rangle}
\newcommand{\Nat}{\mathrm{I}\!\mathrm{N}}
\newcommand{\SO}[1]{\mathrm{SO}{(#1)}}
\newcommand{\DS}{\ensuremath{D_6}}
\newcommand{\ES}{\ensuremath{E_6}}
\newcommand{\wtNS}{{{\scriptstyle \widetilde{N\!S}}}}
\newcommand{\NS}{{{\scriptstyle N\!S}}}
\newcommand{\kkett}[1]{{\| #1 \rangle\!\rangle }}
\newcommand{\cD}{{\mathcal D }}
\newcommand{\cQ}{{\mathcal Q }}
\newcommand{\cB}{{\mathcal B }}
\newcommand{\cV}{{\mathcal V}}

\newcommand{\blank}[1]{}

\def\Integers{\mathbb{Z}}

\def\Reals{\mathbb{R}}
\def\Disk{\mathbb{D}}
\def\pr{\partial}
\def\tr{\mathrm{tr}}
\def\PB{\cB}
\def\PBM{\PB(M)}
\def\IM#1#2{I_{M}\expval{#1,#2}}
\def\expval#1{\langle \, #1 \,\rangle}
\DeclareMathOperator\Hom{Hom}
\DeclareMathOperator\Aut{Aut}

\DeclarePairedDelimiter\norm{\lVert}{\rVert}%
\def\integral{\mathrm{int}}
\def\sing{\mathrm{sing}}
\def\distr{\mathrm{distr}}

\newcommand{\overbar}[1]{\mkern2mu\overbracket[0.25pt][-1pt]{\mkern-2mu#1\mkern-5mu}\mkern 5mu}

\def\Aut{\mathbf{Aut}}
\def\SVIR2{\ensuremath{\mathrm{SVIR}_3{}^{\otimes 2}}}

\def\sVir{\ensuremath{\mathrm{sVir}} }
\def\sTCIM{\ensuremath{\mathrm{SVIR}_3}} 
\def\sTEN{\ensuremath{\mathrm{SVIR}_{10}}}
\begin{document}

%%%% Article title to be placed here
\title{Boundary and Defect CFT: Open Problems and Applications}

\author{%%%% Author details
\fontsize{10pt}{0}{
N.~Andrei$^{1}$, A.~Bissi$^{2}$, M.~Buican$^{3}$, J.~Cardy$^{4,5}$, P.~Dorey$^{6}$, N.~Drukker$^{7}$, J.~Erdmenger$^{8}$, D.~Friedan$^{1,9}$, D.~Fursaev$^{10}$, A.~Konechny$^{11,12}$, C.~Kristjansen$^{13}$, I.~Makabe$^{7}$, Y.~Nakayama$^{14}$, A.~O'Bannon$^{15}$, R.~Parini$^{16}$, B.~Robinson$^{15}$, S.~Ryu$^{17}$, C.~Schmidt-Colinet$^{18}$, V.~Schomerus$^{19}$, C.~Schweigert$^{20}$, and G.M.T.~Watts$^{7}$}}
%Natan Andrei$^{1}$, Agnese Bissi$^{2}$, Matthew Buican$^{3}$, John Cardy$^{4,5}$, Patrick Dorey$^{6}$, Nadav Drukker$^{7}$, Johanna Erdmenger$^{8}$, Daniel Friedan$^{1,9}$, Dmitri Fursaev$^{10}$, Anatoly Konechny$^{11,12}$, Charlotte Kristjansen$^{13}$, Isao Makabe$^{7}$, Yu Nakayama$^{14}$, Andrew O'Bannon$^{15}$, Robert Parini$^{16}$, Brandon Robinson$^{15}$, Shinsei Ryu$^{17}$, Cornelius Schmidt-Colinet$^{18}$, Volker Schomerus$^{19}$, Christoph Schweigert$^{20}$, and Gerard Watts$^{7}$}}

%%%%%%%%% Insert author address here
\address{\fontsize{10pt}{0}{$^{1}$Dept. of Phys., Rutgers Univ., Piscataway, NJ, USA.
\\
$^{2}$Dept. of Phys. and Astro., Uppsala Univ., SE\\
$^{3}$CRST and SPA, Queen Mary Univ. of London, UK\\
$^{4}$Dept. of Phys., Univ. of California, Berkeley, CA, USA\\
$^{5}$All Souls College, Oxford, UK\\
$^{6}$Dept. of Math. Sci., Durham Univ., UK\\
$^{7}$Dept. of Maths, King's College London, UK\\
$^{8}$Julius-Maximilians-Univ. W\"urzburg, DE\\
$^{9}$ Natural Science Inst., Univ. of Iceland, IS\\
$^{10}$Dubna State Univ., Dubna, RU\\
$^{11}$Dept. of Maths, Heriot-Watt Univ., Edinburgh, UK\\
$^{12}$Maxwell Inst. for Math. Sci., Edinburgh, UK \\
$^{13}$Niels Bohr Inst., Copenhagen Univ., , DK\\
$^{14}$Dept. of Phys., Rikkyo Univ., Tokyo, JP\\
$^{15}$STAG Research Centre, Univ. of Southampton, UK\\
$^{16}$Department of Mathematics, University of York, UK\\
$^{17}$J. Franck Inst. \& KCTP., Univ. of Chicago, IL, USA\\
$^{18}$Arnold Sommerfeld Center, Univ. M\"unchen, DE\\
$^{19}$DESY Theory Group, DESY Hamburg, Hamburg, DE\\
$^{20}$Fachbereich Math., Univ. Hamburg, Hamburg, DE}
}

%%%% Subject entries to be placed here %%%%
\subject{Theoretical Physics, Mathematical Physics}

%%%% Keyword entries to be placed here %%%%

%%%% Insert corresponding author and its email address}
\corres{B. Robinson\\
\email{B.J.Robinson@soton.ac.uk}}
\vspace{-.4cm}
%%%% Abstract text to be placed here %%%%%%%%%%%%
\begin{abstract}
Proceedings of the workshop ``Boundary and Defect Conformal Field Theory: Open Problems and Applications,'' Chicheley Hall, Buckinghamshire, UK, 7-8 Sept. 2017.
\end{abstract}
%%%%%%%%%%%%%%%%%%%%%%%%%%%
%%%%%%%%%% Insert the texts which can accomdate on firstpage in the tag "fmtext" %%%%%

\maketitle
\section{Introduction and Overview}
The following proceedings are a compilation of talks presented at the workshop ``Boundary and Defect Conformal Field Theory: Open Problems and Applications'' held at Chicheley Hall in Buckinghamshire, UK on 7 and 8 September 2017.  Each section that follows was authored by the speaker listed in the footnote attached to the corresponding section title. In this introduction we provide context and motivation for these proceedings, summarize the contents of each section, and extract a few general lessons from what follows. We therefore hope to provide a bird's-eye view of the different directions that research into conformal field theories (CFTs) with boundaries or defects is taking.

Quantum field theory (QFT) lies at the heart of much of modern theoretical physics: it describes systems in particle physics, condensed matter physics, and even quantum gravity, via holographic duality. Combined with the machinery of the renormalization group (RG), one can in principle systematically study an enormous variety of phenomena at different length scales in QFT---and therefore in nature.

CFTs occupy privileged places in the space of QFTs---they lie at the endpoints of RG flows. Therefore, they often characterize the ultraviolet (UV) and infrared (IR) limits of QFTs. Moreover, CFTs describe critical phenomena and the worldsheet theory in string theory. They also provide important testing grounds for integrability, duality, and other general phenomena in QFT.

CFTs are powerful because they are highly symmetric. In particular, CFTs are scale-invariant by definition, and so all correlation lengths are infinite. CFTs are also invariant under translations, rotations, boosts, and inversions. These symmetries constrain correlation functions, sometimes completely, providing a powerful non-perturbative approach to many aspects of QFT.

However, no real-world system has infinite size---boundary effects are always important. Moreover, all real-world systems involve impurities, differently-ordered regions separated by domain walls, and other types of defects that break translational and rotational symmetries to subgroups\footnote{In general, a co-dimension one defect between CFTs (i.e. a domain wall) can be mapped to a boundary of the product CFT via the folding trick, as discussed for example in~\sn{watts}.}. In short, the application of CFT to the real world necessarily entails studying boundary CFTs (bCFTs) and defect CFTs (dCFTs).

The reduced symmetry of bCFTs and dCFTs compared to CFTs loosens spectral constraints, making correlation functions richer and more intricate. For example, in a bCFT, scalar operators can have non-zero one-point functions, which are generally forbidden in a CFT. However, bCFTs and dCFTs often retain enough symmetry to provide calculable non-perturbative information about many systems. For example, bCFT provides a solution to the single-impurity Kondo problem~\cite{Affleck:1995ge} and a fully non-perturbative definition of D-branes and other spacetime defects in string theory~\cite{Recknagel:2013uja}.

Remarkably, bCFT and dCFT can also provide insight into RG flows beyond the critical endpoints. Indeed, consider a CFT with a relevant deformation that triggers an RG flow to another CFT. Now, imagine integrating that relevant deformation over half the spacetime. Then, in the IR, the result is an interface between the UV and IR CFTs, called an RG interface or RG domain wall~\cite{BR}. Such a construction, though clearly sacrificing some of the spacetime symmetry, is potentially very powerful: the problem of classifying RG flows between CFTs maps onto the problem of classifying defects between CFTs---which should be simpler, because much of the machinery of CFT can be brought to bear. Similarly, if the relevant deformation produces a mass gap, then the resulting IR will be the UV CFT with a boundary, hence the problem of classifying gapped vacua maps to the simpler problem of classifying conformal boundary conditions. In short, defects and boundaries allow CFT techniques to be extended to a much larger domain in the space of QFTs.

Given the incredible breadth of bCFT and dCFT, the workshop's goal was to bring together a diverse group of leading physicists working on different topics in these fields and search for common themes and unifying ideas. The workshop was also motivated by the desire to understand how recent developments that have reshaped CFT---like the modern conformal bootstrap, non-perturbative results in supersymmetry (SUSY), and recent advances in holography---combined with recent developments that have reshaped condensed matter physics---like topological phases and new results in integrability---have impacted bCFT and dCFT. Conversely, we wanted to know what impact bCFT and dCFT might have on these fields, i.e.what new applications of bCFT and dCFT may be possible. Given the wonderfully diverse and probing talks described below, we believe the workshop was a success on all fronts.

To help orient the reader interested in the various different themes of the workshop, we have organized the contributions thematically. In the next three subsections, we identify these themes and point to a few ideas unifying the contributions within each theme.

\subsubsection*{RG flows: Interfaces and boundaries}
The first six talks (\sn{cardy}-\sn{dorey}) address the behavior of boundaries and/or defects under RG flows. Of these, the first four (\sn{cardy}-\sn{ryu}) study boundaries and defects as maps between UV and IR CFTs, as described above. In particular, in~\sn{cardy} John Cardy describes a variational approach, in~\sn{konechny} Anatoly Konechny uses a combination of analytics and the truncated conformal space approach, in~\sn{schmidt-colinet} Cornelius Schmidt-Colinet employs holography, and in~\sn{ryu} Shinsei Ryu uses ideas from topology (and explores setups in which CFTs are themselves boundaries of higher-dimensional massive theories)\footnote{Note that when the IR phase is massive but contains non-trivial topological degrees of freedom, one still sometimes refers, as in \sn{ryu}, to the object separating the UV and IR as a defect or domain wall.}. The final two talks of the group each have a slightly different focus. In~\sn{andrei} Natan Andrei studies a purely boundary RG flow and describes the behavior of certain observables involving quantum dots coupled to Luttinger liquids, using techniques from integrability. In~\sn{erdmenger} Johanna Erdmenger explores impurity entropy in a holographic version of the Kondo model, involving an RG flow on an impurity coupled to a strongly-interacting CFT at large $N$. In~\sn{dorey} Patrick Dorey discusses boundary conditions that break integrability in a massive theory.

\subsubsection*{Supersymmetric boundaries and defects}
The next four talks (\sn{watts}-\sn{bissi}) discuss aspects of SUSY bCFTs and dCFTs. In~\sn{watts} Gerard Watts finds new non-topological defects in the tri-critical Ising model. In~\sn{drukker} Nadav Drukker discusses the SUSY multiplet for the displacement operator of co-dimension one defects in 4D $\mathcal{N}=1$ theories (the key role of the displacement operator has also been emphasized recently in~\cite{Herzog:2017xha, Herzog:2017kkj}). In~\sn{kristjansen} Charlotte Kristjansen discusses holographic constructions of defects that interpolate between two copies of 4D $\mathcal{N}=4$ super Yang-Mills theory with different rank gauge groups. In~\sn{bissi} Agnese Bissi explains techniques for computing loop corrections in AdS/CFT, which could be applied to study quantum effects in SUSY holographic dCFTs at subleading order in the $1/N$ expansion.

\subsubsection*{A toolbox for bCFTs, dCFTs, and beyond}
The final five talks (\sn{fursaev}-\sn{friedan}) deal with creating new tools to analyze bCFTs, dCFTs, and their generalizations. In~\sn{fursaev} Dmitri Fursaev discusses integrated conformal anomalies of bCFTs and dCFTs, and their use in determining various observables. In~\sn{schomerus}, Volker Schomerus develops new tools to analyze correlation functions of defects via connections to Calogero-Sutherland models (which should allow the application of conformal bootstrap techniques to these correlators). In~\sn{nakayama} Yu Nakayama implements the conformal bootstrap in what turns out to be one of the simplest curved backgrounds: real projective space (states in this space are closely related to boundary states in flat space CFT). In~\sn{schweigert} Christoph Schweigert develops the proper three-dimensional topological field theory framework in which to analyze rational CFTs and, for logarithmic CFTs, a modular functor constructed by a ``Lego-Teichm\"uller game'' (which ultimately leads to a better understanding of boundary conditions in these theories [189]). Finally, in~\sn{friedan} Daniel Friedan explores entirely new species of quantum theories built from defects.

\subsection*{Open Problems and Future Directions}
These proceedings present an incredibly diverse set of results in bCFT, dCFT, and beyond, and suggest various interesting open questions. Here we briefly mention just a few. Are there general principles that select the RG domain wall(s)/boundary state(s) from the possible set connecting the UV and IR endpoints of RG flows? What more can SUSY tell us about bCFTs and dCFTs, for example via SUSY localization? Are there boundary or defect \lq\lq $c$-theorems" generalizing the $g$-theorem~\cite{Friedan:2003yc,Casini:2016fgb} of two-dimensional bCFT to higher dimensions? To what extent can SUSY, the conformal bootstrap, integrability, topological field theory, and the other techniques discussed in these proceedings be used to address such questions? Could they also lead to the discovery of new conformal boundary conditions, defects, or even entirely new perspectives on what QFT is?

We hope that these proceedings will inspire new research into boundaries and defects in CFT, by making connections between sub-fields, finding novel applications, and advancing the state of the art of bCFT and dCFT, all building upon the remarkable progress summarized below.

%%%%%%%%%%%%%%%%%%%%%%%%%%%%%%%%%%%%%%%%%%%%%%%%%CARDY
\section[Bulk Renormalization Group Flows and Boundary States]{Bulk Renormalization Group Flows and Boundary States\protect\footnote{This section was authored by John Cardy from the Department of Physics, University of California Berkeley, Berkeley, CA 94720, USA and All Souls College, University of Oxford, Oxford OX1 4AL, UK.  The support for this section comes in part from the Simons Foundation and from the Perimeter Institute for Theoretical Physics. Research at the Perimeter Institute is supported by the Government of Canada through the Department of Innovation, Science and Economic
Development and by the Province of Ontario through the Ministry of Research and Innovation.}}
\label{sec:cardy}

\subsection{Motivation}

Conformal field theories (CFTs) are supposed to correspond to the non-trivial renormalization group (RG) fixed points of relativistic quantum field theories (QFTs). Such theories typically contain a number of scaling operators of dimension $\Delta<d$ (where $d$ is the space-time dimension), which, if added to the action, are relevant and drive the theory to what is, generically, a trivial fixed point. The points along this trajectory then correspond to a massive QFT. In general there is a multiplicity of such basins of attraction of the RG flows, but enumerating them and determining which combinations of relevant operators lead to which basins, and therefore to what kind of massive QFT, in general requires non-perturbative methods.
This problem is equivalent to mapping out the phase diagram in the vicinity of the critical point corresponding to the CFT. 

Another way of characterizing these massive theories is through the analysis of the possible boundary states of the CFT. Imagine the scenario in which the relevant operators are switched on in only a half-space, say $x_0<0$. This will then appear as some boundary condition on the CFT in $x_0>0$. However the boundary conditions themselves undergo RG lows, with fixed points corresponding
 to so-called conformal boundary conditions. 
Therefore on scales $\sim M^{-1}$, where $M$ is the mass scale of the perturbed theory, the correlations near the boundary should be those of a conformal boundary condition, deformed by \em irrelevant \em boundary operators.

Thus an important problem in all these cases is to determine to which conformal boundary condition a particular combination of bulk operators should correspond. In this talk I describe a simple prescription for doing this, based on a variational approach. A longer account appears in Ref.~\cite{JC}.

More specifically, we consider the action of deformed CFT with hamiltonian
$$
\hat H=\hat H_{CFT}+\sum_j\lambda_j\int\hat\Phi_j(x)d^{d-1}x\,,
$$
where the $\{\Phi_j\}$ are a set of relevant operators,
and a variational state
$$
|\{\alpha_a\},\{\tau_a\}\rangle=\sum_a\alpha_a\,e^{-\tau_a\hat H_{CFT}}|a\rangle\,,
$$
where the sum over all possible conformal boundary states, and minimize
\begin{equation}\label{eq1}
\lim_{L\to\infty}\frac1{L^D}\frac{\langle\{\alpha_a\},\{\tau_a\}|\hat H_{CFT}+\sum_j\lambda_j\int\hat\Phi_j(x)d^{d-1}x|\{\alpha_a\},\{\tau_a\}\rangle}{\langle\{\alpha_a\},\{\tau_a\}|\{\alpha_a\},\{\tau_a\}\rangle}\,,
\end{equation}
with respect to the $\{\alpha_a\}$ and $\{\tau_a\}$.

Eq.~(\ref{eq1}) simplifies because both terms are diagonal in the basis of physical boundary states, and they have a prescribed dependence on the $\{\tau_a\}$. Thus. for each $a$, we need to minimize an expression of the form
$$
E_a=\frac{D\sigma_a}{(2\tau_a)^{D+1}}+\sum_j\lambda_j\frac{A_a^j}{(2\tau_a)^{\Delta_j}}\,,
$$
with respect to $\tau_a$, and then choose the $a$ which gives the absolute minimum. Here $\sigma_a$ is the universal coefficient of the Casimir energy per unit length with boundary conditions $a$, and $A_a^j$ the universal coefficient of the 1-point function $\langle a|\Phi_j|a\rangle$. 

\subsection{Results in two dimensions}

In this case, $\sigma_a=c/12$, where $c$ is the central charge, and for the minimal models \cite{CL}
$$
A^j_a=\frac{S_a^j}{S_a^0}\left(\frac{S_0^0}{S_0^j}\right)^{1/2}\,.
$$
where $S_a^j$ are the known elements of the modular $S$-matrix of the CFT. This allows explicit calculations.
There are several general features of the solutions.
\begin{enumerate}
\item By its nature the method always gives the correct scaling of the energy with the coupling constants.
\item For a prescribed $a$, there is always some combination of couplings $\{\lambda_j\}$ for which this gives the minimum energy. Thus all boundary states correspond to an RG sink, and they all have a finite basin of attraction.
\item For any combination of couplings, there is always at least one $a$ for which a minimum occurs for finite $\tau_a$. This unfortunately rules out the possibility of the approximation correctly describing RG flows to another non-trivial CFT. Instead, in these cases, there are always at least two degenerate minima, indicating phase coexistence rather than a diverging correlation length. However the coexisting phases are physically sensible.
\item In known examples of integrable perturbations, where there is a higher degeneracy of RG sinks than expected on purely physical grounds, the approximation appears to reproduce this.
 \end{enumerate}
 
 An alternative criterion, involving comparing overlaps between numerical approximations to the actual ground and different boundary states, has been suggested by Konechny\cite{Kon}. 
From a numerical point of view our approach cannot be competitive with earlier methods such as the truncated conformal space approach \cite{TCFT1,TCFT2}, but it is much simpler and moreover gives new insight into the physical relationship between conformal boundary states and ground states of gapped theories.

%%%%%%%%%%%%%%%%%%%%%%%%%%%%%%%%%%%%%%%%%%%%%%%%%KONECHNY
\section[RG boundaries and interfaces in Ising field theory]{RG boundaries and interfaces in Ising field theory\protect\footnote{This section was authored by Anatoly Konechny from the Department of Mathematics at Heriot-Watt University, Edinburgh, UK and the Maxwell Institute for Mathematical Sciences, Edinburgh, UK.}}\label{sec:konechny}
This presentation is based on \cite{AK} where the details can be found. Here we only sketch the main idea as applied to the Ising field theory (IFT). 
The critical Ising model is a conformal field theory described in terms of free massless fermions $\psi$, $\bar \psi$. It  has two relevant operators related to temperature and magnetic field. The lFT can be defined 
as a perturbed critical Ising model:
\begin{equation}\label{IFT}
S_{\rm IFT}=\frac{1}{2\pi}\int (\psi\bar \partial \psi + \bar \psi \partial \bar \psi + im\bar \psi \psi )\, d^2 x + h\!\int\!\! \sigma\, d^2 x \, .
\end{equation} 

It is known that RG flows for all real values of the couplings $m$ and $h$ end up in a massive theory. In the far infrared only the vacuum state 
survives in the spectrum.    If we, following the general idea of \cite{BR}, perturb the critical theory only on a half space then in the far 
infrared we obtain a conformal boundary condition which we call an RG boundary. 
Moreover, if we put both (perturbed and unperturbed) theories on a cylinder than the asymptotic vacuum of 
the massive theory will be described as a conformal boundary state of the unperturbed theory. In the critical Ising model there are only three basic 
conformal boundary conditions that correspond to free of fixed boundary spins. Direct sums of these 3 basic conditions are also possible. 
The space of all RG flows corresponding to (\ref{IFT}) then breaks up into a finite number of domains labeled by RG boundaries. 
We can think of these domains as  infrared phases of the massive theories. 

As generic flows correspond to non-integrable QFTs we use the numerical techniques of Truncated Conformal Space approach (TCSA) 
and Truncated Free Fermion Space Approach (TFFSA) of \cite{FZ} to chart out the above domains. 
Calculating numerically the ratios of  low conformal weight components of the vacuum vector we obtained a complete description 
of the assignment of RG boundaries to RG flows. The same answers were also independently obtained 
in \cite{JC} using an analytic approach based on a variational method. 

We also explored RG flows with purely imaginary magnetic field. In this case there is a flow that ends up in a non-trivial infrared fixed 
point descried by Yang-Lee CFT. For this flow perturbing the critical Ising model on a half plane results in a conformal interface between 
the critical Ising and Yang-Lee CFTs. It is known due to \cite{QRW} that there are only 12 such conformal interfaces that makes this flow particularly attractive to 
study. Combining numerical results with symmetry considerations and exactly solvable boundary flows we arrived to a conclusion that the flow in the space of interfaces is non-convergent in this case. Although the theory arrives to a fixed point the interface (that represents how the infrared states sit in the ultraviolet theory)
perpetually  oscillates never approaching any single conformal interface. This is possible due to the lack of unitarity.

%%%%%%%%%%%%%%%%%%%%%%%%%%%%%%%%%%%%%%%%%%%%%%%%%SCHMIDT-COLINET
\section[Double trace interfaces]{Double trace interfaces\protect\footnote{This section was authored by Cornelius Schmidt-Colinet from the Arnold Sommerfeld Center for Theoretical Physics, Ludwig-Maximilians-Universit\"at M\"unchen, Theresienstr. 37, 80333 Munich, Germany. }}\label{sec:schmidt-colinet}
\subsection{Double trace perturbation}
For a scalar operator $\varphi$ on $d+1$ dimensional AdS space, with 
mass $-\tfrac{d^2}{4}\leq m\leq-\tfrac{d^2}{4}+1$, the expansion
\begin{equation}
\phi(X)\sim \psi_+(x)\rho^{\Delta_+}\,+\,
\psi_-(x)\rho^{\Delta_-}\,,\quad \rho\rightarrow 0
\end{equation}
close to the boundary $\rho=0$ of $AdS_{d+1}$ admits two different 
unitary boundary CFTs~\cite{Klebanov}: Either we take $\Delta_+=\tfrac{d}{2}+\nu$ 
with $\nu^2=\tfrac{d^2}{4}+m^2$ as the conformal dimension of the dual 
CFT scalar operator $\varphi=\varphi_+$ --- in which case 
$\psi_+(x)$ is proportional to the one-point function 
$\langle\varphi_+\rangle$ in the presence of the source~$\psi_-$ --- 
or we pick $\Delta_-=\tfrac{d}{2}+\nu$ as the dimension. 
The two choices lead to different CFTs. $CFT_-$, corresponding to the 
choice $\Delta_-$, serves as the UV fixed point of an RG flow triggered 
by the relevant ``double trace'' operator $\varphi_-^2$, obtained (for large $N$) 
as the operator of lowest non-trivial dimension in 
the $\varphi_-\times\varphi_-$ OPE. The RG flow takes the theory in the IR to $CFT_+$, 
the theory corresponding to the choice $\Delta_+$.

\subsection{Bulk Green's function and CFT correlators}
In order to study\footnote{Details for the following computation 
can be found in~\cite{us}} the RG interface~\cite{BR} between 
$CFT_-$ and $CFT_+$, we consider the Euclidean bulk, {\it i.e.} 
hyperbolic space $H_{d+1}$, in the Janus-type~\cite{Bak} coordinates
\begin{equation}\label{coordinates}
ds^2_{H^{d+1}}=\frac{dz^2}{4z^2(1-z)^2}+\frac{ds_{H^d}^2}{4z(1-z)}\,, 
\end{equation}
where the radial coordinate $z$ runs from $0$ to 1. This choice of global 
coordinates foliates $H_{d+1}$ into slices of $H_d$,
where the slice at any value of the coordinate $z$ meets the slice
at $1-z$ at its conformal boundary.
For the scalar bulk field, we impose $\Delta_-$ boundary conditions
at $z=0$, and $\Delta_+$ boundary conditions at $z=1$. The interface
is then located at the asymptotic intersection of the two $H^d$
patches covering the boundary of $H^{d+1}$.\\

The scalar Green's function in the bulk must then fall off as 
$G(z,x;z',x')\sim z^{\Delta_-}$ for $z\rightarrow 0$ (such that $x$ 
gives a point in $\cal A_-$), and as $G(z,x;z',x')\sim (1-z)^{\Delta_+}$ 
at $z\rightarrow 1$ (where $x$ is in $\cal{A}_+$). In the 
coordinates~\eqref{coordinates}, the Green's function can be found from 
harmonic analysis. Starting from eigenfunctions $\psi_s$ of the Laplacian $-\nabla_{H^d}$ 
on the slices $H^d$, one may separate variables and write the bulk scalar field
as a product $\phi(z,x)=\int d\mu(s)\psi_s(x)\Phi(s,z)$, 
where $d\mu(s)$ is an appropriate spectral measure.
The space of solutions for the radial component $\Phi$ is 
two-dimensional, but in order to satisfy the boundary conditions
($a$ at $z=0$ and $b$ at $z=1$, where $a,b=\pm$) we may use two different 
bases $\Phi^a_L$ and $\Phi^b_R$ with definite asymptotics
\begin{equation}
\Phi^a_L\sim z^{\Delta_a}\quad (z\rightarrow 0)\,,\qquad
\Phi^b_R\sim (1-z)^{\Delta_b}\quad (z\rightarrow 1)\,.
\end{equation}
The solutions involve hypergeometric functions ${}_2F_1$, and
a change between the bases $\Phi^a_L$ and $\Phi^b_R$ can be obtained
from Kummer's relations of ${}_2F_1(z)$ to ${}_2F_1(1-z)$.
The bulk Green's function is then found from a suitable ansatz
involving the solutions of the bulk scalar field equation, which
manifestly satisfies the correct asymptotics. One finds that $G^{ab}(z,x;z',x')$,
with boundary condition $a$ for $z\rightarrow 0$ and $b$ for $z\rightarrow 1$,
is of the form
\begin{equation}
G^{ab}(z,x;z',x')=\int_0^\infty d\sigma {\cal A}^{ab}_{\sigma}J_{\sigma}(x,x')
\Phi_L^a\big(\sigma,{\rm min}(z,z')\big)\,\Phi_R^b\big(\sigma,{\rm max}(z,z')\big)\,,
\end{equation}
where $\sigma$ is a continuous part of the label $s$, ${\cal A}^{ab}_{\sigma}$ is 
a constant involving Kummer's connection coefficients, $J_\sigma$ is a particular 
combination of products $\psi_s(x)\bar{\psi}_s(x')$ at fixed $\sigma$, and we use 
the fact that $\Phi_{L,R}$ actually only depend on $\sigma$.
Specifying to the case $(a,b)=(-,+)$, the bulk-boundary propagator can be obtained
as the limit
\begin{align}
K^{-+}(z,x;x')=\tfrac{1}{d-2\Delta_+}\lim_{z'\rightarrow1}
(4z'(1-z'))^{-\Delta_+/2}G^{-+}(z,x;z',x')\,,
\end{align}
where $x'$ is in ${\cal A}^+$. In the expression for the bulk-boundary propagator,
the integration over $\sigma$ can be performed explicitly. 
Moving the bulk insertion at $(z,x)$ to the boundary, standard expansion 
methods then yield the scalar two-point correlation functions.
In this way, we obtain the (flat-space) conformal correlation function
\begin{equation}\label{++correlator}
\cor{\varphi_+(x)\varphi_+(x')}=\frac{1}{|x-x'|^{2\Delta_+}}
\left(1+B\,\xi^{\Delta_+}\hg{d/2}{\Delta_+}{\Delta_++1}{-\xi}\right)\,,
\end{equation} 
where the interface is planar and situated at $y=0$ for $x=(\vec{x},y)$, 
$\xi=(x-x')^2/(4yy')$ is the conformal cross ratio, and
\begin{equation}
B=\frac{\Gamma(d/2)\Gamma(\nu+1)}{\Gamma(\Delta_++1)}\frac{\sin(\pi\nu)}{\pi}\,.
\end{equation}
We also obtain, in the same conformal frame as in~\eqref{++correlator},
\begin{equation}\label{+-correlator}
\cor{\varphi_-(x)\varphi_+(x')}=\sqrt{\frac{\sin(\pi\nu)}{\pi\nu}}
\frac{\Gamma(d/2)}{\sqrt{\Gamma(\Delta_+)\Gamma(\Delta_-)}}
\frac{(-\xi)^{-\frac{d}{2}}}{(2y')^{\Delta_+}(-2y)^{\Delta_-}}\,,
\end{equation}
where $CFT^+$ is situated in the half-space $y>0$.\\

The $\varphi_+\times\varphi_+$ OPE yields a conformal block decomposition
into (quasi-)primary contributions (\cite{McAvity}, see also C. Kristjansen's 
contribution elsewhere in these proceedings). The result~\eqref{++correlator} 
implies that the OPE contains primary operators of dimension 
$\Delta_n=2\Delta_++2n$ for $n\in\Nat_0$, which we can identify as the double 
trace operators of $\varphi_+$ descendents, with OPE coefficients
\begin{equation}
C_{++}{}^n\,B_n{}^{0}=\frac{\sin(\pi\nu)}{\pi}\frac{\Gamma(\frac{d}{2})
\Gamma(\nu+n+1)}{n!\Gamma(\Delta_++n+1)}\frac{(\nu)_n(\Delta_+)_n}{(\Delta_++\nu+n)_n}\,.
\end{equation}
In this equation, $C_{++}{}^n$ is the OPE coefficient for the primary operator
of dimension $\Delta_n$, and $B_n{}^0$ is the coefficient of the interface identity 
operator when the insertion is close to the interface.\\
From \eqref{+-correlator}, we deduce that the bulk-to-defect OPE of $\phi_+$ contains 
primary defect operators of dimension $\Delta_\alpha=\tfrac{d}{2}+\alpha$, 
$\alpha\in\Nat_0$, where the squares of the OPE coefficients are
\begin{equation}
(B_+{}^\alpha)^2=\frac{\sin(\pi\nu)}{\pi\nu}\frac{\alpha!}{(2\alpha)!}
\frac{\Gamma(\frac{d}{2})}{\Gamma(\Delta_+)}\frac{\Gamma(\nu+1+\alpha)}{(1-\nu)_\alpha}\,.
\end{equation}
These operators form the single-trace part of the full interface spectrum at large
$N$, which can be anticipated by methods explained in~\cite{Aharony}.

\subsection{Interface free energy}
The contribution of the interface to the free energy of the system can be calculated 
from considering the product theory $CFT_+\otimes CFT_-$ and inserting a product of the 
interface in such a way that the two factors are linked. Comparing to the product CFT
without any interface we have
\begin{equation}
2F_{\rm interface}=F_{+-}+F_{-+}-F_{++}-F_{--}=\frac{1}{2}
\log\left(\frac{\det{\cal D}^{+-}\det{\cal D}^{-+}}%
{\det{\cal D}^{++}\det{\cal D}^{--}}\right)\,,
\end{equation}
which is a UV-finite expression. Regularising the IR divergence~\cite{Diaz} leads to
\begin{equation}
\frac{d}{d\nu}2F_{\rm interface}\;\big(=\frac{d}{d\nu}\log g^2\big)\;= 
- \nu\, \frac{\cos\pi\nu}{\cos\tfrac{\pi d}{2}} 
\frac{\Gamma(\Delta_+)\Gamma(\Delta_-)}{\Gamma(1+d)}\,.
\end{equation} 
Since the sphere in two dimensions is conformally equivalent to the cylinder,
this yields the $g$ factor of the interface for $d=2$.

\subsection{Conformal field theory tests}
The reason for the simplicity of the interface and the corresponding RG 
flow is the large $N$ limit (see also the contributions by A. Bissi and 
C. Kristjansen in these proceedings). In the gravitational 
bulk, the limit in particular suppresses quantum gravity fluctuations. 
On the CFT side, large $N$ theories are ``generalised free'', which 
means in particular that correlation functions are obtained from 
Wick contractions, and also that the diverging parts of the OPEs 
of the scalar and the double trace operators are of the particular form
\begin{align}
\varphi\times\varphi\sim 1+C\varphi^2\,,\qquad 
\varphi^2\times\varphi^2\sim 1+2C\varphi^2\,,
\end{align} 
for some constant $C$. From this simple behaviour, the constants 
in~\eqref{++correlator} and~\eqref{+-correlator} can be verified 
perturbatively, since the renormalised coupling constant of the
double trace deformation is of order~$\nu$ for small~$\nu$. This
observation also entails that the expansion in~$\nu$ of our
bulk results organises the loop expansion of conformal perturbation
theory, at least in an appropriate scheme.\\

In the special case of a two-dimensional boundary~$d=2$, the
higher spin/CFT correspondence~\cite{Gaberdiel} provides a testing ground 
for our results. In particular, the UV-IR operator overlaps 
can be obtained exactly on the CFT side, using the interface
construction of~\cite{Gaiotto}. The coefficients of~\eqref{++correlator} 
and~\eqref{+-correlator} are precisely matched.

\subsection{Outlook}
The tools of harmonic analysis used to obtain the bulk results
can be replaced by those suitable for mixed boundary value 
problems~\cite{Sneddon} in more complicated situations. In particular, 
our aim for the near future is to investigate the fusion process 
of the presented double trace interfaces by these methods.

%%%%%%%%%%%%%%%%%%%%%%%%%%%%%%%%%%%%%%%%%%%%%%%%%RYU
\section[Boundary conformal field theory and topological phases of matter]{Boundary conformal field theory and topological phases of matter\protect\footnote{This section was authored by Shinsei Ryu from the James Franck Institute and Kadanoff Center for Theoretical Physics at the University of Chicago, Chicago, Illinois 60637, USA . This work was supported in part by the National Science Foundation grant DMR-1455296.}}\label{sec:ryu}

\subsection{Introduction}
%%%% Insert A head here

Boundary conformal field theory (BCFT)
has played a pivotal role in modern theoretical physics,
with its application ranging from problems in condensed matter physics such as 
the Kondo (impurity) problem, to high-energy physics, such as D-branes
\cite{2004hep.th11189C}.
%\cite{Cardy1989}.
In this article, we give an overview of the use of BCFT (and defect CFT)
in the context of topological phases of condensed matter.

Topological phases are fully {\it gapped} states of matter,
which are (topologically) distinct from trivial states that can be written as
product states.
Specifically, we can distinguish the following two kinds of topological phases.
Topologically-ordered (TO) phases
are phases which support non-trivial anyonic excitations,
and are characterized, for example, by
non-trivial topological ground state degeneracy
on spatial manifolds with non-trivial topology.
Symmetry-protected topological (SPT) phases
are phases which cannot be adiabatically deformed
to a trivial product state in the presence of certain symmetry,
although they do not support anyonic excitations, and are trivial
once we do not take the symmetry into account. 
In the long wavelength limit, 
topological phases (of both kinds) are 
expected to be described by topological quantum field theories

%
%On the other hand, CFTs describe {\it gapless} states such
%as quantum critical points. 
%So, why CFTs have something to do with topological phases?

\subsection{Boundary CFT and topological phases}

It is noted that BCFT in (1+1)d (BCFT$_2$)
can be used to discuss properties of  
topological phases of matter both in (1+1)d and (2+1)d.
More specifically, the following connections between BCFT$_2$
and topological phases in (1+1) and (2+1) dimensions
have been discussed in the literature:
\begin{itemize}
\item
  SPT$_2$/BCFT$_2$ \cite{ChoLudwigRyuarXiv2016, 2017JPhA50D4002C} \\
  (topological invariants and entanglement spectrum)

\item
  SPT$_3$/BCFT$_2$ \cite{2017PhRvB96l5105H} \\
  (diagnosing anomalies of boundaries of bulk topological phases)
\item 
  TO$_3$/BCFT$_2$
  \cite{2016PhRvB93x5140W,2012PhRvL.108s6402Q, 2017JHEP09056F,
    2017arXiv170604666W}
  \\
  (computation of various entanglement measures such as
  entanglement entropy, mutual information, negativity, etc.)

\end{itemize}

CFTs describe {\it gapless} states such as quantum critical points --
why CFTs have something to do with topological phases, which are fully gapped?
The basic reason is that CFTs can be in proximity to
gapped phases (i) in the phase diagram or (ii) in space. 

In Case (i), BCFT and SPT phases in the same spacetime dimensions are related,
and this is the case for the SPT$_2$/BCFT$_2$ connection,
More specifically,
the connection can be made by
considering a renomalization group (RG) domain wall
\cite{BR}
between an SPT phase and a CFT, which is adjacent to
the SPT phase in the phase diagram. 
From the point of view of the CFT,
the SPT phase serves as a conformal invariant boundary,
and hence the RG domain wall is described by a BCFT.
It is then possible, from the properties of
the corresponding boundary state,
\footnote{
  By construction, this boundary state preserves
  the symmetry of the SPT phase.
  }
to extract
the group cohomology class --
This is consistent with the known classification of
(1+1)d SPT phases protected by unitary on-site symmetry $G$,
which is given by $H^2(G,U(1))$
\cite{2013PhRvB87o5114C}.
\footnote{
  By on-site, we mean
  group elements $g$ in $G$ are all non-spatial symmetry operation.
  Furthermore, if the total Hilbert space $\mathcal{H}$
  is decomposed into the local Hilbert
  space $\mathcal{H}_i$ defined for each lattice site as
  $\mathcal{H}=\mathcal{H}_1 \otimes \cdots \otimes \mathcal{H}_N$,
  then the on-site symmetry operation $g$ is also factorized as $g=g_1 \otimes g_N$ where $g_i$ acts exclusively on $\mathcal{H}_i$.
}
It was also noted that if one considers
the entanglement spectrum of an SPT phase,
which is in the vicinity of a critical point,
the RG domain wall naturally appears.

The connection of type (ii) connecting topological phases and CFT in one-lower dimensions
is commonly called the bulk-boundary correspondence,
and applies to SPT$_3$/BCFT$_2$
and TO$_3$/BCFT$_2$ connections. 
As known in the context of the quantum Hall effect,
topologically non-trivial bulk phases are necessarily
accompanied by gapless edge states, which suffer from
a quantum anomaly of certain kind (gauge and/or modular anomalies).
For a bulk SPT phase, the corresponding edge theory
has a modular anomaly if the symmetry of the SPT phase is
gauged or orbifolded 
\cite{2012PhRvB85x5132R}.
The modular anomaly can be diagnosed, alternatively,
by considering possible boundaries (boundary states)
in the edge theory.
If the bulk is a non-trivial SPT phase,
it should not be possible to make a boundary to the edge theory,
since it lives already on a boundary of a bulk system
in one higher dimensions, as far as the symmetry of
the SPT is strictly enforced.
This is the SPT$_3$/BCFT$_2$ connection.

Finally, in the TO$_3$/BCFT$_2$ connection,
boundary states (Ishibashi boundary states)
appear when we partition a (2+1)-dimensional topological liquid
into two or more parts, 
and consider the reduced density matrix
for a subregion of the total system
by tracing out its complement.
The explicit expression of the density matrix
can be used to compute the various entanglement measures,
such as the von Neumann and Renyi entanglement entropy,
the mutual information, and the entanglement negativity. 
There are contributions to these quantities, which can be
used to indentiy topological order, e.g., 
the topological entanglement entropy.

\subsection{Boundary states as gapped ground states}

All these applications of BCFTs to topological phases
share one common idea: 
boundary states as gapped (1+1) states
\cite{2015JHEP05152M,2017JPhA50n5403K, JC}.

To be more concrete,
let us start from
a CFT described by the (Euclidean) action $S_*$.
One can then perturb $S_*$ by adding a perturbation:
$
S = S_* - \lambda \int d^2x\, \mathcal{O}(x).
$
In the context of the SPT/BCFT connections,
we assume the perturbation respects the symmetry of
the SPT phase in question.
The perturbation, when relevant,
may drive the system into a gapped phase
without breaking the symmetry.
Alternatively, the theory may flow to a different critical point,
or may break the symmetry. 
In the current context, we disregard these possibilities;
In the entanglement spectrum setup considered in
\cite{ChoLudwigRyuarXiv2016, 2017JPhA50D4002C},
these scenarios are known not to occur.

One may then ask
what the ground state of the gapped theory is;
how does the ground state look like within the
Hilbert space of the CFT?
It is postulated that it is given by a boundary state of the CFT.

We should note that, in general,
it may not be easy to determine
which relevant perturbation leads to
which boundary state,
or which boundary state is obtained
by which relevant perturbation. 
In the context of (1+1)d SPT phases, however,
one can identify boundary states
by assigning/computing an SPT topological invariant.
More specifically,
let us consider a (1+1)d SPT phase
protected by unitary on-site symmetry $G$.
Following the above discussion, 
we expect that ground states of
(1+1)d SPT phases may be represented by boundary states in a CFT.
This CFT is not unique, but should be proximate to the SPT phase;
The CFT is either a critical point separating the SPT phase from other gapped
phases or it represents a critical intermediate phase proximate to the SPT phase.
Then, in \cite{2017JPhA50D4002C},
it is claimed that
$
	g |B\rangle_h
	=
	\varepsilon_B(g|h)
	| B\rangle_h.
  $
Here, $g,h\in G$,
$gh = hg$,
and
$| B\rangle_h$ is the
boundary state in the $h$-twisted sector.
The set of phases $\varepsilon_B(g|h)$ is
the topological invariant of the SPT phase,
and identical to the group cohomology phase.
Furthermore, in \cite{2017JPhA50D4002C},
the topological invariant
$\varepsilon_B(g|h)$ is related to the
symmetry-protected degeneracy of the entanglement spectrum
of SPT phases.

\subsection{Conclusion}

(1+1)-dimensional boundary/defect CFT can provide a convenient tool 
to diagnose and characterize topological phases of matter
both in (1+1) and (2+1)-dimensions.
It would be interesting to explore if
there is a higher dimensional analogue
of the approaches discussed in this article.
In this regard, it should be noted that
the surfaces theories of (3+1)-dimensional SPT phases can be,
when they respect the same symmetry as the bulk,  
either gapless or gapped with topological order.
It is also worth mentioning that,
in addition to boundary states, 
cross cap states are also useful to
study topological phases protected by and enriched with
an orientation reversing symmetry
\cite{2015PhRvB91s5142C}.

%%%%%%%%%%%%%%%%%%%%%%%%%%%%%%%%%%%%%%%%%%%%%%%%%ANDREI

\section[Quantum Dot in Interacting Environments]{Quantum Dot in Interacting Environments\protect\footnote{This section was authored by Natan Andrei from the Department of Physics at Rutgers University, Piscataway, New Jersey, USA 08854. This work was funded by NSF Grant DMR 1410583.}}\label{sec:andrei}
\begin{figure}[h]
\centering
\includegraphics[trim = 20mm 15mm 0mm 10mm, clip, width=0.5\textwidth]{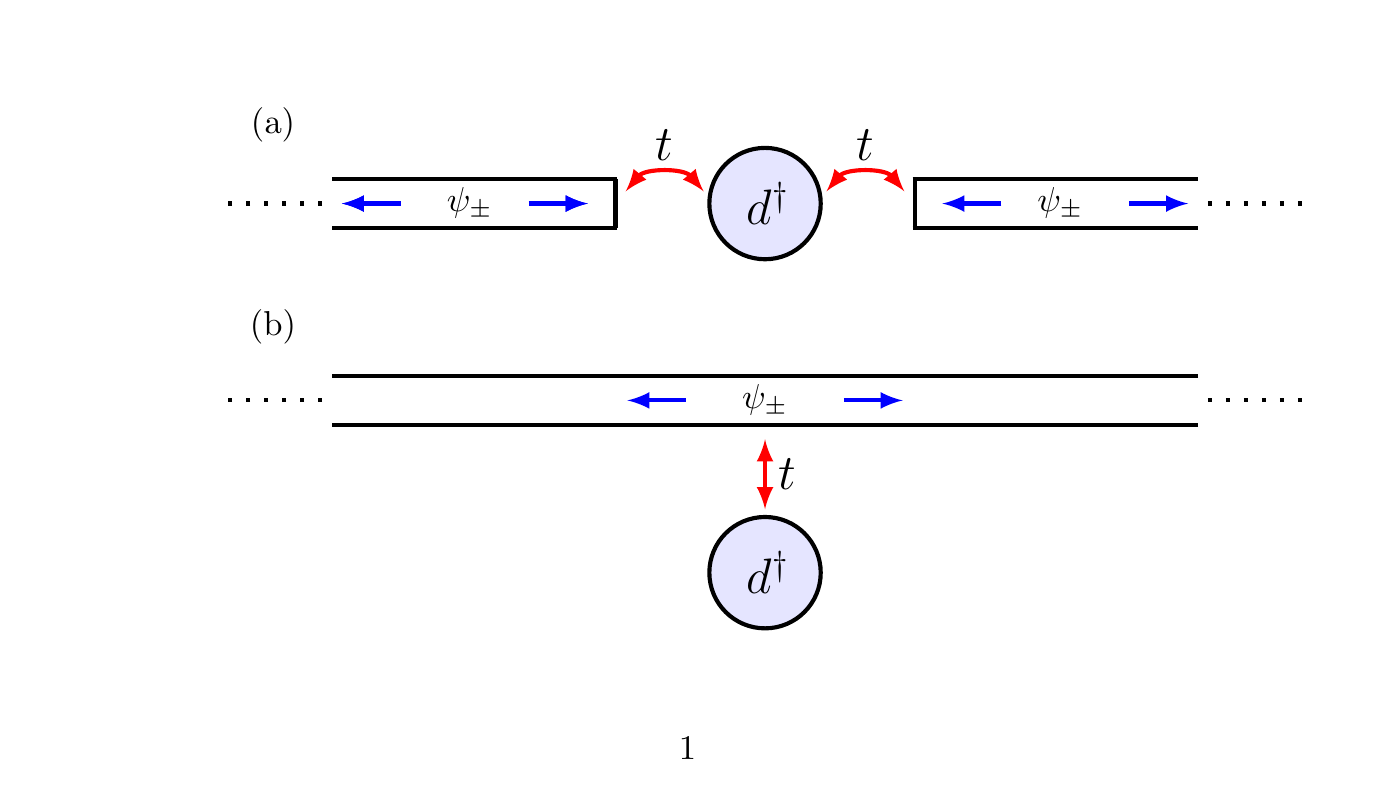}
\caption{We consider two geometries of Luttinger dot system; (a) embedded and (b) side-coupled. The embedded geometry also includes a Coulomb interaction between the dot and leads. Once unfolded the side-coupled and embedded geometries are the same but with the latter containing non local interactions. %\eqref{HLLem}
}
\end{figure}
Coupling a quantum impurity to an interacting one dimensional lead produces some of the most striking phenomena of low dimensional physics. A simple backscattering  impurity is known to cause the wire to be split if the interactions are repulsive while a junction between two leads can lead to perfect conductance in the presence of attractive interactions \cite{KF}. More interesting still are scenarios in which the impurity has internal degrees of freedom. These allow for richer and more exotic phases to appear \cite{TG}. Among these, systems of quantum dots coupled to interacting leads have attracted much attention \cite{KF, FurNag, GogoKom, FurMat, Furusaki, GWB, Duality, Cap, Paata, Meden, Lerner}. The low energy description of the leads is typically given by Luttinger liquid theory which is the effective low energy description of a large number  of  interacting systems  \cite{haldane, TG}.  Here the  individual electrons are dissolved and the excitations are bosonic density modes. In contrast,  the relevant degrees of freedom on the dot are electronic. A competition ensues between the tunneling from lead to dot which is carried out by electrons and the energy cost of reconstituting an electron from the bosons in the lead. 

Such systems are readily achievable in many experimental settings allowing for confrontation of theory with experiment. Luttinger liquids provide the effective description of  carbon nano tubes \cite{LLCNT, JCNT}, fractional quantum Hall edges \cite{LLRMP, FQHLL, FQHLL2}, cold atomic gases  \cite{LLcold, ColdIRL, Jiang, CA} or $^4$He flowing through nano pores \cite{DelMae, Duc} to name but a few. Additionally they are known to describe tunneling processes in higher dimensional resistive leads \cite{Safi, Uni} and more generally are the archetype of a non-Fermi liquid. Luttinger liquid-quantum dot systems have successfully been realized in a number of experiments \cite{qpt, Meb}. These realize the embedded geometry, see Fig. 1(a) of a dot placed between two otherwise disconnected leads.  Measurement of the conductance has revealed interesting non-Fermi liquid scaling as well as Majorana physics.

%\subsection{Models and eigenstates}

We  study the theory of these systems which consist of a  quantum dot attached  to an interacting lead,  a Luttinger liquid,  the attachment being either in   the embedded or the side-coupled geometry. The Hamiltonian of a Luttinger liquid is given by,
\begin{eqnarray}\label{HLLsc}
   H_{\text{LL}}&=&-i  \int  dx (\psi^\dagger_+ \partial_x 
\psi_+   -\psi^\dagger_- \partial_x \psi_-)+ 4g \int   dx  \,\psi_+^\dag(x)\psi^\dag_-(x)\psi_-(x)\psi_+(x)
\end{eqnarray}
where $\psi^\dag_{\pm}$ are right and left moving fermions which interact with a point like interaction of strength $g$ \cite{TG}.  For the side-coupled geometry  we have $x\in [-L/2, L/2]$
while for the embedded geometry we take two Luttinger liquids restricted to  $x\in [-L/2, 0]$ and $x\in [0, L/2]$. The Luttinger model is typically presented in a bosonized form,
\begin{eqnarray}\nonumber
H_{LL}= \frac{1}{4 \pi} \int( K (\nabla \varphi)^2+   \frac{1}{K}\Pi^2) \,dx
 \end{eqnarray}
where $\varphi(x)$ and $\Pi(x)$ are canonically conjugate bosonic fields and $K$ is related to the coupling $g$,
\begin{eqnarray}\label{K}
K=\begin{cases}
1+\frac{\phi}{\pi} &\text{side-coupled}\\
\frac{1}{1-\frac{\phi}{\pi}}& \text{embedded}
\end{cases}
\end{eqnarray}
where  $\phi=-2\arctan{(g)}$. Thus repulsive interaction in the wire, $g>0$, corresponds to $K<1$, attractive interaction to $K>1$, and no interaction to $K=1$.

The quantum dot is modelled by a resonant level with energy $\epsilon_0$ described by,
 \begin{eqnarray}\label{Hd}
  H_{\text{dot}}=\epsilon_0d^\dag d,
  \end{eqnarray}
 coupled to Luttinger liquid via a tunnelling term,
\begin{eqnarray}\label{Ht}
H_t&=& \frac{t}{2} (\psi^\dagger_+(0)+\psi^\dagger_-(0)) d +\text{h.c}
\end{eqnarray}

 \begin{figure}
 \centering
\includegraphics[trim=30 30 30 30,width=.7\textwidth]{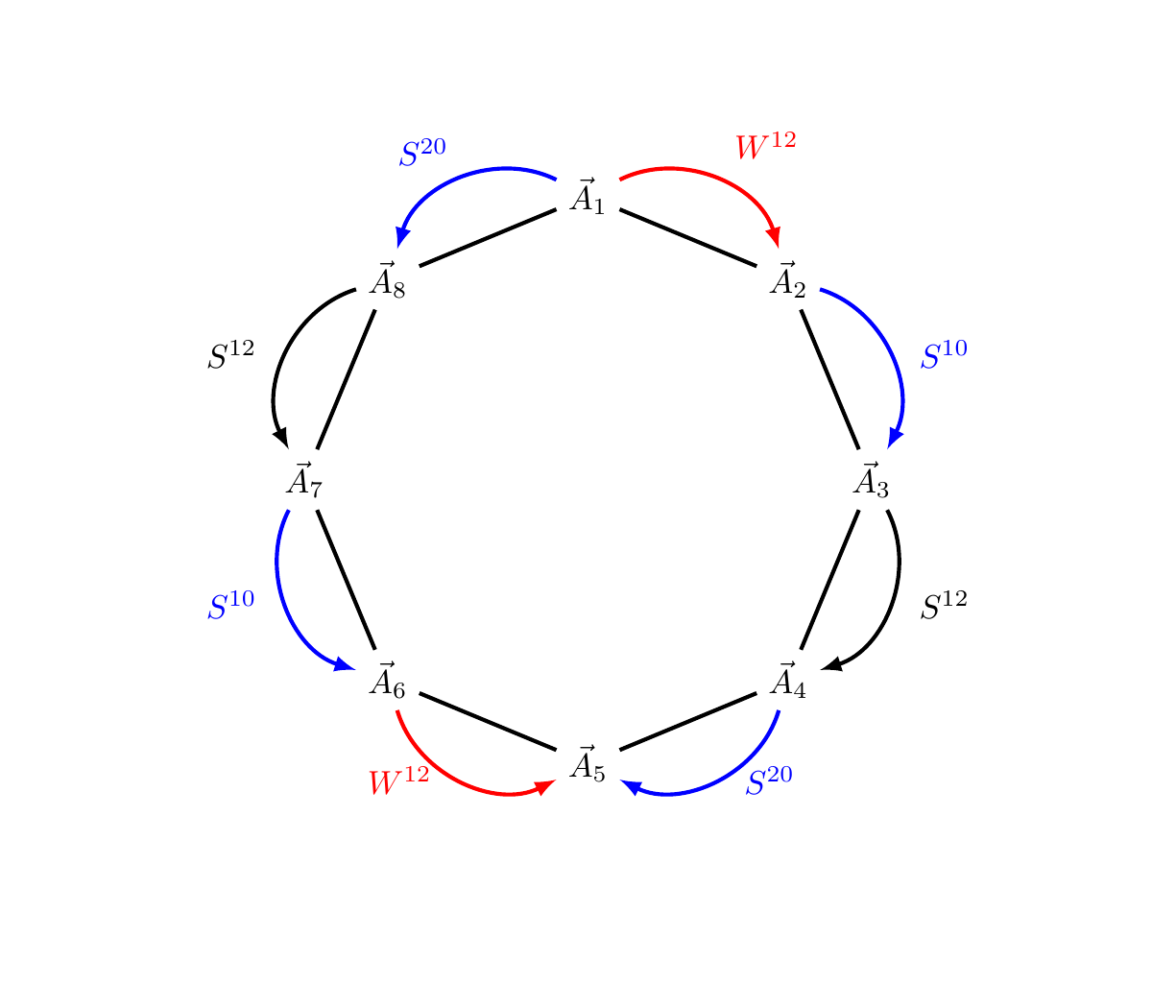}\label{fRE}
\caption{(Color Online) The amplitudes are related by a consistent set of $S$-matrices.}
\end{figure}

The fully interacting Hamiltonian   $H=H_{LL}+H_t+H_{\text{dot}}$ can be studied exactly by means of the Bethe Ansatz approach which allows the  construction of   the full set of  exact eigenstates ,and spectrum, and from it  the determination  of the ground state ($T=0$)  and thermodynamic  ($T>0$)  properties of the system.  It is important to note here that a novel type of Ansatz was necessary,  distinct from that which has been typically used for quantum impurity models \cite{4lectures, TWAKM}. As the problem contains both forward and back scattering we must formulate it in an in-out basis  with the configuration space being partitioned in regions labelled by both the order of the particles and by their  closeness to the origin. One thus partitions the configuration space not to $N!$ regions (orderings of particles) as is conventionally done but to $2^N N!$ regions. The large degeneracy present in the bulk system due to the linear derivative is then used to find  a consistent set of wave functions \cite{ryl}, the consistency assured by a generalised Yang-Baxter relation. The $ 2^2 2!= 8$ amplitudes, for example, describing 2-particle configuration space are related by a consistent set  $S$-matrices deduced from the Hamiltonian and requirement of uniqueness,  see Fig. 2.
 
 \smallskip
 
 Having obtained a consistent and complete set of eigenfunctions one proceeds to impose periodic boundary conditions which lead in turn to a set of equations that determine the momenta $k_j$ and  the energies, $E= \sum_j  k_j$. For details see: C. Rylands and N. Andrei, cond-mat arXiv:1708.07212. Here we present some results  for the dot occupation $n_d= \langle d^{\dagger} d \rangle$ as a function of the level energy $\epsilon_0$ and the temperature $T$, both expressed  with respect to the width $\Gamma$ which serves as the strong coupling scale of the model:
 
   The ground state dot occupation $n_d$ for  attractive interaction is given for for $\epsilon_0<\Gamma$ or $\epsilon_0>\Gamma$ by,
\begin{eqnarray}\label{ng1}
n_d=\begin{cases}
\frac{1}{2}-\left[\sum_{n=0}^\infty a_n\left(\frac{\epsilon_0}{\Gamma}\right)^{2n+1}+b_n\left(\frac{\epsilon_0}{\Gamma}\right)^{(2n+1)/(K-1)}\right]\\
\sum_{n=0}^\infty c_n\left(\frac{\Gamma}{\epsilon_0}\right)^{n+1}~~~~~~~~~~~~~~~~~~~~{\rm for~~}  \Gamma< \epsilon_0
\end{cases}
\end{eqnarray} 
while for repulsive interaction we have,
\begin{eqnarray}\label{nl2}
n_d=\begin{cases}
\frac{1}{2}-\sum_{n=0}^\infty a_n\left(\frac{\epsilon_0}{\Gamma}\right)^{2n+1}       ~~~~~~~~~~~~~{\rm for~~}  \Gamma> \epsilon_0\\
\sum_{n=0}^\infty c_n\left(\frac{\Gamma}{\epsilon_0}\right)^{n+1}+b_n\left(\frac{\Gamma}{\epsilon_0}\right)^{(2n+1)/(1-K)}
\end{cases}
\end{eqnarray} 
where $a_n,b_n$ and $c_n$ are constants. See  FIG 3 and 4.

Beginning with the attractive case, $K>1$ we see that at low energy, $\epsilon_0<\Gamma$ the system is strongly coupled with the dot  becoming hybridized with the bulk. At the low energy fixed point $(\epsilon_0=0)$ the dot is fully hybridized and has $n_d=1/2$. The leading term in the expansion about this is $\epsilon_0/\Gamma$ which indicates that the leading irrelevant operator has dimension 2. We identify it as the stress energy tensor \cite{ZAM2}. The next order term $(\epsilon_0/\Gamma)^{1/(K-1)}$ is due to the backscattering which is generated at low energies but is irrelevant for $K>1$. At high energies, $\epsilon_0>\Gamma$, the system becomes weakly coupled with the fixed point ($\epsilon_0\to\infty$) describing a decoupled empty dot, $n_d=0$. The expansion about this fixed point is in terms of integer powers  indicating  that the tunnelling operator $d^\dag \psi(0)$ has dimension $1/2$. The first few terms of the expansion are plotted in Fig. 3 from which we see that the dot occupation is suppressed as a function of $\epsilon_0$ for $K>1$ as compared to the non interacting case due to the backscattering. 
For the repulsive case,  $K<1$, we see again as for $K>1$, the dot is strongly coupled at low energy and weakly coupled at high energy with the same leading terms in the expansion about these points, however the term generated by the backscattering now appears in the expansion about the high energy fixed point. This stems from the fact that backscattering is relevant for $K<1$ and leads to an enhancement of the dot occupation as compared to the $K=1$ case,  Fig. 4.

  \begin{figure}
  \centering
\includegraphics[width=.475\textwidth]{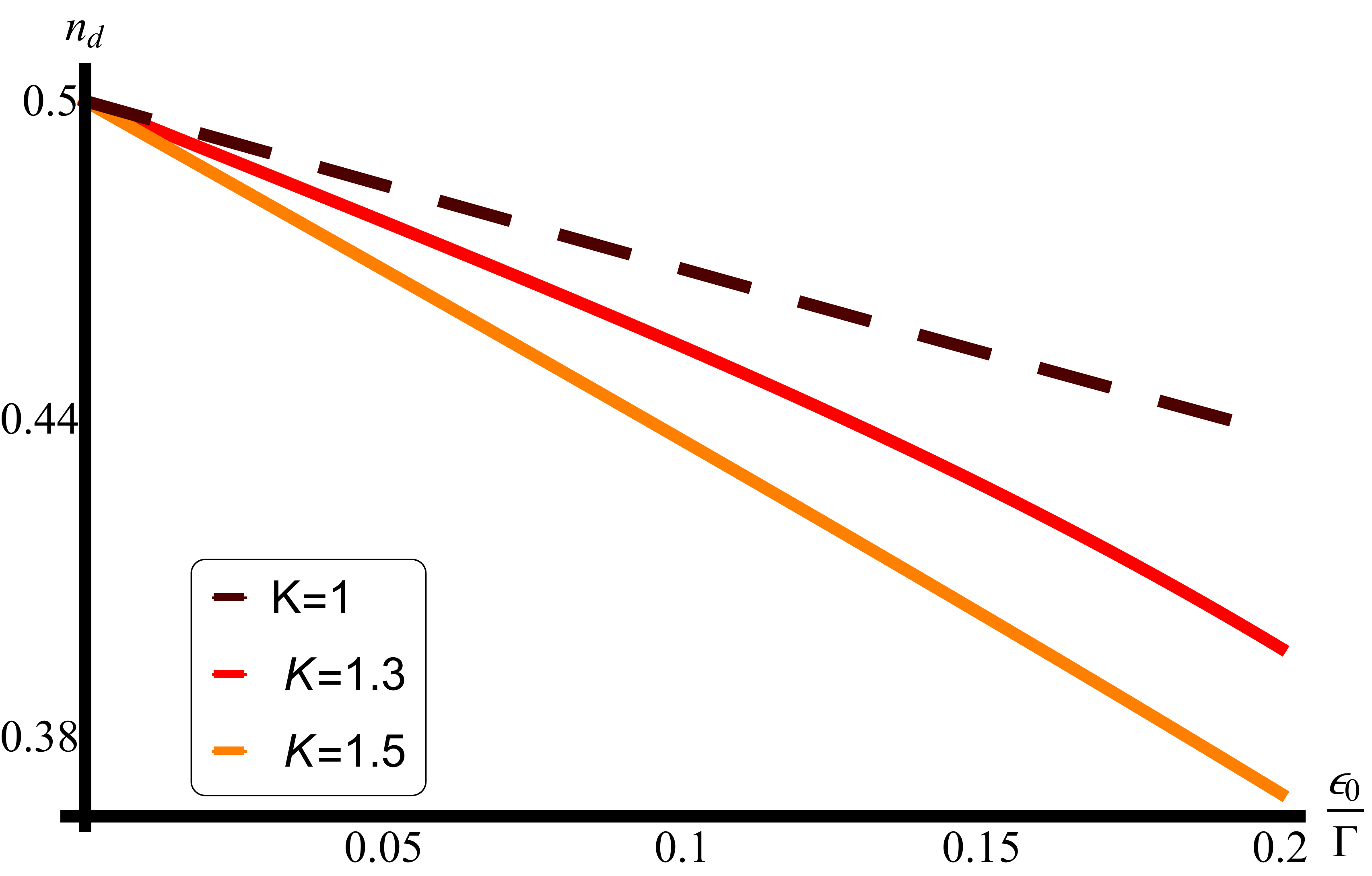}
\includegraphics[width=.475\textwidth]{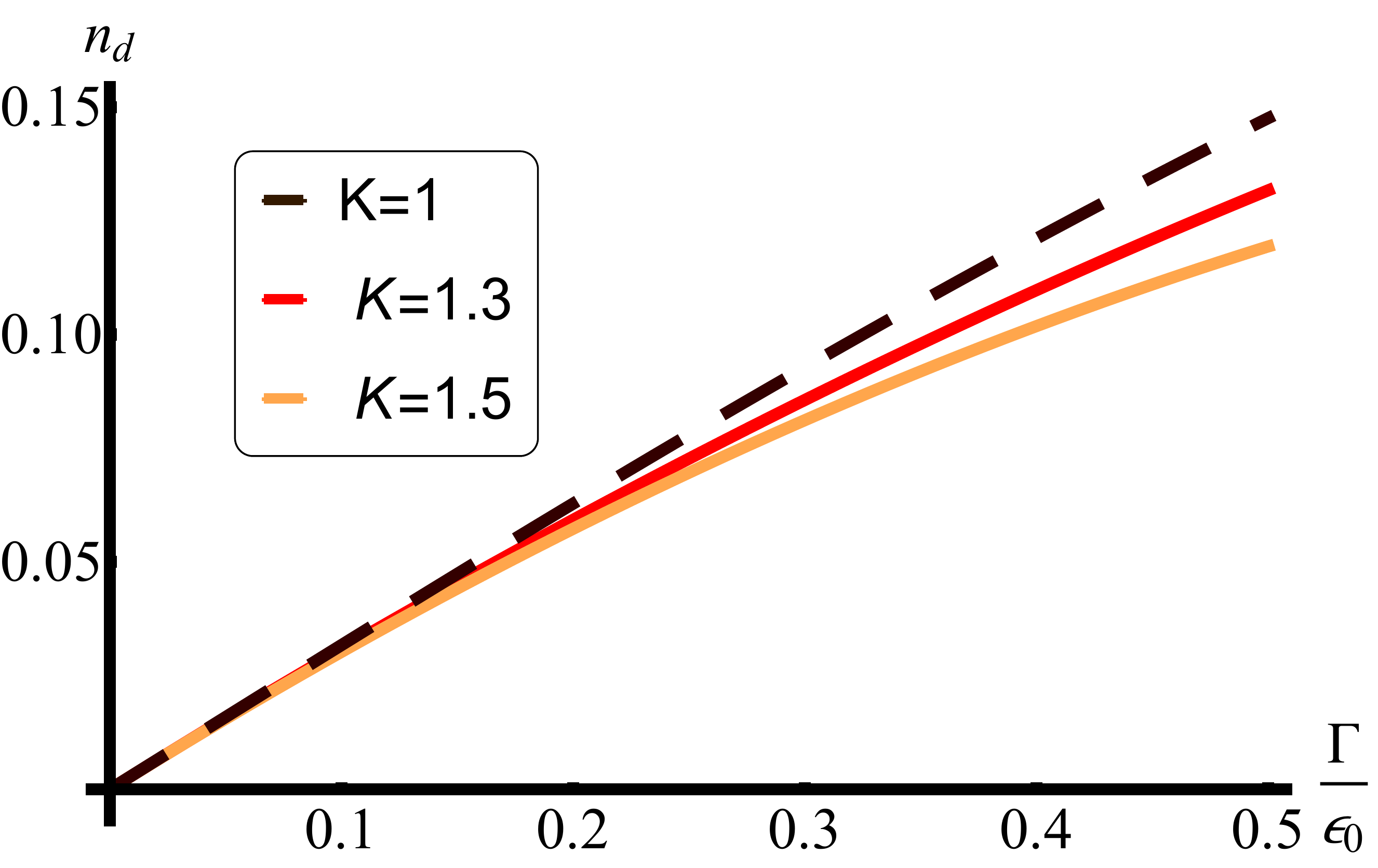}
\caption{(Color Online). The dot occupation at small (left) and large (right) dot energy, $\epsilon_0/\Gamma$, for different values of $K>1$. The effect of attractive interactions is to suppress the dot occupation as compared to the non interacting case (dashed line). This effect becomes stronger for increasing $K$.}
\end{figure} 
\begin{figure}
\centering
\includegraphics[width=.475\textwidth]{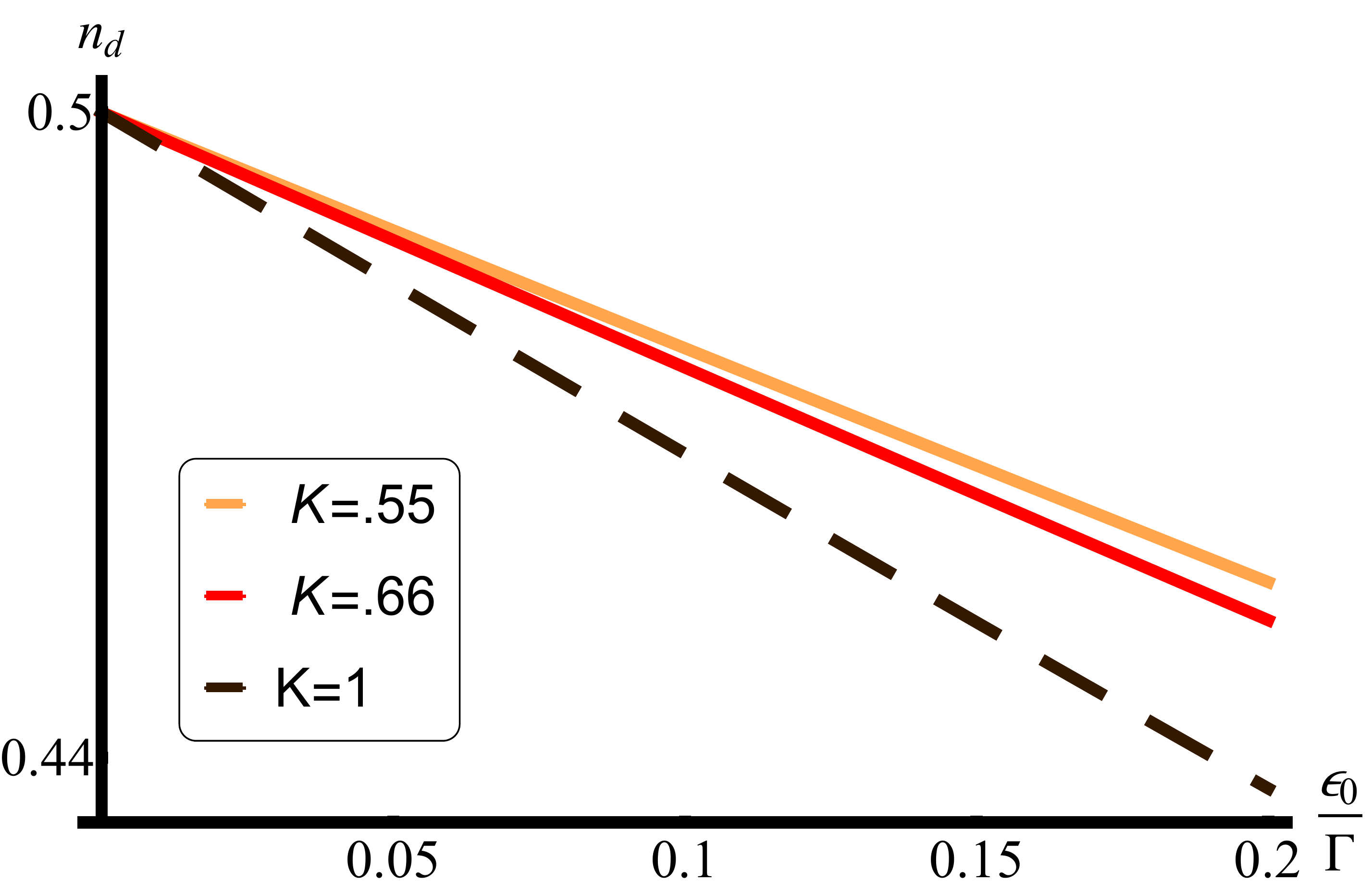}
\includegraphics[width=.475\textwidth]{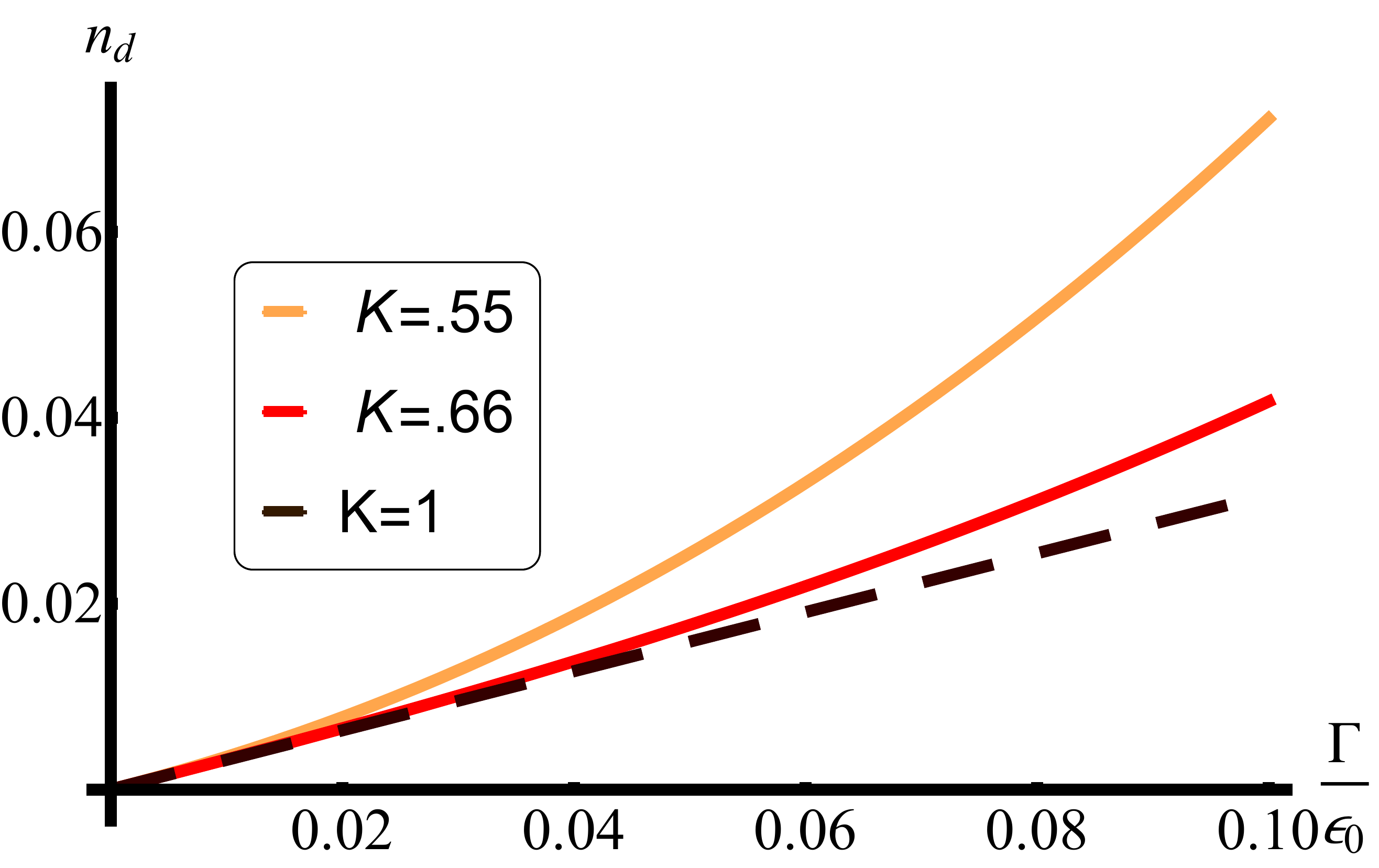}
\caption{(Color Online).The dot occupation at small (left) and large (right) dot energy for different values of $K$. The effect of repulsive interactions $K<1$ is to enhance the dot occupation as compared to the non interacting case (dashed line) with the effect increasing as $K$ decreases.}
\end{figure}

The following RG picture emerges for the side coupled dot: For all $K\in[0,2]$ the system flows from weak coupling at high energy to strong coupling at low energy. The low energy fixed point describes a dot which is fully hybridized with the bulk and has the fixed point occupation $n_d=1/2$. The hybridized dot then acts as a backscattering potential via co-tunnelling. The leading irrelevant operator which perturbs away from the fixed point is the stress energy tensor and results in odd integer powers of $\epsilon_0/\Gamma$ in the dot occupation. For $K>1$ the backscattering is irrelevant which gives rise to odd powers of  $(\epsilon_0/\Gamma)^{1/(K-1)}$ resulting in a suppression of the dot occupation at $\epsilon_0>0$.  For $K<1$ on the other hand it is relevant and generates no other terms in the expansion. The high energy fixed point describes a decoupled dot which has $n_d=0$ for $\epsilon_0\to \infty$ or $n_d=1$ for $\epsilon_0\to-\infty$. By reducing the energy scale we flow away from the fixed point with the tunnelling operator $d^\dag\psi_\pm(0)$ which is the leading relevant operator and has dimension $1/2$ as in the free model. This give rise to integer powers of $\Gamma/\epsilon_0$ in $n_d$. Additionally when $K<1$ backscattering is relevant and causes  odd powers of $(\Gamma/\epsilon_0)^{1/(1-K)}$ to appear resulting in an enhancement of the dot occupation .

\begin{figure}
\centering
\includegraphics[width=.8\textwidth]{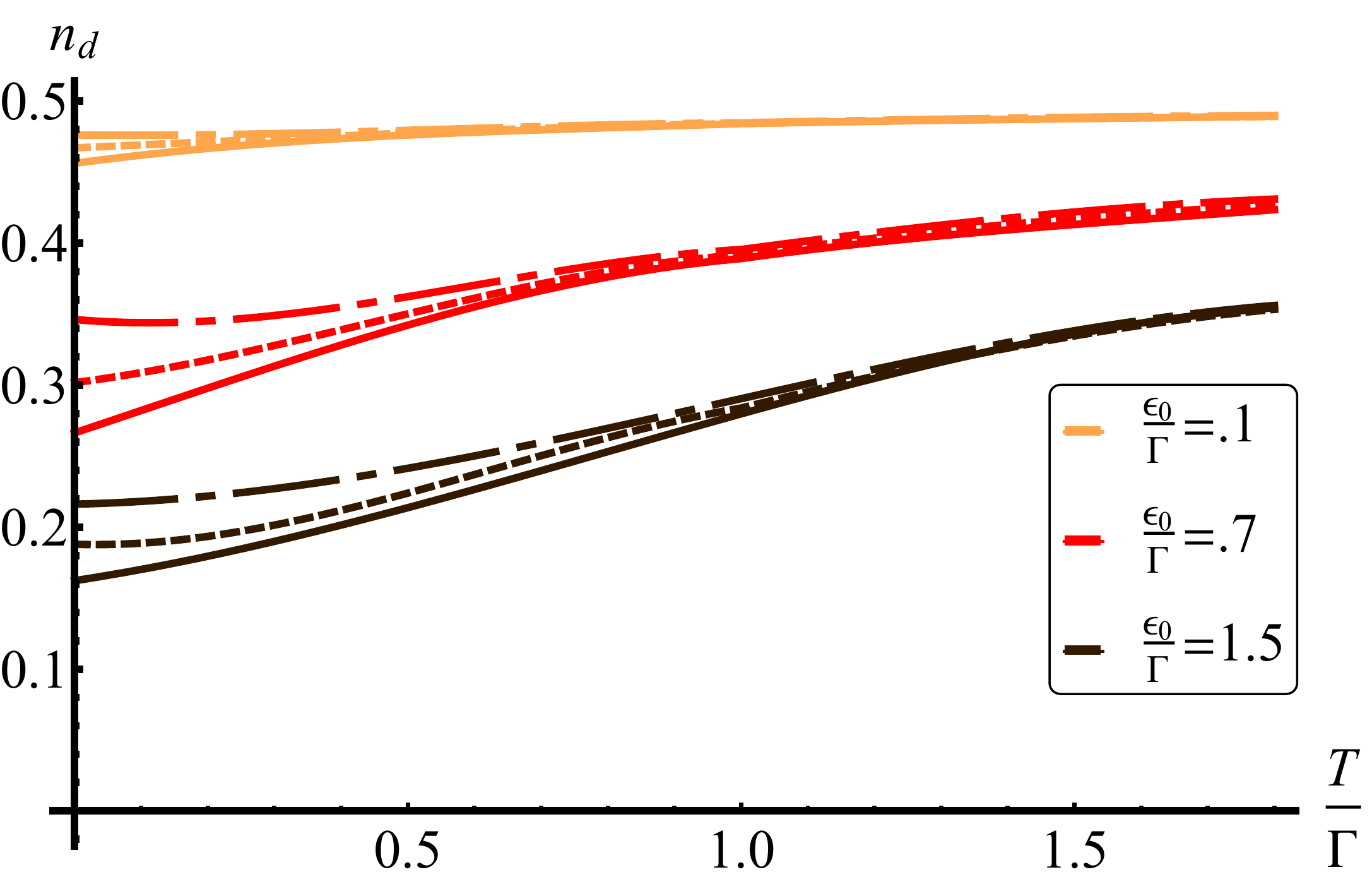}
\caption{(Color Online): The dot occupation for fixed $\epsilon_o/\Gamma$ as a function of temperature. The interaction is taken to be $K=\frac{4}{3}$ (dot-dashed lines), $K=1$ (dashed lines) and $K=\frac{2}{3}$ (solid lines). We see the enhancement and suppression of the dot occupation for repulsive and attractive interaction with the effect most pronounced as the temperature is lowered.}
\end{figure}
 
 Now turning to the study of the system at finite temperature we obtained the thermodynamic Bethe Ansatz equations following the approach of Yang-Yang and Takahashi. 
Using these we can  check the RG picture we arrived at earlier using the ground state dot occupation still holds true at finite temperature. We find that   the temperature  again  is measured with respect to  the level width for the model $\Gamma$ which serves as the strong coupling scale . Thus the system is strongly coupled at  low temperature $T\ll\Gamma$ and weakly coupled at high temperature $T\gg\Gamma$. We also obtained  $g$-function of the model, defined to be the difference in the UV and IR entropy of the impurity
\begin{eqnarray}
g=S_{\text{UV}}-S_{\text{IR}}=\log{2}+\frac{1}{2}\log{\left(\frac{1}{K}\right)}.
\end{eqnarray}
This is always positive for the range of values considered in agreement with the requirement that as we move along the RG flow by lowering the temperature,  massless degrees of freedom are integrated out. The first term comes from the charge degrees of freedom and corresponds to the entropy of a decoupled dot at high temperature which is fully hybridised at low temperature. The second term comes from the chiral degrees of freedom and is the same as for the Kane-Fisher model of a back scattering impurity\cite{FSW, ryl}. Note however that although $g>0$,  the second term which is due  to the backscattering,  is negative for $K>1$. This relative sign between the charge and chiral terms is related to the competition between the tunnelling and the backscattering.We see from this that at high temperature the dot is decoupled and as $T$ is lowered it becomes hybridised with the dot whereupon it acts as a back scattering impurity. In the non interacting limit the $K\to1$ this last term disappears and we recover the expected result. 

We may go beyond the fixed point behaviour to get the leading order corrections and determine the specific heat. The low temperature specific heat is  found to be
\begin{eqnarray}
C_v\sim \frac{T}{\Gamma} + \rm{corrections}
\end{eqnarray}
 which agrees with the expectation that the irrelevant operator is the stress energy tensor. The corrections are still to be calculated but the exponents are expected to agree with those found earlier.

The lack of fine tuned parameters in the side-coupled model make it a good candidate for experimental realizations. Such a system may be created placing a quantum dot near a carbon nanotube,  the edge of a quantum Hall sample or a topological insulator. The dot occupation can then be measured by means of a quantum point contact and compared to our results.

%%%%%%%%%%%%%%%%%%%%%%%%%%%%%%%%%%%%%%%%%%%%%%%%%ERDMENGER
%%%% Article title to be placed here
\section[A Holographic Kondo model]{A Holographic Kondo model\protect\footnote{This section was authored by Johana Erdmenger from the Institute for Theoretical Physics and
  Astrophysics at Julius-Maximilians-Universit\"at W\"urzburg, Am Hubland 97074 W\"urzburg, Germany. The talk was based on joint work  with Andy O'Bannon, Carlos Hoyos, Mario Flory, Max Newrzella, Jackson Wu and  Ioannis Papadimitriou. }}\label{sec:erdmenger}

\subsection{The Kondo model and its gravity dual}

The original model of Kondo \cite{Kondo} describes a SU(2) spin impurity
interacting with a free electron gas. It is given by the Hamiltonian
\begin{equation}
\label{KondoHamiltonian} 
H = \frac{v_F}{2 \pi} i \psi^\dagger \pr_x \psi \, + \,
\frac{v_F}{2}  \lambda_K \delta(x)  {J}  \cdot {S} \,
,
\end{equation}
with $v_F$  the Fermi velocity, $\psi$ the electron field,
${J}$ the electron current, ${S}$ the  impurity spin and $\lambda_K$
the Kondo coupling.
 The Kondo Hamiltonian explains
the logarithmic rise of the resistivity with decreasing temperature
observed in metals with magnetic impurities: at low
energies, the impurity is screened by the electrons. From a theoretical
perspective, this behaviour is due to the fact that the beta function
of $\lambda_K$ is negative.

The holographic Kondo model of  \cite{Erdmenger:2013dpa} differs from the
original condensed matter model in that the ambient electrons are strongly coupled among
themselves even before the interaction with the magnetic impurity is turned on. Moreover, the impurity is an $SU(N)$ spin with $N
\rightarrow \infty.$ The ambient degrees of freedom are dual to a
gravity theory in an AdS$_3$ geometry at finite temperature. The
impurity degrees of freedom are dual to an AdS$_2$ subspace. 
 The dual gravity
model corresponds to a holographic RG flow dual to a UV fixed point
perturbed by a marginally relevant double-trace operator, which flows to an IR
fixed point. In addition, in the IR a condensate forms, such that the
model has some similarity to a holographic superconductor
\cite{Hartnoll:2008vx}.  

As found in
\cite{ReadNewns,Coleman}, the Kondo
model simplifies considerably when the rank $N$ of the spin group is
taken to infinity.  In this limit the interaction term ${J} \cdot
{S}$, which involves  two vectors in spin space,
 reduces to a product $\mathcal{OO}^\dagger$ of a
scalar operator $\mathcal{O}$ and its conjugate. $\mathcal{O}$  involves an
electron $\psi$ and an auxiliary 0+1-dimensional fermion field $\chi$.
The latter is
introduced by writing the spin $S$ as a fermion bilinear. It
is found that the operator $\mathcal{O}$ condenses below a critical temperature.

\subsection{Gravity action: properties and applications}

The motivation for establishing a gravity dual of the Kondo model is
twofold. First, it will provide a new application of
gauge/gravity duality relevant to condensed matter physics. Second,
it will provide a gravity dual of a well-understood condensed matter model
with an RG flow, which may provide new insights into the duality's working
mechanisms. Crucially, our
holographic Kondo model will have some features that are distinctly
different from the well-known field theory Kondo model described
above. Most importantly, the 1+1-dimensional electron gas will be
strongly coupled even before considering the interaction with the
impurity. The model thus has some resemblance to a Luttinger liquid coupled to
an impurity spin. Furthermore, in our model the $SU(N)$ spin symmetry will be gauged. 

The holographic Kondo model of  \cite{Erdmenger:2013dpa} is motivated
by a D-brane construction in string theory involving D3-, D7- and
D5-branes. However, a simpler, phenomenological gauge/gravity model captures the essential physics, namely the model with gravity action
\begin{align}
S = &\frac{1}{8 \pi G_N} \int dz dx dt \sqrt{-g}\,  (R - 2 \Lambda) - \frac{N}{4
  \pi} \int\limits_{\mathrm{AdS}_3}  A \wedge {d} A \nonumber\\ & - N \int dx dt
                                                    \sqrt{-g} \left(
                                                    \frac{1}{4}
                                                    f^{mn}
                                                    f_{mn}  + (D^m 
                                                    \Phi)^\dagger (
                                                    D_m \Phi) -
                                                    V(\Phi) 
                                                    \right) \, . \label{action}
\end{align}
Here, $z$ is the AdS holographic coordinate, $x$ is the spatial coordinate
along the boundary and $t$ is time. The bulk defect dual to the impurity is at $x=0$. The
first term is the Einstein-Hilbert action with negative
cosmological constant $\Lambda$. The second term is a Chern-Simons
term for the gauge field $A_\mu$ dual to the electron current
$J^\mu$. We take $A_\mu$ to be Abelian, which implies
that we have only one flavour (or channel) of electrons.  $f_{mn}$ is the field strength of the defect Abelian gauge field $a_m$ with $m \in \{t,z \}$. Its time component $a_t $ is dual to the charge $\chi^\star \chi$, which at the boundary takes the value $Q=
q/N$ with $q$ the dimension of the antisymmetric representation of
the impurity spin. $D_m$ is a covariant derivative, $D_m =
\pr_m + i A_m \Phi - i a_m \Phi$. 
For the complex scalar, we assume its potential takes the simple
form
%\begin{equation} \label{potK}
$V (\Phi^\dagger \Phi) = M^2 \Phi^\dagger \Phi$.
%\end{equation}
We choose $M^2$ such that $\Phi^\dagger \Phi$ is a
relevant operator in the UV. It becomes marginally relevant
when perturbing about the UV fixed point. For the background geometry we take the solution
to the gravity equations of motion the AdS BTZ
black hole, i.e.
\begin{gather} ds^2_{\mathrm{BTZ}} = \frac{1}{z} \left( \frac{1}{h(z)} dz^2 - h(z)
  dt^2 \right)\, , \qquad h(z) = 1 - \frac{z^2}{z_h^2} \, \nonumber,
\end{gather}
where we choose units with AdS radius equal to one, $L=1$, and the horizon location
$z_h$ is related to the temperature $T$ by
%\begin{gather}
$T = {1}/({2 \pi z_h})$.
%\end{gather}
The holographic dictionary for this model appears in table
\ref{table2}. 
\begin{table}
\begin{center}
\begin{tabular}{|l|c|l|}
\hline
{Operator} & & {Gravity field} \\ \hline
Electron current $J$ & $\Leftrightarrow$ & Chern-Simons gauge field $A$
in $AdS_3$ \\  \hline Charge density $q =
\chi^\dagger \chi$  & $\Leftrightarrow$ & 2d gauge field $a$ in $AdS_2$
\\ \hline
Operator ${\cal O} =
\psi^\dagger \chi$  & $\Leftrightarrow$ & 2d complex scalar $\Phi$ in
                                          AdS$_2$  \\
\hline
\end{tabular}
\caption{Field-operator map for the holographic Kondo model.}
\label{table2}
\end{center}
\end{table}
The charge $q$ measures the dimension of the totally antisymmetric spin
representation and thus the number of impurity degrees of freedom.
The equations of motion derived from the action \eqref{action} can be solved
numerically. The results reveal that the operator $\mathcal{O}$ dual to the
field $\Phi$ indeed condenses below a critical temperature. This leads
to the expected screening of impurity degrees of freedom: the impurity charge indeed
decreases as $T$ decreases below the critical temperature.

\subsection{Impurity Entropy}

The concept of holographic entanglement entropy
has proved to be an important ingredient to the
holographic dictionary \cite{Ryu}.  In general, the entanglement entropy
is defined for two Hilbert spaces $\mathcal{H}_A$ and
$\mathcal{H}_B$. In the AdS/CFT correspondence, it is useful to
consider $A$ and $B$ to be two disjunct space regions in the
CFT. Defining the reduced density matrix to be
\begin{gather}
\rho_A = \tr_B \rho \, ,
\end{gather}
where $\rho$ is the density matrix of the entire space, the
entanglement entropy is given by its von Neumann entropy
\begin{gather}
S = - \tr_A \, \rho_A \ln \rho_A \, .
\end{gather}
Ryu and
Takayanagi proposed the holographic dual of the entanglement entropy
to be
\begin{gather} \label{RT}
S = \frac{ \mathrm{Area} \gamma_A}{4 G_{d+1}}  \, ,
\end{gather}
where $G_{d+1}$ is the Newton constant of the dual gravity space and
$\gamma_A$ is the area of the minimal bulk surface whose boundary
coincides with the boundary of region A. For a field theory in 1+1
dimensions, the region A may be taken to be a line of length $\ell$,
and the bulk minimal surface $\gamma_A$  becomes a bulk geodesic joining the two
endpoints of this line. We note that for a 1+1-dimensional CFT
at finite temperature, with the BTZ black hole as gravity dual, it is
found both in the CFT \cite{Calabrese:2004eu} and on the gravity side
\cite{Ryu}  that the
entanglement entropy for a line of length $\ell$ is given by
\begin{gather} \label{SBH}
S_\mathrm{BH} (\ell) = \frac{c}{3} \ln \left( \frac{1}{\pi \epsilon T} \sinh (2 \pi
\ell T) \right) \, ,
\end{gather}
with $\epsilon$ a cut-off parameter.

For the Kondo model, a useful quantity to consider is the {\it impurity
  entropy} which is given by the difference of the entanglement
entropies in presence and in absence of the magnetic impurity,
\begin{gather} \label{Simp}
S_\mathrm{imp} = S_\mathrm{ impurity \; present} - S_\mathrm{impurity
  \; absent} \, . 
\end{gather}

In the previous section we considered the probe limit of the holographic Kondo model,
in which the fields on the AdS$_2$ defect do not backreact on the
AdS$_3$ geometry. However, including the backreaction is necessary in
order to calculate the effect of the defect on the Ryu-Takayanagi
surface. A simple model that achieves this
\cite{Erdmenger:2014xya,Erdmenger:2015spo} consists
of cutting the 2+1-dimensional geometry in two halves at the defect at
$x=0$ and joining these  back together subject to the {\it Israel
  junction condition} \cite{Israel:1966rt}
\begin{gather} \label{Israel}
  K_{\mu \nu}  - \gamma_{\mu \nu} K  = - \frac{\kappa_G}{2} T_{\mu \nu}
  \, ,
\end{gather}
We refer to the joining hypersurface as `brane'. 
In \eqref{Israel}, $\gamma$ and $K$ are the induced metric and extrinsic curvature
at the joining hypersurface extending in $(t,z)$ directions. $T_{\mu \nu}$
is the energy-momentum tensor for the matter fields $a$ and $\Phi$
at the defect, and $\kappa_G $ is the gravitational constant with
$\kappa_G^2 = 8 \pi G_N$. 
The matter fields $\Phi$ and $a$ lead to a non-zero tension on the brane, which
varies with the radial coordinate. The higher the tension on this
brane, the longer the geodesic joining the two endpoints of the
entangling interval will be. A numerical solution of the Israel
junction condition reveals  that the brane tension decreases with
decreasing temperature, which leads to a shorter geodesic. This in
turn leads to a decrease of the impurity entropy \eqref{Simp}. This
decrease is  in agreement with the screening of the
impurity degrees of freedom.

For large entangling
regions $\ell$, we may approximate the
impurity entropy to linear order by noting that the length decrease of
the Ryu-Takayanagi geodesic $\gamma_A$ translates into a decrease of the
entangling region $\ell$ itself. To linear order, this implies that
the entangling region is given by $\ell +D$ in the UV and by $\ell$ in
the IR, for $D \ll \ell$. Using \eqref{SBH} we may thus write for the
difference of the impurity between its UV and IR values
\begin{align}
\Delta S_\mathrm{imp} & = S_\mathrm{BH} (\ell +D) - S_\mathrm{BH}
                        (\ell) \nonumber\\
& \simeq D \cdot \pr_\ell S_\mathrm{BH} (\ell)  = \frac{2 \pi DT}{3}
  \coth (2 \pi \ell T)  \label{diff}
  \, .
\end{align}
It is a non-trivial result that subject to identifying the scale $D$
with the {\it Kondo correlation length} of condensed matter
physics, $D \propto \xi_K$, then the result agrees with previous
field-theory results for the Kondo impurity entropy
\cite{Laflorencie}. 

A more extensive review of the holographic Kondo model presented may
be found in \cite{Erdmenger:2018xqz}. Quantum quenches were studied in \cite{Erdmenger:2016msd}
and correlators and spectral functions in  \cite{Erdmenger:2016vud,Erdmenger:2016jjg}.

%%%%%%%%%%%%%%%%%%%%%%%%%%%%%%%%%%%%%%%%%%%%%%%%%DOREY
\section[Integrability breaking on the boundary]{Integrability breaking on the boundary\protect\footnote{This section was authored by Patrick Dorey and Robert Parini. Patrick Dorey is from The Department of Mathematical Sciences, Durham University, Durham, UK. This work is funded in part by STFC consolidated grant number ST/P000371/1. Robert Parini is in the Department of Mathematics, University of York, York, UK.}}\label{sec:dorey}

The sine-Gordon equation is an initial-value problem for a function $u(x,t)$ in 1+1 dimensions, obeying the equation of motion
\[
u_{tt}-u_{xx}+\sin(u)=0\,.
\]
If space is the full line, $-\infty<x<\infty$, this equation  is very well-known to be
integrable, with a classical spectrum of kinks, antikinks and breathers. If instead space is cut down to the half-line  $-\infty<x\le 0$, a boundary condition must be given at $x=0$ to make the problem well-posed. A natural question is to ask which of these boundary conditions are compatible with integrability, in the sense of preserving the full set of energy-like conservation laws. Even specialising to the case of no additional boundary degrees of freedom, the full answer was only found surprisingly recently, by Ghoshal and Zamolodchikov in 1994 \cite{Ghoshal1993}:
\[
\left. \left[
        u_x + 4K \sin\left(\frac{u-\widehat u}{2}\right)
        \right] \right|_{x=0}
        = 0\,,
\]
where $\hat u$ and $K$ are two free parameters. A number of special cases, including (zero) 
Dirichlet ($u|_{x=0}=0$)
and Neumann ($u_x|_{x=0}=0$),
had been known to be integrable before.
Ghoshal and Zamolodhikov  found their more-general
set via a consideration of the
lowest-spin extra conserved charges in the full-line model; soon after, in 1995,
MacIntyre \cite{MacIntyre1994} showed the existence of an infinite set of
conservation laws for these boundaries.

Scattering from a Ghoshal-Zamolodchikov boundary is simple: kinks and antikinks reflect perfectly, as either kinks or antikinks, with no loss of energy, and hence no production of radiation or breathers.

However, real life is not integrable, and it might be interesting to explore other, non-integrable, possibillities, outwith the Ghoshal-Zamolodchikov set. This is a `minimal' way to break integrability -- just at one point -- and one could be forgiven for thinking that it wouldn't make much difference. This turns out not to be the case.

\begin{figure}[!h]
\centering
\includegraphics[width=2in]{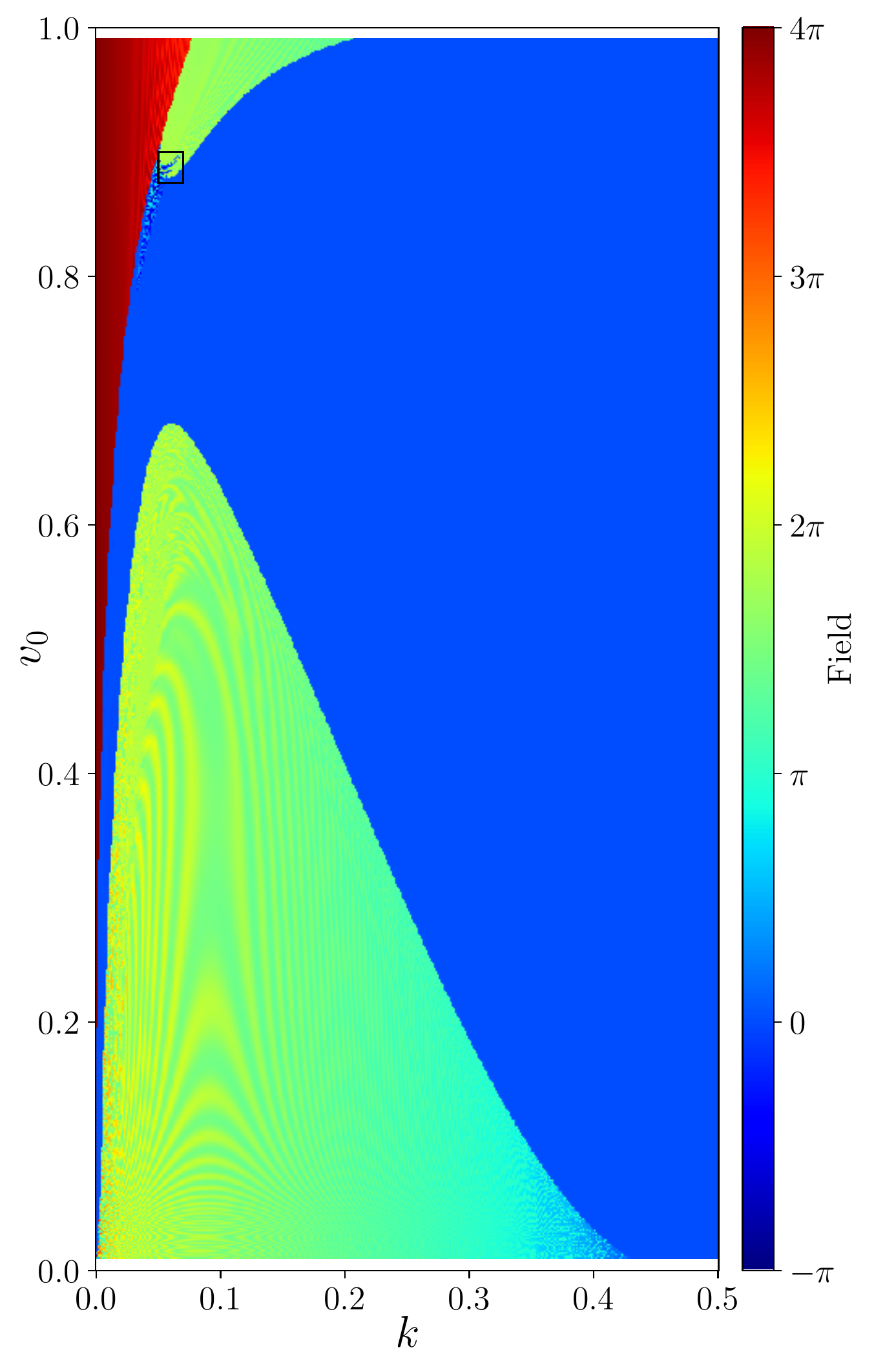}~~~~~
\raisebox{60pt}{\includegraphics[width=2.05in]{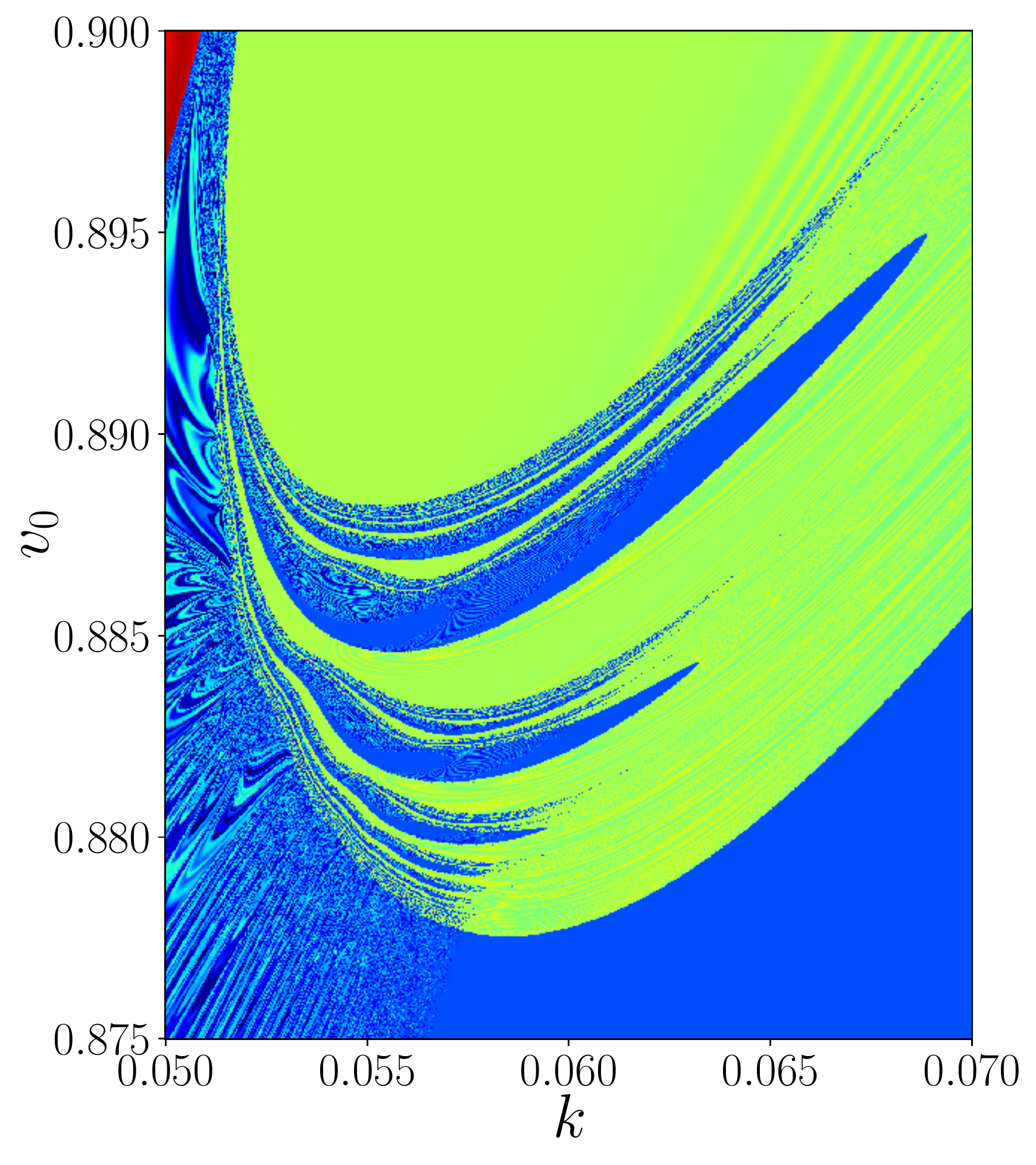}}
%%% where xxxxxx name represents "figurename.eps"
\caption{A `phase diagram' of late-time field values at $x=0$: on the left, a full scan; on the right, a zoomed-in view of the small rectangle in the top left quadrant of the full plot.}
\label{bndry}
\end{figure}

In \cite{Arthur2015} we looked at a natural set of boundary conditions which interpolates between Dirichlet and Neumann in a non-integrable way, namely the one-parameter family of homegeneous Robin boundaries, found by linearising the homogeneous ($\widehat u = 0$) cases of the Ghoshal-Zamolodchikov boundary:
\[
\left. \left[
        u_x + 2k u
        \right] \right|_{x=0}
        = 0.
\]
Setting $k=0$ gives Neumann, while $k\to\infty$ is Dirichlet. Away from these limits, the Robin boundary does not interact nicely with the higher sine-Gordon conserved charges, integrability is lost, and scattering becomes much more complicated. As a first sign of this, figure~\ref{bndry}, taken with small modifications from \cite{Arthur2015}, shows the late-time values $u_{\rm late}$ of the field at $x=0$ for the scattering of an initial sine-Gordon antikink against the Robin boundary, for different initial velocities $v_0$ and parameter values $k$.

For the chosen initial condition, the value of the field at $x=-\infty$ is $2\pi$, and the half-line topological charge of the final configuration is $\tau_{\rm final}=(u_{\rm late}-2\pi)/2\pi$. Roughly speaking, in the red regions, $u_{\rm late}\approx 4\pi$, $\tau_{\rm final}=1$, and the final configuration contains a kink; in the blue regions, $u_{\rm late}\approx 0$, $\tau_{\rm final}=-1$, and the final configuration contains an antikink; and in the light green regions, $u_{\rm late}\approx 2\pi$, $\tau_{\rm final}=0$, and the final configuration contains neither kink nor antikink, or else both of them. Looking down the left and right sides of the left-hand plot shows this is consistent with the known behaviours of the two integrable limits: on the left, $k=0$, the boundary condition is Neumann, and an antikink reflects as a kink; while on the right, $k=0.5$, which is near enough to the $k\to\infty$ (Dirichlet) limit for the initial antikink to be reflected as another antikink. However it is clear that the story is much more complicated in between these two limits. Zooming in to the small rectangular region in the top left corner of the full plot starts to show the extent of this complexity: for a given value of the boundary parameter $k$ in this region, a very small change in the initial velocity of the antikink can cause it to be reflected back as an antikink, or not.

At this stage there are two immediate questions: first, how can we disentangle the full soliton content of the final state? So far we only looked at the net topological charge, but this is blind both to breathers and to additional kink-antikink pairs. Second, what's going on? What is the reason for the complicated, almost fractal, structures present in some parts of figure \ref{bndry}?

For the first question, we found that the `direct' part of the full-line inverse scattering method allowed us to make progress. The key idea is that if we wait sufficiently long after the impact of the initial right-moving antikink on the boundary, all excitations will again be far from the boundary, travelling leftwards in a spacial region where integrability still holds. In other words, some sort of `asymptotic integrability' is at work, whereby integrability is only broken for a finite amount of time. (Unfortunately -- or perhaps not, since this is the ultimate origin of the fine fractal-like structures in figure \ref{bndry} -- this finite amount of time can be arbitrarily large, depending on the initial conditions.) Once all excitations have exited stage left from the boundary region -- something that can be checked numerically by monitoring the amount of energy left near to the boundary -- we can patch the numerically-obtained late-time boundary solution onto a full line, and then compute the scattering data for the linear problem associated with the full-line Lax pair to extract the content of kinks, antikinks and breathers. Full details are in \cite{Arthur2015}, while programs implementing our approach can be found at \cite{Parini2017}.

Concerning the second question, similar `fractal' structures had been observed a long time ago in a nonintegrable field theory in 1+1 dimensions, namely the $\phi^4$ model on the full line (see for example \cite{Campbell:1983xu,Anninos:1991un,Goodman:2007}). In this model, kinks and antikinks attract each other, and also lose some energy to both radiational and vibrational modes when they scatter. This loss of energy means that below some critical impact velocity $v_c$, an incident kink and antikink pair does not have sufficient energy to reseparate after their initial collision, or `bounce'. However if enough of the lost energy has been `parked' in a vibrational mode, then on recollision, subject to a suitable resonance condition, it can be returned to the translational modes, allowing the kink and antikink to escape after one or more subsequent bounces, and leading to an elaborate hierarchy of windows of kink-antikink escape, all lying below $v_c$. This `resonant scattering' mechanism has been generalised in various ways, including to the $\phi^6$ model where the energy is parked not in the vibrational modes of single kinks and antikinks but rather in the vacuum between a suitably-ordered kink-antikink pair \cite{Dorey:2011yw}, and to the $\phi^4$ model on a half-line \cite{Dorey2015}. However it is not immediately obvious how any of this can apply to the sine-Gordon model, where the kinks and antikinks famously do not have any internal vibrational modes. Nevertheless it turns out that a resonance mechanism \textit{is} at work in this model, and is the reason behind the intricate structures visible in figure \ref{bndry}. Even though the sine-Gordon kink has no vibrational modes, the breather \textit{does} oscillate, and in some regimes it is both produced in the initial boundary collision, and also attracted back to the boundary afterwards. This idea was backed up by both numerical and analytical calculations in
\cite{Arthur2015}, but there is plenty of room for further work before a full understanding can be claimed.

There are a number of conclusions to be drawn from all of this. We have seen that classical boundary scattering in the sine-Gordon model is surprisingly rich once integrability is broken at the boundary, though many features of the `phase diagram' of figure \ref{bndry} remain to be understood. To this end, it would be nice to develop a more effective collective-coordinate description of the boundary situation. This is challenging, as the boundary interaction tends to force the excitation of many other modes, but at least while everything is far from the boundary it is possible that integrability will help. So far in our work, this integrability has only been used in a particularly simple-minded way, just to disentangle the final state after the complicated interactions with the non-integrable boundary have been handled numerically. It is possible that more can be done to exploit integrability in the quarter-plane $\{(x,t) : x<0, t>0\}$ which is the domain of our initial-value problem. The so-called Fokas method (see \cite{Peloni2015} for a recent review) is one possible avenue for further progress in this regard. Finally, at the back of our minds throughout this work was the thought to look at the corresponding quantum field theory. In the bulk, there has been some work on the treatment of non-integrable quantum field theories as deformations of integrable theories \cite{DMS1996}. In our non-integrable boundary model, there are some features which may allow for additional progress. In particular,
the `asymptotic integrabilty' mentioned above suggests that the space of asymptotic quantum \textit{in} and \textit{out} states should be the same as that for integrable boundary scattering, making the model a promising half-way house to a full treatment of integrability breaking in quantum field theory, and one where some of the tools from quantum integrability might be relevant to the study of its breakdown.

%%%%%%%%%%%%%%%%%%%%%%%%%%%%%%%%%%%%%%%%%%%%%%%%%WATTS
%%%% Article title to be placed here
\section[Defects in the tricritical Ising Model]{Defects in the tricritical Ising Model\protect\footnote{This section was authored by Gerard M. T. Watts and Isao Makabe from the Department of Mathematics at King's College London, The Strand, London WC2R 2LS, UK. The talk was based on joint work, published in \cite{MW}. }}\label{sec:watts}

\subsection{The problem}
%%%% Insert A head here

The tri-critical Ising model is a minimal model for the Virasoro
algebra with central charge $c=7/10$. A conformal defect in this model
is characterised by an operator $D$ that satisfies   
$$ (L_m - \bar L_{-m})D = D(L_m - \bar L_{-m})$$
The topological defects (which satisfy 
$ L_m D = D L_m\;,$ $\bar L_{-m}D = D\bar L_{-m}$) and factorised
defects (which satisfy $ (L_m - \bar L_{-m}) D = D (L_m - \bar L_{-m})
= 0$) have been classified for all Virasoro minimal models \cite{PZ}; the
challenge is to find non-topological 
non-factorising conformal defects.  

There is evidence for the existence of such defects from both
perturbation theory and from numerical studies using the truncated
conformal space approach but these are not exact results. 
One way to get exact descriptions
is the folding trick: this identifies 
conformal defects in a CFT with  conformal boundary conditions on
CFT$^{\otimes 2}$.  
The problem is that boundary conditions are only classified for $c<1$ and
the central charge of TCIM$^{\otimes 2}$ is $7/5 > 1$.
One way round this is to 
look for a larger symmetry which simplifies the problem.
The tricritical Ising model does not itself have a larger symmetry but it can
be constructed from a superconformal minimal model ($\sTCIM$) with
the same central charge. The folded model $\SVIR2$
again has $c=7/5$ but  
this is now a minimal value for the super Virasoro algebra and the
folded model is itself a minimal model $\sTEN$.
The superconformal boundary conditions for $\sTEN$ should be
classifiable and this could lead to new defects in the tri-critical
Ising model.

This idea was proposed by Gang and Yamaguchi in
\cite{GY}: 
they proposed boundary states for $\sTEN$ which would lead to
defects in TCIM. 
There are however two problems.
Firstly, the defects predicted by Gang and Yamaguchi do not
satisfy the Cardy constraint; secondly, we believe that the defects
are not correctly GSO projected and so are not actually defects in
TCIM.  We have a  new proposal, based in part on ideas of Gaiotto for the
construction of Renormalisation Group defects.
This idea is to use a topological interface $I$ between the TCIM and
the Neveu-Schwarz sector of $\sTCIM$, and then use the folding trick to
obtain defects in $\sTCIM$ as boundary conditions
for the Neveu-Schwarz sector of $\sTEN$. 
In pictures, we obtain a defect $D'$ in \sTCIM\ from a boundary
condition in \SVIR2$ = \sTEN$,

\centerline{\includegraphics[width=3.7in]{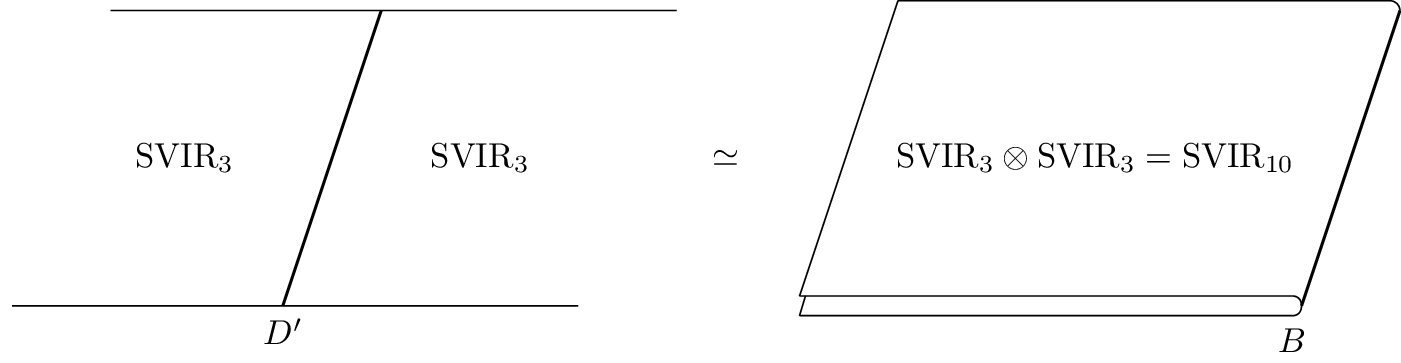}}

%\vspace{3mm}
%\scalebox{0.75}{\begin{tikzpicture}
%	% parameters to shift diagrams
%	\pgfmathsetmacro{\hshift}{4}
%	\pgfmathsetmacro{\vshift}{5}
%
%	\node at (3.25,2-\vshift) {$\simeq$};
%
%	
%	% bottom left part
%	\draw (1-\hshift,3.5-\vshift) -- (6.75-\hshift,3.5-\vshift);
%	\draw (0-\hshift,0.5-\vshift) -- (5.75-\hshift,0.5-\vshift);
%
%	\draw[thick] (2.75-\hshift,0.5-\vshift) -- (3.75-\hshift,3.5-\vshift);
%	
%	\node at (1.75-\hshift,2-\vshift) {\sTCIM};
%	\node at (4.75-\hshift,2-\vshift) {\sTCIM};	
%
%	\node[below] at (2.75-\hshift,0.5-\vshift) {$D'$}; 
%	
%	% bottom right part
%	\draw (1+\hshift,3.6-\vshift) -- (6+\hshift,3.6-\vshift);
%	\draw (0+\hshift,0.6-\vshift) -- (5+\hshift,0.6-\vshift);
%	\draw (0+\hshift,0.4-\vshift) -- (5+\hshift,0.4-\vshift);
%	\draw (0+\hshift,0.4-\vshift) -- (0.06+\hshift,0.6-\vshift);
%	
%	\draw plot [smooth,tension=2] coordinates {(5+\hshift,0.6-\vshift) (5.1+\hshift,0.5-\vshift) (5+\hshift,0.4-\vshift)};
%	\draw plot (6+\hshift,3.6-\vshift) to[out=0,in=90] (6.1+\hshift,3.5-\vshift);
%	
%	\draw (0+\hshift,0.6-\vshift) -- (1+\hshift,3.6-\vshift);
%	\draw[thick] (5.1+\hshift,0.5-\vshift) -- (6.1+\hshift,3.5-\vshift);
%	
%	\node at (3+\hshift,2-\vshift) {$\sTCIM \otimes \sTCIM = \sTEN$};
%	\node[below] at (5+\hshift,0.4-\vshift) {$B$}; 
%	
%\end{tikzpicture}}%

\noindent and, from that, obtain
a defect in TCIM using topological interfaces:

\vspace{3mm}
\centerline{\includegraphics[width=3.7in]{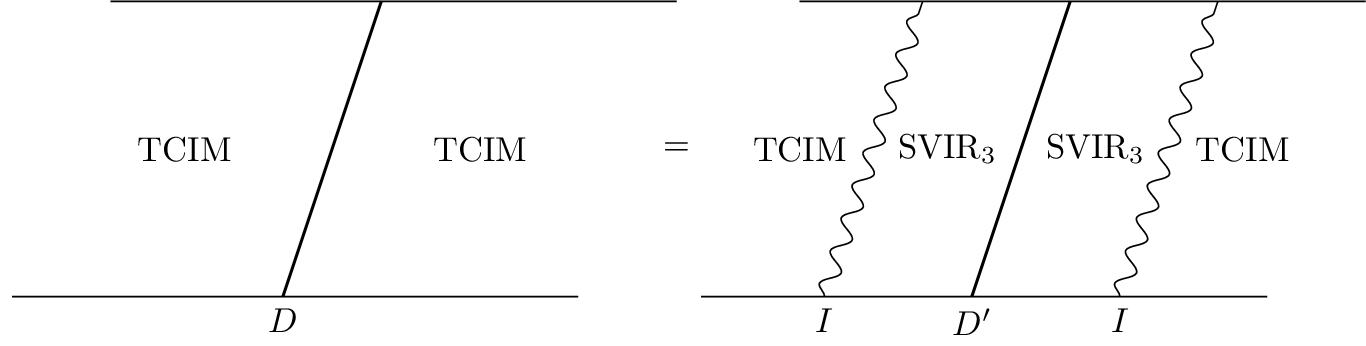}}

%\scalebox{0.75}{\begin{tikzpicture}
%	% parameters to shift diagrams
%	\pgfmathsetmacro{\hshift}{3.5}
%	\pgfmathsetmacro{\vshift}{5}
%
%	% top left part
%	\draw (1-\hshift,3.5) -- (6.75-\hshift,3.5);
%	\draw (0-\hshift,0.5) -- (5.75-\hshift,0.5);
%
%	\draw[thick] (2.75-\hshift,0.5) -- (3.75-\hshift,3.5);
%	
%	\node at (1.75-\hshift,2) {TCIM};
%	\node at (4.75-\hshift,2) {TCIM};	
%
%	\node[below] at (2.75-\hshift,0.5) {$D$}; 
%
%	% equal signs
%	\node at (3.25,2) {$=$};
%	\node at (3.25,2-\vshift) {$\simeq$};
%
%	% top right part
%	\draw (1+\hshift,3.5) -- (6.75+\hshift,3.5);
%	\draw (0+\hshift,0.5) -- (5.75+\hshift,0.5);
%	
%	\draw[snake it] (1.25+\hshift,0.5) -- (2.25+\hshift,3.5);
%	\draw[thick] (2.75+\hshift,0.5) -- (3.75+\hshift,3.5);
%	\draw[snake it] (4.25+\hshift,0.5) -- (5.25+\hshift,3.5);
%	
%	\node at (1+\hshift,2) {TCIM};
%	\node at (2.5+\hshift,2) {\sTCIM};
%	\node at (4+\hshift,2) {\sTCIM};
%	\node at (5.5+\hshift,2) {TCIM};	
%	
%	\node[below] at (1.25+\hshift,0.5) {$I$}; 
%	\node[below] at (2.75+\hshift,0.5) {$D'$}; 
%	\node[below] at (4.25+\hshift,0.5) {$I$}; 
%	
%\end{tikzpicture}
%}
%\end{center}

This has the advantage that we can use the Neveu-Schwarz sector of the
boundary states found by Gang and Yamaguchi (and do not need the
problematic Ramond sector) but has the 
disadvantage that we cannot expect to produce elementary defects in TCIM.

As a check, we first considered the Ising model (the Virasoro minimal
model with $c=1/2$) which is related to
the free fermion in the same way that TCIM is related to \sVir.
This resulted in four fundamental defects in the free fermion, two
topological interfaces between the free fermion and the Ising model,
and the reconstruction of a two-dimensional subspace of the known
defects in the Ising model. 

For $\sTEN$, we found that we could
use the Neveu-Schwarz sectors of the boundary states found by Gang
and Yamaguchi and, 
with some changes to choices of representatives (amongst other
choices) we could make a uniform presentation of these states.
There are 48 boundary states,
\[
\kkett{(a,b)_\NS}
\;,\;\;
  \kkett{(a,b)_\wtNS} 
= (-1)^F \kkett{(a,b)_\NS}
\]
{
They are labelled by a pair of nodes from the Dynkin diagrams of \DS\
and \ES, bi-coloured as here, and are invariant under $b\mapsto r(b)$,
the symmetry of the 
\ES\ diagram. This would be broken by the addition of Ramond sectors,
as originally proposed by Gang and Yamaguchi.

\centerline{\includegraphics[width=3.7in]{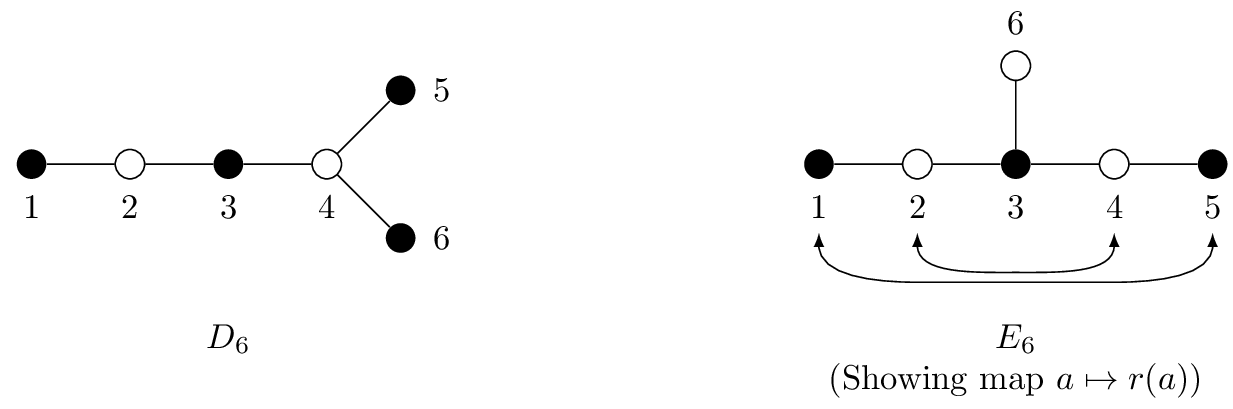}}

We also found that these split into two distinct sets, depending on whether
the colours of the nodes $(a,b)$ are the same or not: we need to use
different embeddings of the superconformal algebra for the two
sets. 
Furthermore, we could show that these boundary states are exactly
those which preserve a super-W-algebra symmetry, $SW(3/2,10)$.
%\newpage

We can calculate easily two properties of the defects:
Firstly, the entropy or $g$-value. 
This decreases along RG flows. The entropy of a superposition of
defects is the sum of their energies.
Secondly,  the transmission coefficient. 
This takes values $0\leq T\leq 1$ in a unitary theory. 
It is 1 for a topological defect and 0 for a factorising defect.
A non-topological non-factorising defect will have $0<T<1$.

We found that defects labelled $(a,1)$ and $(a,6)$ are either
topological or factorising; those labelled $(a,2)$ and $(a,3)$ are
neither. 
The defect of lowest entropy is $(1,2)$ with $g=1.015..$ and
$T=0.633..$ 
This last $g$ value is not the sum of the $g$-values of any
combination of known topological or factorising defects.

There was a problem though: the interfaces that convert a defect in
\sTCIM\ into a defect in TCIM are not nicely related to the
superconformal structure. This means that 
we cannot easily find the overlaps of the new defects. 
Even the overlaps of their constituent Ishibashi states are hard to find.
As an example, when the interfaces are included, a typical overlap
would be 
\[
q^{-\frac 7{120}}\,[
1 + \tfrac 12 q^{\frac 32}+ q^2 + \tfrac 12 q^{\frac 52} + q^3 +
q^{\frac 72} + 
\tfrac{ 49123 }{17689}q^4 + \tfrac 32 q^{\frac 92} + 
\tfrac{ 49123 }{17689} q^5 + 
\tfrac 52 q^{\frac{11}2} + 
\tfrac{102941115}{16999129} q^6 + ...
]
\]
\subsection{Conclusions}

We have found convincing evidence for non-topological non-factorising
conformal defects in TCIM.
We have not been able to find the partition function owing the
breaking of superconformal symmetry by the Interface operators.
We are currently calculating the perturbative corrections to the
transmission coefficients in minimal models to compare with the values
found here.
We would like to look further into the use of extended algebras to
construct new conformal defects.
It would be interesting to derive the boundary conditions for the free
fermion and \sTCIM\ from first principles or using the methods of
Novak and Runkel \cite{NR}.

\enlargethispage{20pt}

%\aucontribute{
%Talk given by G.M.T. Watts, 
%based on joint work, published in \cite{MW}. This work was fully joint
%work and both authors gave final approval for publication.
%}

%%%%%%%%%%%%%%%%%%%%%%%%%%%%%%%%%%%%%%%%%%%%%%%%%DRUKKER
\section[Energy-momentum multiplets for supersymmetric 
defects and the displacement operator]{Energy-momentum multiplets for supersymmetric 
defects and the displacement operator\protect\footnote{This section was authored by Nadav Drukker from the Department of Mathematics, King's College London, The Strand, London WC2R 2LS, UK. This talk was based on work with D.~Martelli and I.~Shamir \cite{DMS}.}}\label{sec:drukker}
\subsection{Bosonic example}
%%%% Insert A head here

Let us start with a simple bosonic example, as motivation. 
We take a 4d scalar $\phi$ and a 3d scalar $a$, with the latter 
confined to a planar submanifold $\Sigma$. The 
4d and 3d actions are
\begin{align}
\int {\cal L}^{(4)}
&
=\int \left(-\frac12\partial^\mu\phi\partial_\mu\phi-V_4(\phi)\right),
\\
\int_\Sigma {\cal L}^{(3)}
&
=\int_\Sigma\left(-\frac12\partial^ia\partial_ia-V_3(a)\right).
\end{align}

The energy-momentum tensor includes contributions from both 4d and 3d fields. 
and requires the embedding 
$\delta_\mu^i$ on the directions tangent to $\Sigma$
\begin{equation} \label{Tmn_comb_app}
T_{\mu\nu}=T^{(4)}_{\mu\nu}+\delta(x^n)\delta_\mu^i\delta_\nu^jT^{(3)}_{ji},
\end{equation}
where $x^n$ is the coordinate normal to $\Sigma$ and 
\begin{align}
T^{(4)}_{\mu\nu}
&
=\partial_\mu\phi\partial_\nu\phi + \eta_{\mu\nu} {\cal L}^{(4)}
\\
T^{(3)}_{ij}
&
=\partial_ia\partial_ja + \eta_{ij}{\cal L}^{(3)}.
\end{align}
Using the classical equations of motion one finds
\begin{equation}
\partial^\mu T_{\mu\nu} =0\,.
\end{equation}

To make the system interesting, we need to couple the 3d and 4d fields. The simplest 
way to do that is
\begin{equation}
\int_\Sigma {\cal L} ^{(I)}
=-\int_\Sigma V_I(\phi,a),
\end{equation}
with an arbitrary coupling potential $V_I$.

Now the 3d term in the energy-moment tensor is
\begin{equation}
T^{(3)}_{ij}=\partial_ia\partial_ja + \eta_{ij}({\cal L}^{(3)}+{\cal L}^{(I)}).
\end{equation}
Repeating the calculation, we now find a violation of conservation
\begin{equation}
\label{bos}
\partial^\mu T_{\mu\nu} = n_\nu \delta(x^n)\,d,
\qquad
d=\partial_n V_I(\phi,a).
\end{equation}

\subsection{The $\cal S$ multiplet and defects}

Any 4d theory with ${\cal N}=1$ supersymmetry has an ${\cal S}$-multiplet, containing the 
energy momentum tensor \cite{KS}. In the presence of a defect we propose a modification to its 
equation by the addition of the last term as 
\begin{equation}
\bar D^{\dot\alpha} {\cal S}_{\alpha\dot\alpha}
=2(\chi_\alpha - D_\alpha X)+\delta(\tilde y^n){\cal Z}_\alpha\,,
\qquad
{\cal Z}_\alpha=
\Sigma_\alpha + \tilde\Theta_\alpha \Sigma + (\Gamma^i \tilde\Theta)_\alpha \Sigma_i
\end{equation}
Here ${\cal S}_{\alpha\dot\alpha}$ is a real vector superfield, 
$X$ and $\chi_\alpha$ are chiral superfields and 
$D^\alpha \chi_\alpha = \bar D_{\dot\alpha} \bar \chi^{\dot\alpha}$. 
The coordinates in the direction normal to the defect are 
$\tilde{y}^n = x^n - 2 i \tilde{\Theta} \Theta - 2 \tilde{\Theta}^2$ and 
$\tilde{\Theta}_A = \frac 1 2 (\lambda_A^\alpha \theta_\alpha 
+ \bar{\lambda}_{A}^{\dot\alpha} \bar{\theta}_{\dot{\alpha}})$. 
Finally $\Gamma$ is a 3d gamma matrix.

In the $\tilde\Theta$ expansion of $\cal S$ one finds at the linear level
\begin{equation}
\tilde \Delta_A {\cal S}_\mu |_{\tilde\Theta=0}
=
-(\lambda_A^\alpha S_{\alpha \mu}
  + \bar{\lambda}_{A}^{\dot{\alpha}} \bar{S}_{\mu \dot{\alpha}}) -4 i (\Gamma^j \Theta)_A T_{j \mu} + \dots
\end{equation}
with $S$ and $T$ the supercurrent and energy-momentum tensor, respectively. 
The components appearing here are those that should be preserved even in the presence 
of the defect. 
A straightforward computation then leads to 
\begin{equation}
\begin{aligned}
%\tilde{\Delta}_A \partial^\mu \mathcal{S}_\mu |_{\tilde\Theta=0}
%&=
\delta(x^n) \left. \left(- \frac i 4 \Delta^B \Delta_A (\Sigma_B + \bar{\Sigma}_B) - \frac 1 4 (\Gamma^i \Delta)_A (\Sigma_i + \bar{\Sigma}_i) \right)\right|_{\tilde\Theta=0}
%\\
%&\quad
+ \frac i 2 \partial_n \left. \left( \delta(x^n) (\Sigma_A - \bar{\Sigma}_A) \right) \right|_{\tilde\Theta=0}.
\end{aligned}
\end{equation}
So for consistency we should sent the first parenthesis to zero, which imposes conditions 
on $\Sigma$.

The sub-multiplet of ${\cal S}$ containing the violations is given by
\begin{equation}
\mathcal{S}_\mu\vert_{\tilde{\Theta} = 0}
= -i \Theta^A (\lambda_A^\alpha S_{\alpha \mu}
- \bar{\lambda}_{A\dot{\alpha}} \bar{S}_\mu^{\dot{\alpha}})
- 2 i \Theta^2 T_{n \mu} + \dots
\end{equation}
The derivative is
\begin{equation}
\partial^\mu \mathcal{S}_\mu\vert_{\tilde{\Theta} = 0}
=
-\frac 1 4 \delta(x^n) \Delta^A (\mathcal{Z}_A - \bar{\mathcal{Z}}_A)\vert_{\tilde{\Theta} = 0}
- \frac i 4 \delta(x^n) (\bar D \bar \sigma_n {\cal Z} - D \sigma_n\bar {\cal Z})\vert_{\tilde{\Theta} = 0}.
\end{equation}
Since $T_{n \mu}$ appears in the $\Theta^2$ component of ${\cal S}_\mu$, only the 
second term contributes to the displacement and is given by the scalar $\Sigma$ piece
\begin{equation}
\partial^\mu T_{n \mu} \sim \delta(x^n) \Delta^2 \Sigma
\end{equation}
like in the bosonic case (\ref{bos}). In fact, if we supersymmetrise the above 
model we will find exactly the same expression (plus the fermion contribution).

%%%%%%%%%%%%%%%%%%%%%%%%%%%%%%%%%%%%%%%%%%%%%%%%%KRISTJANSEN
\section[Boundary and Defect Conformal Field Theory]{Boundary and Defect Conformal Field Theory\protect\footnote{This section was authored by Charlotte Kristjansen from Niels Bohr Institute at Copenhagen University, DK-2100 Copenhagen \O, Denmark. This work was supported by the Sapere Aude Top Researcher program of the DFF-FNU through grant number DFF-4002-00037 }}\label{sec:kristjansen}
\subsection{The defect set-up}

An interesting 4D defect or boundary conform\-al field theory can be found within  AdS/CFT~\cite{Karch:2000ct,DeWolfe:2001pq,Erdmenger:2002ex,Constable:1999ac}.
The theory is a special version of ${\cal N}=4$ super Yang-Mills theory (SYM) where a co-dimension one defect has been inserted at $x_3=0$
and separates two regions of space-time, $x_3>0$ and $x_3<0$, where the rank of the gauge group is $SU(N)$ and $SU(N-k)$ respectively. The difference in gauge group is implemented by considering 
${\cal N}=4$ SYM  with a classical solution where three of
the scalar fields are non-vanishing and space-time dependent on one side of the defect, $x_3>0$
\begin{align}\label{classical}
\phi_i^{\text{cl}} = -\frac{1}{x_3} t_i \oplus 0_{(N-k)\times (N-k)}, \hspace{0.5cm} i=1,2,3,
\end{align}
where the $t_i$ constitute a $k$-dimensional irreducible representation of $SU(2)$.
The non-trivial classical solution leads to a partial breaking of conformal as well as super symmetry. The conformal symmetry is reduced
from  $SO(2,4)$  to $SO(2,3)$ and the R-symmetry is reduced from $SO(6)$ to $SO(3)\times SO(3)$. The full symmetry group of the dCFT is $OSp(4|4)$.

The above dCFT has a dual string theory description where a probe D5-brane has been embedded in the usual $AdS_5\times S^5$ background, the embedding being such that the D5-brane geometry is $AdS_4\times S^2$ and there is a background
gauge field with $k$ units of magnetic flux through the $S^2$.

\subsection{Integrable one-point functions}
Defect conformal field theories exhibit novel features compared to conformal field theories without defects, one of these being
the possibility of non-trivial one-point functions.  Symmetries constrain the one-point functions to be of the form
\begin{align}
\label{eq: form of the one-point function}
 \langle \mathcal{O}_{\Delta}(x)\rangle=\frac{C}{x_3^{\Delta}},
\end{align}
where $\Delta$ is the conformal dimension.  One-point functions of the above described defect 
version of ${\cal N}=4$ SYM show very strong signs of integrability. Making use of a  boundary state in the form of a specific
matrix product state (MPS) and invoking the tools of  integrable spin chains one can derive a closed expression for the
one-point functions at tree level in the full scalar sector.  As is well-known, conformal  operators built from only two complex fields, 
which we will choose as $\phi_1+i\phi_4$ and $\phi_2+ i\phi_5$, can be identified with the eigenstates $|\mathbf{u}\rangle$ of the Heisenberg spin chain. Accordingly, one can express the corresponding one-point functions
(up to a trivial field theoretical pre-factor) as~\cite{deLeeuw:2015hxa}
\begin{align}
C_k = \frac{\langle\text{MPS}|\mathbf{u} \rangle}{\sqrt{\langle\mathbf{u} |\mathbf{u} \rangle}}, \hspace{0.5cm}
 |\text{MPS}\rangle= \text{tr} \prod_{n=1}^L  \Big[t_1 \otimes |\!\uparrow\rangle_n  +  t_2 \otimes|\!\downarrow\rangle_n\Big],
\end{align}
where $k$ refers to the dimension of the representation for the $t_i$ and $L$ to the length of the operator (spin chain state). One
then finds that the one-point function for $k>2$ can be expressed as 
\begin{align}\label{eq:SU2genK}
C_k =i^L T_{k-1}(0)\sqrt{\frac{Q(\frac{i}{2})Q(0)}{Q^2(\frac{ik}{2})} }\sqrt{\frac{\det  G_+}{\det  G_-}},
\end{align}
where $T_{k-1}$ is the transfer matrix of the Heisenberg spin chain in the $k$-dimensional representation, $Q(u)$ is the Baxter polynomial, and 
the matrices $G_\pm$ are related to the Gaudin norm of the Bethe state in question as $\langle \mathbf{u}|\mathbf{u}\rangle
\propto\det G_+ \det G_-$. For $k=2$ one can show that the matrix product state is cohomologically equivalent to a raised version
of the 
N\'eel state~\cite{deLeeuw:2015hxa,Buhl-Mortensen:2015gfd} and the result for $C_2$ can be read off from~\cite{Brockmann}.  The result~(\ref{eq:SU2genK}) for general $k$ can then be proven by
recursion~\cite{Buhl-Mortensen:2015gfd}. The formulas above can be extended to the $SU(3)$ sector~\cite{deLeeuw:2016umh} as well as to the full scalar $SO(6)$ sector~\cite{deLeeuw:2018mkd}. Furthermore, in the case of the $SU(2)$ sector, one can likewise find a closed expression for the
one-loop contribution to the one-point function~\cite{Buhl-Mortensen:2017ind}. 
Finally, based on the observed integrability structure at tree level and one-loop 
one can make an educated guess for an all loop asymptotic formula for the one-point function of a certain chiral primary,
a formula which turns out to agree up to wrapping order 
with the result of a supergravity calculation carried out in a certain double scaling limit~\cite{Buhl-Mortensen:2017ind}, cf.\  section~\ref{Quantum-check}. Overlaps of the type above between a spin chain
eigenstate and an initial state in the form of a matrix product state 
are of relevance for the study of quantum quenches. Recently a possible characterisation of  integrable initial states
was given in~\cite{Piroli:2017sei}.

\subsection{A quantum check of AdS/dCFT\label{Quantum-check}}
Whereas the usual AdS/CFT system is described in terms of only two parameters, the 't Hooft coupling, $\lambda$, and the rank of the gauge group, $N$,
one has in the defect set-up an extra tunable parameter, $k$. This fact makes it possible to impose on top of the
usual planar limit,  $N\rightarrow \infty$, a certain double scaling limit~\cite{Nagasaki:2011ue} 
\begin{align}
\lambda\to\infty, \hspace{0.5cm} k\to \infty,  \hspace{0.5cm} \frac{\lambda}{k^2}\,\,\text{ fixed}.
\end{align}
On the string theory side, considering $\lambda\rightarrow \infty$ makes possible a supergravity approximation in which it turns out that the expectation values of certain simple observables organise into an expansion in positive powers the double scaling parameter 
$\lambda/k^2$. Hence, these expectation values can be compared to the result of a standard perturbative field theoretical
computation thus making possible a comparison of string and field theory in a situation where both supersymmetry and conformal
symmetry are partially broken. A comparison based on tree-level results in the field theory gives a match to leading order in the
double scaling parameter~\cite{Nagasaki:2012re} 
but the string theory computation also gives rise to predictions about higher loop field theory. For instance,
in the case of the chiral primary operator ${\cal O}_L=\mbox{Tr} (\phi_1+i\phi_4)^L$ string theory predicts
\begin{align}\label{st}
%\left.
\frac{\langle {\cal O}_L\rangle_{\text{1-loop}}}{\langle {\cal O}_L\rangle_{\text{tree-level}}}
%\right|_{\text{string}}
=\frac{\lambda}{4\pi^2 k^2} \frac{L(L+1)}{L-1}.
\end{align}
Setting up the program for perturbative computations in the defect CFT is quite non-trivial due to the classical fields. This
task was carried out in~\cite{Buhl-Mortensen:2016pxs,Buhl-Mortensen:2016jqo}. Expanding around the classical fields one gets not only new cubic interaction terms but also quadratic terms which mix both colour and flavour of the ${\cal N}=4 $ SYM fields and which in addition carry $x_3$-dependence.  The diagonalisation of the quadratic
terms can be dealt with using fuzzy spherical harmonics, except for a few terms which can be eliminated through
gauge-fixing. The resulting masses are still $x_3$-dependent but this dependence can be dealt with by considering
the corresponding propagators as propagators in an auxiliary $AdS_4$ space. This $AdS_4$-space has the defect
as its boundary and  $x_3$ as its radial coordinate. In order to preserve supersymmetry special care is needed in the 
choice of boundary conditions at the defect for the fields that stay massless. Furthermore, it is crucial that one chooses a
supersymmetry preserving regulator when carrying out higher loop computations. When these various precautions are 
met, one indeed obtains a match with the string theory prediction~({\ref{st}) thus achieving   a positive quantum
check of AdS/CFT in a system where both conformal symmetry and supersymmetry are partially broken~~\cite{Buhl-Mortensen:2016pxs,Buhl-Mortensen:2016jqo}.

\subsection{Data-mining using defect one- and two-point functions}
 
Two novel types of two-point functions are possible in the defect set-up namely two-point functions between 
bulk operators of unequal
conformal dimensions and two-point functions involving a bulk field and a defect field.  These two-point functions are constrained
by symmetries to be of the form
\begin{align}
\langle \mathcal{O}_i(x) \mathcal{O}_j(y) \rangle = \frac{f_{ij}(\xi)}{(x_3)^{\Delta_i}(y_3)^{\Delta_j}},
\hspace{0.5cm}
 \langle \mathcal{O}_i(x) \hat{\mathcal{O}}_j(\vec{y}) \rangle =\frac{\mu_{ij}}{(x_3)^{\Delta_{i}-\Delta_{j}}|x-(\vec{y},0)|^{2\Delta_{j}}},
\end{align}
where $f_{ij}(\xi)$ is a function of the conformal cross ratio $\xi=\frac{|x -y|^2}{4 x_3 y_3}$ and where $\mu_{ij}$ is denoted as
a bulk-to-boundary coupling.
The bulk two-point functions obey a crossing relation much like four-point functions in an ordinary CFT and this relation gives rise
to  a boundary conformal bootstrap equation~\cite{Liendo:2012hy,Liendo:2016ymz}. In one channel the bulk two point function
in the presence of the defect can be expressed in terms of one-point functions and  
three-point functions of the theory without the defect. In the other channel the two-point function can be expressed in terms  of one-point functions  and bulk-to-boundary couplings. This in particular means that from the knowledge of just one- and two-point functions
of the defect set-up one can extract three-point functions of the theory without defect as well  as bulk-to-boundary couplings.
A number of examples of this type of data-mining was given in~\cite{deLeeuw:2017dkd}.

\subsection{Outlook}

The idea of using one- and two-point functions as input for the boundary conformal bootstrap equations is only in its infancy.
Taking it further would potentially be very interesting but in order to fully exploit it for the present  dCFT a better understanding of the 3D theory living on the defect is needed. So far the action on the defect has only been explicitly written down for
$k=0$\cite{DeWolfe:2001pq}. As regards the integrability properties of one-point functions an interesting question is whether
a closed formula can be found even at tree-level for the closely related $SO(5)$ symmetric D3-D7-brane based dCFT.  For this field theory one-point
functions have been shown to vanish in the $SU(2)$ and $SU(3)$ sub-sectors. One-point functions are non-trivial when the
full scalar $SO(6)$ sector is considered but a closed expression has not yet been found~\cite{deLeeuw:2016ofj} except
in the case of chiral primaries~\cite{Kristjansen:2012tn}. Finally, an 
open problem is to perform a string theoretical computation of the one-point function of a non-protected operator.

%%%%%%%%%% Insert bibliography here %%%%%%%%%%%%%%

\vskip2pc

%\noindent {\bf Please follow the coding for references as shown below.}

%%%%%%%%%%%%%%%%%%%%%%%%%%%%%%%%%%%%%%%%%%%%%%%%%BISSI
\section[Loop Corrections to Supergravity on $AdS_5 \times S^5$]{Loop Corrections to Supergravity on $AdS_5 \times S^5$\protect\footnote{This section was authored by Agnese Bissi from the Department of Physics and Astronomy at Uppsala University, Box 516, SE-75120 Uppsala, Sweden. This work was supported in part by the Knut and Alice Wallenberg Foundation under grant KAW 2016.0129.} }\label{sec:bissi}

The $AdS/CFT$ correspondence relates four-dimensional ${\cal N}=4$ Super Yang-Mills (SYM) to type IIB string theory on $AdS_5 \times S^5$ \cite{Maldacena:1997re,Gubser:1998bc,Witten:1998qj}. In particular, single-trace chiral primary operators (CPO) of weight $p$ and transforming under the representation $[0,p,0]$ of the $SU(4)$ $R-$symmetry group, ${\cal O}_p$, map to supergravity fields with mass $m^2=p(p-4)$. Two and three-point correlators of arbitrary CPOs are non renomalized, thus are determined by the free field content of the $\mathcal{N}$ = 4 SYM and do not acquire quantum corrections. Four-point correlators of CPOs are much richer observables, they contain dynamical and generically coupling dependent information. In the limit of large $\lambda=g_{YM}^2 N$ four-point correlator of the stress-tensor multiplet (corresponding to $p=2$) were computed long ago \cite{Freedman:1998tz, DHoker:1998vkc}. Recently, an elegant algorithm based on symmetries and consistency conditions to determine the four-point correlator of arbitrary CPOs was proposed in \cite{Rastelli:2016nze, Rastelli:2017udc}, see also \cite{Dolan:2006ec,Uruchurtu:2011wh}. In this short note we present subleading corrections in $1/N$ to correlators in the large t' Hooft coupling regime, corresponding to quantum corrections on the gravity side \cite{Alday:2017xua}. \\

The analytic bootstrap was initiated in \cite{Komargodski:2012ek,Fitzpatrick:2012yx} and developed into a powerful algebraic machinery in \cite{Alday:2015eya, Alday:2015ewa, Alday:2016njk,Caron-Huot:2017vep}. This method allows determining all orders of the large spin expansion of dynamical quantities, such as anomalous dimensions and OPE coefficients, by knowing some specific singularities of the four point correlator. We apply this procedure to compute $\frac{1}{N^4}$ corrections to the anomalous dimension of double trace operators of bare dimension $4+\ell$ and spin $\ell$. For the first few values of the spin the results are
\begin{eqnarray*}
\Delta_{0,2}&=& 6 - \frac{4}{N^2}- \frac{45}{N^4} + \cdots\\
\Delta_{0,4}&=& 8 - \frac{48}{25}\frac{1}{N^2}-\frac{12768}{3125}\frac{1}{N^4}+\cdots
\end{eqnarray*}

\subsection{Details}
The main ingredient is the four point function of CPOs ${\cal O}_{2}$ 
\begin{equation}
\nonumber
\langle {\cal O}_{2}(x_1) {\cal O}_{2}(x_2) {\cal O}_{2}(x_3) {\cal O}_{2}(x_4)\rangle = \sum_{\cal R} \frac{{\cal G}^{({\cal R})}(u,v)}{x_{12}^4x_{34}^4}
\end{equation}
where the sum runs over representations in the tensor product $[0,2,0] \times [0,2,0]$ and $u= \frac{x_{12}^2 x_{34}^2}{x_{13}^2 x_{24}^2}, v= \frac{x_{14}^2 x_{23}^2}{x_{13}^2 x_{24}^2}$ are the cross-ratios. 
Superconformal symmetry allows writing all contributions ${\cal G}^{({\cal R})}(u,v)$ in terms of a single non-protected function ${\cal G}(u,v)$ satisfying the following crossing relation
\begin{equation}
v^2 {\cal G}(u,v) -u^2 {\cal G}(v,u) + 4(u^2-v^2) + \frac{4(u-v)}{c}=0
\end{equation}
where $c=\frac{N^2-1}{4}$ is the central charge, see \cite{Beem:2013qxa,Beem:2016wfs} for a detailed discussion. It is possible to write ${\cal G}(u,v)=\mathcal{H}(u,v)+{\cal G}^{short}(u,v)$ in such a way to disentangle the contribution from operators belonging to long and semishort- short multiplets ($1/4$ and $1/2$ BPS operators), respectively. Since both the dimensions and OPE coefficients of the latter are non renormalizable, the function ${\cal G}^{short}(u,v)$ is completely fixed and known in closed form. Instead $\mathcal{H}(u,v)$ admits a decomposition in superconformal blocks
\begin{equation}
\mathcal{H}(u,v) = \sum_{\tau,\ell} a_{\tau,\ell} u^{\tau/2} g_{\tau+4,\ell}(u,v)
\end{equation}
where the sum runs over superconformal primary operators in long multiplets, which are singlet of $SU(4)$, with twist (dimension minus the spin) $\tau$ and even spin $\ell$, and $a_{\tau,\ell}$ denotes the squared OPE coefficients. In the limit of infinite central charge $\mathcal{H}(u,v)$ reduces to the generalised free fields result $\mathcal{H}^{(0)}(u,v)$, which agrees with the large $c$ result in the Born approximation (free theory). The intermediate operators correspond to double-trace operators of twist $\tau_n=4+2n$ and the OPE coefficients 
\begin{equation}
a^{(0)}_{n,\ell}=\frac{\pi (\ell+1)  (\ell+2 n+6) \Gamma (n+3) \Gamma (\ell+n+4)}{2^{2 \ell+4 n+9} \Gamma \left(n+\frac{5}{2}\right) \Gamma \left(\ell+n+\frac{7}{2}\right)}
\end{equation}
The four-point correlator, as well as the twist and the OPE coefficient, can be expanded for large central charge $c$, 
\begin{eqnarray}
\label{dataexpansion}
\tau_{n,\ell} &=& 4+2n + \frac{1}{c} \gamma^{(1)}_{n,\ell} + \frac{1}{c^2} \gamma^{(2)}_{n,\ell} + \cdots\\
a_{n,\ell}&=&a^{(0)}_{n,\ell}+\frac{1}{c} a^{(1)}_{n,\ell}+\frac{1}{c^2} a^{(2)}_{n,\ell}+\cdots \nonumber
\end{eqnarray}
At order $1/c$ and in the limit of large $\lambda$, there is no new operators appearing in the OPE and $\mathcal{H}^{(1)}(u,v)$ can be computed from the classical supergravity result \cite{Dolan:2006ec,Rastelli:2016nze}, with correction to the spectrum and OPE coefficients \cite{DHoker:1999mic, Arutyunov:2000ku, Alday:2014tsa}
\begin{eqnarray}
 \gamma^{(1)}_{n,\ell} &=& -\frac{\kappa_n}{(1+\ell)(6+\ell+2n)},\\
 a_{n,\ell}^{(1)} &=& \frac{1}{2} \partial_n\left(  a_{n,\ell}^{(0)}  \gamma^{(1)}_{n,\ell} \right)
\end{eqnarray}
where $\kappa_n=(n+1)(n+2)(n+3)(n+4)$. For a given $n$ and $\ell$ there is more than one superconformal primary in the singlet of $SU(4)$, except for $n=0$. The above corrections should then be interpreted as (weighted-)averages. At order $1/c^2$ since ${\cal G}^{short}(u,v)$ receives contributions only up to order $1/c$, the crossing equation for $\mathcal{H}^{(2)}(u,v)$ reads
\begin{equation}
\label{crossing}
v^2 \mathcal{H}^{(2)}(u,v) = u^2 \mathcal{H}^{(2)}(v,u).
\end{equation}
We will follow the strategy in \cite{Aharony:2016dwx}: determine the piece proportional to $\log^2u$ in $\mathcal{H}^{(2)}(u,v)$ from the CFT-data at order $1/c$. By crossing symmetry this leads to a precise divergence proportional to $\log^2 v$. Matching this divergence  fixes  $\gamma^{(2)}_{n,\ell}$ and  $a^{(2)}_{n,\ell}$ to all orders in $1/\ell$. Plugging (\ref{dataexpansion}) into the conformal block decomposition and expanding up to order $1/c^2$, we find that the piece proportional to $\log^2u$ is
\begin{align}
\label{H2log}
\left. \mathcal{H}^{(2)}(u,v) \right|_{\log^2u} = \sum_{n,\ell}  \frac{1}{8}a^{(0)}_{n,\ell}  (\gamma^{(1)}_{n,\ell} )^2  u^{2+n} g_{n,\ell}(u,v).
\end{align}
where $g_{n,\ell}(u,v) \equiv g_{\tau_{n}^{(0)}+4,\ell}(u,v)$. %
An obstacle to compute this is the mixing among double-trace operators $[{\cal O}_2,{\cal O}_2]_{n,\ell}, [{\cal O}_3,{\cal O}_3]_{n-1,\ell}, \dots$. They have the same twist and spin at zeroth order and transform under the same representation of $SU(4)$, hence quantities above should be interpreted as averages, weighted by their respective OPE coefficient at zeroth order. Therefore, the weighted average $\langle (\gamma^{(1)}_{n,\ell} )^2\rangle$ does not follow from the leading order result, except for $n=0$, for which there is a unique state. This problem can be solved by considering the complete family of four-point correlators $\langle {\cal O}_p{\cal O}_p{\cal O}_q{\cal O}_q \rangle$ in generalised free field theory and in the supergravity approximation. This allows computing $\langle (\gamma^{(1)}_{n,\ell} )^2\rangle$ which is given by
\begin{equation}
\langle (\gamma^{(1)}_{n,\ell} )^2\rangle =\sum_{p=2}^\infty \frac{\alpha_p \kappa_n^2}{(J^2-(n+2)(n+3))^2}
 \prod_{k=2}^{p-1}\frac{ (n-k+2)(n+k+3)}{(J^2-k(k+1))}
\end{equation}
where $\alpha_p=p^2(p^2-1)/12$ and $J^2=(\ell+n+3)(\ell+n+4)$. Each term inside the sum represents the contribution from the $p-$th KK mode, or more precisely the intermediate double-trace operators $[{\cal O}_p,{\cal O}_p]$. Crossing symmetry relates this term to a specific combination involving $\gamma_{n,\ell}^{(2)}$. More precisely, it is possible to compute all the terms in the large spin expansion of $\gamma_{0,\ell}^{(2)}$ and to resum them \footnote{Results for $n >0$ are discussed in \cite{Alday:2017vkk}}. The resulting contribution for each KK-mode is
\begin{eqnarray}
\left. \gamma_{n,\ell}^{(2)}-\frac{1}{2}\gamma_{n,\ell}^{(1)}\partial_n \gamma_{n,\ell}^{(1)}\right|_{n=0, p} =\alpha_p \frac{P^{(14+2\ell)}(p)}{(p^2-4)(p^2-1)p}+\alpha_p (p^2-4)(p^2-1)p^3Q^{(4+2\ell)}(p) \psi^{(2)}(p) \nonumber
\end{eqnarray}
for some polynomials $P,Q$. At large $p$ we find
\begin{eqnarray}
\nonumber
\left. \hat \gamma^{(2)}_{0,\ell} \right|_{p} \sim \frac{\alpha_p}{p^{3+2\ell}}
\end{eqnarray}
since $\alpha_p \sim p^4$, this implies the sum over $p$ is actually divergent for spin zero. This agrees with the presence of a quadratic divergence in the 10d supergravity computation \cite{Metsaev:1987ju}, see a detailed discussion in \cite{Alday:2017vkk}. For spin two and higher we get a convergent sum which leads to the results quoted in the introduction.

%%%%%%%%%%%%%%% End of first page %%%%%%%%%%%%%%%%%%%%%

\subsection{Conclusion}
We have reported results for the CFT-data of unprotected operators in ${\cal N}=4$ SYM to order $1/N^4$ and at large 't Hooft coupling. These results appeared in \cite{Alday:2017xua}, while related discussions are contained in \cite{Alday:2017vkk, Aprile:2017bgs, Aprile:2017xsp, Aprile:2017qoy}. There are open questions that would be nice to address. 
It would be interesting to discuss $1/\lambda$ corrections. At leading order this would amount of studying $1/\lambda$ corrections to the correlators $\langle {\cal O}_p{\cal O}_p{\cal O}_q{\cal O}_q \rangle$ which correspond to the addition of the first truncated solution with arbitrary coefficients, which can be fixed in principle along the line of \cite{Goncalves:2014ffa}. At order $1/N^4$ one would have to 'square' the supergravity contribution plus these contributions. One could also consider the exchange of a finite number of single-trace operators, combining the results of \cite{Alday:2017gde} with the methods of this note. Other interesting directions would be to combine the method presented in this note with recent direct computations of Witten diagrams at one loop \cite{Yuan:2017vgp} and with integrability based approaches \cite{Bargheer:2017nne}.

\vskip6pt

\enlargethispage{20pt}

%\ack{I would like to thank Fernando Alday for collaboration on \cite{Alday:2017xua} and related topics. The work of the author is supported in part by the Knut and Alice Wallenberg Foundation under grant KAW 2016.0129.}

%%%%%%%%%%%%%%%%%%%%%%%%%%%%%%%%%%%%%%%%%%%%%%%%%FURSAEV
\section[BCFT: Anomalies, Entanglement, Holography]{BCFT: Anomalies, Entanglement,
Holography\protect\footnote{This section was authored by Dmitri Fursaev from Dubna State University, Universitetskaya str. 19 141 980, Dubna, Russian Federation.}}\label{sec:fursaev}
\subsection{Introduction}

Physical boundaries are known to result in observable quantum
effects, such as Casimir forces between conducting surfaces, for 
example. Studying and understanding boundary effects, in general, may be a complicated problem
which should take into account properties of real materials, interactions and etc.
Conformal fields theories (CFTs) with idealized boundaries and boundary conditions (BC) are a useful 
test bed to develop ones intuition about quantum physics near boundaries.
This article is related to recent studies of integrated conformal anomalies (ICA) in CFT's with boundaries (BCFT).
ICA are completely determined by a set  of bulk and boundary charges.  We show that the same charges also fix other important observables 
in BCFTs near (or on) the boundary. Examples we briefly consider
include quantum stress-energy tensor (Casimir energy density), correlations functions of the displacement operator, the entanglement
entropy.  The entanglement entropy, when the entangling surface crosses the boundary, is a natural quantity to probe the strength of quantum 
correlations on the boundaries.  Interestingly, all above examples allow a holographic description in terms
of AdS gravity, at least in a bottom-up approach.  For the sake of brevity we consider 4D BCFT's. In odd dimensions ICA consist of pure boundary terms.
However analysis of 3D BCFT shows that boundary charges in ICA similarly determine physical observables.

\subsection{Integrated conformal anomaly in 4D BCFT}

Consider a 4D BCFT on a background manifold $\cal M$ with a boundary $\partial {\cal M}$. 
The integral conformal anomaly, ICA, is defined as the variation of the effective 
action $W$,
\begin{align}\label{1.1}
\begin{split}
{\cal A}\equiv \partial_\sigma W[e^{2\sigma}g_{\mu\nu}]_{\sigma=0}
\end{split}
\end{align}
under scaling with a constant factor $\sigma$. If imposed BC do not 
break the conformal invariance, ICA has the following universal structure
\cite{Dowker:1989ue,Herzog:2015ioa,Fursaev:2015wpa} :
\begin{align}\label{1.2}
\begin{split}
{\cal A}=-2a~\chi_4-c~i+q_1j_1+q_2j_2
\end{split}
\end{align}
Quantities $\chi_4$, $i$,
$j_1$ and $j_2$ are scale invariant functionals,
$\chi_4$ is the Euler characteristics of $\cal M$, $i$ is the integral over $\cal M$ of a square of the Weyl tensor $C_{\mu\nu\lambda\rho}$ of $\cal M$, $a$ and $c$ are bulk charges.  The boundaries result in 2 terms with boundary charges $q_1$ and $q_2$, 
\begin{align}\label{1.3}
\begin{split}
j_1={1 \over 16\pi^2} \int_{\partial {\cal M}}\sqrt{H}d^3x ~C_{\mu\nu\lambda\rho}N^\nu N^\rho\hat{K}^{\mu\lambda}
~~,~~j_2={1 \over 16\pi^2} \int_{\partial {\cal M}}\sqrt{H}d^3x~
\mbox{Tr}(\hat{K}^3)~~.
\end{split}
\end{align}
Here  $N^\nu$ is a unit normal vector to $\partial {\cal M}$, and $\hat{K}^{\mu\lambda}$ is the traceless part of extrinsic curvature of
$\partial {\cal M}$. 

Let us dwell on some general properties of the charges. Bulk charges $a$ and $c$ are well known for free CFT's. Recently ICA on squashed spheres 
were  used to find these charges for higher spin theories \cite{Beccaria:2017lcz}.
As was suggested in \cite{Cardy:1988cwa} and proved in \cite{Komargodski:2011vj} $a$-charge monotonically decreases from UV to IR.
Actual values for boundary charges  are listed in \cite{Fursaev:2015wpa} for a number of free BCFT (spins $s=0,1/2,1$) with Dirichlet or Robin BC.
For these models $q_1$ do not depend on BC, and are related to $c$-charge: $q_1=8c$. The origin and implications of this relation were discussed
in \cite{Solodukhin:2015eca,Herzog:2017xha}. It was argued in \cite{Herzog:2017xha} that the relation may receive quantum corrections due to interactions, at least in certain models.  Charges $q_2$  depend on BC, except for spin $s=1$.  
 
%\section{Correlation functions}

It is important that $c$-charge is connected with coefficient $C_T$ in two-point correlation functions of the stress-energy tensor \cite{Osborn:1993cr}, 
\begin{align}\label{2.1}
\begin{split}
\langle T_{\mu\nu}(x)T_{\lambda\rho}(0)\rangle \sim {C_T \over x^8}I_{\mu\nu,\lambda\rho}(x).
\end{split}
\end{align}
One can show that $c=(\pi^2 /40)C_T$.  Analogously, one can consider correlation functions of the displacement operator $D^n$ (which generates 
variation of the action under a shift of the boundary). The 2-point correlator on $\partial {\cal M}$ has the structure:
\begin{align}\label{2.2}
\begin{split}
\langle D^n({\bf x})D^n(0)\rangle \sim {c_{nn} \over ({\bf x}^2)^4},
\end{split}
\end{align}
where ${\bf x}$ are coordinates on $\partial {\cal M}$.
It was shown in \cite{Herzog:2017xha}  that $q_1=4/3 c_{nn}$. Remarkably, $q_2$ yields a coefficient at a 3-point correlation
function of the displacement operator, see \cite{Herzog:2017kkj}.

\subsection{Quantum energy density}

Another set of observables related to ICA is the local energy density near the boundary. It was known long time ago \cite{Deutsch}
that the renormalized expectation value of the stress-energy tensor near the boundary allows an asymptotic expansion in terms of inverse powers of the 
geodesic distance from  a point to the boundary.  Recently it was shown \cite{Miao:2017aba} that for free 4D BCFT ($s=0,1$) this expansion
is fixed by boundary charges:
\begin{align}\label{3.1}
\begin{split}
\langle T_{\mu\nu}({\bf x},y)\rangle \sim {1 \over 16\pi^2}\left({q_1 \over y^3}\hat{K}_{\mu\nu}({\bf x})+{t_{\mu\nu}({\bf x}) \over y^2}+....\right)
\end{split}
\end{align}
$$
t_{\mu\nu}({\bf x})=\frac 13 q_1\left(-2N_{(\mu} H^\lambda_{\nu)}(\partial_\lambda K+R_{\lambda\alpha} N^\alpha\right)+
\frac 16 q_1 K\hat{K}_{\mu\nu}-
$$
\begin{align}\label{3.2}
\begin{split}
(2q_1+3q_2)\left(\hat{K}_{\mu\beta}\hat{K}_{\nu}^\beta-\frac 13 H_{\mu\nu}\hat{K}_{\alpha\beta}
\hat{K}^{\alpha\beta}\right)
\end{split}
\end{align}
Here we use geodesic coordinates, where a point 
on $\cal M$ has coordinates $(y,{\bf x})$, $y$ being  a geodesic distance from it to a point on $\partial {\cal M}$.  Curvatures
in (\ref{3.2}) are calculated at $\partial {\cal M}$.

\subsection{Entanglement entropy}

Consider a BCFT specified by a density matrix  $\hat{\rho}$.
Entanglement entropy between states located in
spatially separated parts, $A$ and $B$, with a common boundary $\cal B$ is determined
by a reduced density matrix (for the region $A$) $\hat{{\rho}}_A=\mbox{Tr}_B \hat{\rho}$.
The entanglement R\'{e}nyi entropy of an order $q$ is defined as
$S_q({\cal B})=\ln \mbox{Tr}_A \hat{{\rho}}^{~q}_A / (1-q)$,
where $q$ is a non-negative parameter, $q\neq 1$. The corresponding entanglement entropy
$S({\cal B})=-\mbox{Tr}_A \hat{{\rho}}_A\ln \hat{{\rho}}_A$  follows from $S_q({\cal B})$ in the limit $q\to 1$. One can consider integer values $q=2,3,...$,
then $S_q=(qW-W_q)/(1-q)$, where $W_q$ is the effective action of the theory on
a manifold ${\cal M}_q$, $W_1=W$. ${\cal M}_q$ is constructed from $q$ copies of the initial background space $\cal M$ and has conical
singularities at $\cal B$.
Correspondingly, $S=\partial_{q=1}W_q-W$. 
The entanglement entropy is a divergent quantity, 
\begin{align}\label{4.3}
\begin{split}
S({\cal B})\simeq {A({\cal B}) \over \varepsilon^2}+ {P({\cal B}) \over \varepsilon^2}
 +s_{log}\ln \varepsilon+...~~,
\end{split}
\end{align} 
where $\varepsilon$ is a UV cutoff, $A({\cal B})$ and $P({\cal B})$ are the area and perimeter of $\cal B$, 
respectively (we omit numerical prefactors in the first two terms).  The perimeter term appears when $\cal B$ 
crosses $\partial {\cal M}$ \cite{Fursaev:2006ng}.
This is the case we are interested in.
Let ${\cal A}_q$ be ICA for $W_q$. The ICA determines logarithmic 
divergences of the entropy, $s_{log}=\partial_{q=1}{\cal A}_q-{\cal A}$. By using this one can show that \cite{Fursaev:2013mxa}
\begin{align}\label{4.4}
\begin{split}
s_{log}=aF_a+cF_c+bF_b+dF_d+eF_e.
\end{split}
\end{align} 
Here $F_a,F_c,F_b$ are some conformal invariants on $\cal B$ which depend on internal geometry of $\cal M$, topology and external geometry 
of $\cal B$. $F_d$ and $F_e$ are some other set of conformal invariants on $\cal C=\partial {\cal M} \cap {\cal B}$
which appear only when $\cal B$  crosses $\partial {\cal M}$. One can show that $b=c$, see \cite{Fursaev:2013fta} and references therein.
Therefore, bulk charges $a$, $c$ fix uniquely the logarithmic divergences of the entropy in the absence of boundaries.

$F_d$ is a conformally invariant integral on $\cal C$ of a projection,  $\hat{K}_{vv}$, of 
components of the extrinsic curvature of $\partial {\cal M}$ on a tangent vector to $\cal C$. Another invariant $F_e$ depends on extrinsic curvatures 
of $\cal B$. $F_d$  and $F_e$ can be defined up to conformally invariant dimensionless prefactors depending on a tilt angle between $\cal B$ and 
$\partial {\cal M}$. Fortunately, when $\cal B$ is orthogonal to
$\partial {\cal M}$, $F_e$ vanishes and $F_d$ is unique. So one can fix $d$ for a set of free BCFT  \cite{Fursaev:2013mxa,FarajiAstaneh:2017hqv} .
Remarkably,  $d$-charges appear to be related to ICA as $d=3a-14c-35/12~q_2$, see \cite{FarajiAstaneh:2017hqv}, thus ICA determines $s_{log}$ in BCFT.

\subsection{Holography}

A first attempt to develop a holographic approach to BCFT was done in \cite{Takayanagi:2011zk}. The basic idea of \cite{Takayanagi:2011zk} 
is that a holographic dual of a BCFT is  an AdS gravity one dimension higher in a domain restricted by a  hypersurface $\cal S$ ($\cal S$  
being
dual to $\partial {\cal M}$). $\cal S$  is a dynamical surface whose equations depend on boundary  conditions.
Two types of equations were suggested for $\cal S$ in  \cite{Miao:2017gyt} and in \cite{Astaneh:2017ghi} with the aim to reproduce 
ICA in supersymmetric BCFT. For instance,  $\cal S$  was treated in \cite{Astaneh:2017ghi} as a minimal surface. This allowed the authors to derive
holographically the boundary charges in ICA for ${\cal N}=4$  Super  $SU(N)$ YM theory (assuming that the charges are protected from quantum corrections by the supersymmetry). BC were chosen to break half of supersymmeties .

Holographic Ryu-Takayanagi formula \cite{Ryu:2006bv} reproduces successfully the logarithmic part, see (\ref{4.4}), of entanglement entropy \cite{Schwimmer:2008yh} in the absence of boundaries. The holographic formula was used in \cite{FarajiAstaneh:2017hqv} for
of BCFT's (the Super  $SU(N)$ YM) by using prescription of \cite{Astaneh:2017ghi} and reproduced (\ref{4.4}) in case when $\cal B$ is orthogonal to the boundary.

%%%%%%%%%%%%%%%%%%%%%%%%%%%%%%%%%%%%%%%%%%%%%%%%%SCHOMERUS
\section[Blocks for Conformal Defects]{Blocks for Conformal Defects\protect\footnote{This section was authored by Volker Schomerus from  the DESY Theory Group at DESY Hamburg, Notkestrasse~85, D-33607 Hamburg, Germany and the Department of Particle Physics and Astrophysics at the Weizmann Institute of Science, Rehovot 76100, Israel.}}\label{sec:schomerus}
\subsection{Introduction}
Extended objects such as line or surface operators, defects,
interfaces and boundaries are important probes of the dynamics in
quantum field theory. They give rise to observables that can detect
a wide range of phenomena including phase transitions and non-perturbative
dualities. In 2-dimensional conformal field theories, they also turned out
to play a vital role for modern formulations of the bootstrap programme.
In fact, in the presence of extended objects, the usual crossing symmetry
becomes part of a much larger system of sewing constraints \cite{CL}.
While initially the 2-dimensional bootstrap started from the crossing
symmetry of bulk four-point functions to gradually bootstrap correlators
involving extended objects, better strategies were adopted later which 
depart from some of the sewing constraints involving extended objects. 
The usual crossing symmetry constraint is then solved at a later stage 
to find the bulk spectrum and operator product expansion, see e.g. 
\cite{Runkel:2005qw}.

The bootstrap programme, whether in its original formulation \cite{Polyakov:1974gs} or
in the presence of extended objects, relies on conformal partial wave expansions
\cite{Mack:1973cwx,Ferrara:1973vz} that decompose physical correlation functions into
kinematically determined blocks/partial waves and dynamically determined coefficients.

In the original bulk bootstrap program, the conformal blocks for a four-point correlator
are functions of two cross ratios and the coefficients are those that appear in the
operator product expansion of local fields. Such conformal partial wave expansions
thereby separate very neatly the dynamical meat of a conformal field theory from its
kinematical bones.

In order to perform a conformal partial wave expansion one needs a good
understanding of the relevant conformal blocks. While they are in principle
determined by conformal symmetry alone, it is still a highly non-trivial
challenge to identify them in the zoo of special functions, determine
their analytical properties or to develop efficient expansion formulas. In
the case of scalar four-point functions much progress had been made in the
conformal field theory literature starting with \cite{Dolan:2000ut,Dolan:2003hv,
Dolan:2011dv}. If the dimension $d$ is even, one can actually construct the conformal
blocks from products of two hypergeometric functions each of which depends on one
of the cross ratios. For more generic dimensions many important properties of
the scalar blocks had been understood. These include their detailed analytical
structure and various series expansions.

Extended objects give rise to new families of blocks. There are a few
cases, such as boundary or defect 2-point functions, which involve only
one or two cross ratios. In these cases it turns out that the relevant
blocks are made from the same functions as for bulk four-point functions,
\cite{Liendo:2012hy,Billo:2016cpy}. Hence all the knowledge on blocks that
had been assembled in the context of bulk four-point functions carries over
to the case of two bulk fields in the presence of a boundary or defect. But
what about more general situations such as e.g. the correlation function of
two (Wilson- or 't Hooft) line operators in a $d$-dimensional conformal
field theory? Such configurations often possess more than two conformal
invariant cross ratios. Two conformal line operators in a 4-dimensional
theory, for example, give rise to three cross ratios. For a configuration
of a $p$- and a $q$-dimensional object in a $d$-dimensional theory, the
number of cross ratios is given by $N = \textit{min\/}(d-p,q+2)$ if $p\geq q$
\cite{Gadde:2016fbj}. So clearly, the study of such defect correlation
functions involves new types of special functions which depend on more
than two variables.

In order to explore the features of these new functions, understand
their analytical properties or find useful expansions one could try to
follow the same route that was used for scalar four-point blocks, see e.g.
\cite{Fukuda:2017cup} for some recent work in this direction. It is
the central message of this short paper, however, that there is another
route that gives much more direct access to defect blocks. It relies
on a generalization of an observation in \cite{Isachenkov:2016gim} that
4-point blocks are wave functions of certain integrable 2-particle
Hamiltonians of Calogero-Sutherland type \cite{Calogero:1970nt,Sutherland:1971ks}.
The solution theory for this quantum mechanics problem is an important
subject of modern mathematics, starting with the seminal work of
Heckman-Opdam \cite{Heckman:1987}, see \cite{Isachenkov:2017qgn} for
a recent review in the context of conformal blocks. Much of the
development in mathematics is not restricted to the 2-particle
case and it has given rise to an extensive branch of the modern
theory of multi-variable hypergeometric functions which are quite
well understood.

This proceedings is based on the works \cite{Isachenkov:2017qgn,Schomerus:2016epl, Schomerus:2017eny, Isachenkov:2018pef}.
\subsection{Main Results}

In order to put all this mathematical knowledge to use in the context
of defect blocks, we need to link the corresponding conformal blocks,
which depend on $N$ variables, to wave functions of an $N$-particle
Calogero-Sutherland model. That such a link exists is the main claim of
this work. Following a general route through harmonic analysis on the
conformal group \cite{Schomerus:2016epl}, one can indeed construct
systematically the relevant Calogero-Sutherland Hamiltonian. For all
values of the dimensions $d,p,q$ it takes the form
\begin{align}
H_\textrm{CS} &= -\sum_{i=1}^{N} \frac{\partial^2}{\partial \tau_i^2} +
\frac{k_3(k_3-1)}{2}\sum_{i<j}^N\left[ \sinh^{-2}\left(\frac{\tau_i-\tau_j}{2}\right)+
\sinh^{-2}\left(\frac{\tau_i+\tau_j}{2}\right) \right]
\nonumber \\[2mm]
&\qquad
+ \sum_{i=1}^{N}\left[ k_2(k_2-1)\sinh^{-2}\left(\tau_i\right) +
\frac{k_1(2k_2+k_1-1)}{4}\sinh^{-2}\left(\frac{\tau_i}{2}\right) \right] \ .
\label{eq:HCS} \end{align}
The coupling constants $k_i, i= 1,2,3$ that appear in the potential can
assume complex values though we will mostly be interested in cases in which
they are real. The coordinates $\tau_i$ may also be complex in general as long
as the potential remains real.

Our claim is that all Casimir equations for defect blocks can be recast into an
eigenvalue equation for the Calogero-Sutherland Hamiltonian \eqref{eq:HCS} with
coupling constants depending on the choice of defects and the dimension $d$. In
the case of two defects of dimension $p \geq q$ with $q \neq 0$, the relevant
couplings read
\begin{align} \label{eq:Caspqpar}
 k_1=N-\frac{d}{2} \,,\quad k_2=\frac{p-q}{2} \,,
\quad k_3=\frac12 \, .
\end{align}
The Calogero-Sutherland problem is to be considered on the semi-infinite cuboid
that is parametrized by a real coordinate $\tau_1 \in \mathbb{R}_+$ and $N-1$ angles
$\varphi_i, i=2, \dots, N,$ such that $\tau_i = i \varphi_i$. When interpreted in
terms of the geometry of two spherical defects, $\tau_1$ parametrizes the ratio of
their radii while $\tau_2, \dots, \tau_N$ relative rotation angles. Precise formulas
will be presented in our forthcoming paper.

The formulas \eqref{eq:Caspqpar} still apply when $q =0$, but in this case there
exists an important extension. A zero-dimensional defect
may also be thought of as the insertion of two local bulk scalar fields. When
these possess the same conformal weight, the coupling constants of the
associated Calogero-Sutherland model are given by eq.\ \eqref{eq:Caspqpar}.
If the conformal weight of the two bulk fields differs by $2a=\Delta_2-
\Delta_1\neq 0$ the coupling constants become
\begin{align}\label{eq:Caspq0}
k_1=2-\frac{d}{2} \,,\quad k_2=\frac{p}{2} \,,\quad k_3=\frac12 + a \, .
\end{align}
These results have direct implications on relations between different types of
blocks. One such example concerns the case of two bulk fields in the presence of
a defect that we have just discussed. Comparison of the Calogero-Sutherland model
with couplings \eqref{eq:Caspq0} with that for bulk four-point functions 
\cite{Isachenkov:2016gim} shows that the defect blocks are
related to the blocks for four scalar fields with conformal weights $\Delta'_i$
satisfying $2a = \Delta'_2 - \Delta'_1$ and $0 = \Delta'_3 -\Delta_4'$. This
generalizes an observation in \cite{Billo:2016cpy}.  Several similar relations
will be discussed in the forthcoming paper.

Our formulas for the coupling constants may be derived by restricting the Laplace
operator on the conformal group manifold to the quotient space $G_p \backslash G / G_q$
where $G = \SO{1,d+1}$ is the conformal group and $G_p = \SO{1,p+1} \times \SO{d-p}$ the
subgroup that leaves the defect invariant. It is not difficult to see that the action of
$G_p \times G_q$ on the conformal group $G$ is stabilized by the diagonal subgroup
$B_{pq} = \SO{p-q} \times \SO{|d-p-q-2|}$. Once this is taken into account, it is
straightforward to compute the dimension of the double coset space,
$$ \dim G_p \backslash G / G_q = \dim G - \dim G_p - \dim G_q + \dim B_{pq} = N \ .
$$
Hence, after reduction to the double coset, the Laplace operator on the conformal
group becomes a second order operator in $N$ variables. The latter can be transformed
into a multi-particle Hamiltonian of Calogero-Sutherland type, see
\cite{Olshanetsky:1983wh,Feher:2009wp,Schomerus:2016epl,Schomerus:2017eny} for
related discussions and our forthcoming paper for a derivation in the case of
defect blocks.

\subsection{Outlook}

Once the connection between defect blocks and Calogero-Sutherland
models is established, one can exploit the extensive mathematical theory. In fact,
the study of Calogero-Sutherland models became an important subject in
mathematics, starting with the work of Heckman and Opdam \cite{Heckman:1987}
that initiated much of the modern theory of multi-variable hypergeometric
functions. A lightning review of some central results can be found in
\cite{Isachenkov:2017qgn,Isachenkov:2018pef}. The relevant defect
blocks may be constructed explicitly in terms of the well studied class of
Harish-Chandra functions. Many profound results such as series expansions,
poles and their residues in momentum space, as well as global analytical
properties including the position of cuts and their monodromies can be
found in the literature or at least be derived from published results.

%%%%%%%%%%%%%%%%%%%%%%%%%%%%%%%%%%%%%%%%%%%%%%%%%NAKYAMA

\section[CFTs on real projective spaces]{CFTs on real projective spaces\protect\footnote{This section was authored by Yu Nakayama from The Department of Physics, Rikkyo University, Tokyo, Japan.}}\label{sec:nakayama}
Suppose we know everything about a conformal field theory (CFT) in flat Minkowski space-time, how much can we solve the same CFT in non-trivial curved space-time?  We have a gut feeling that we should be able to do it because in the Lagrangian formalism, once we know the action in flat space-time, the action in the curved background is more or less fixed (up to some arbitrariness from curvature induced interactions), and in principle we should able to solve it.

In practice, however, it seems very hard to accomplish this, when the theory is strongly coupled with or without the Lagrangian formalism. The aim of this talk is to discuss how much we can solve the CFT on a certain curved background, a real projective space, given the CFT data on the flat Euclidean space.

The real projective space is defined by the identification of anti-podal points on the $d$-dimensional sphere. Let us do the conformal map from the sphere to a plane $R^d$, then the anti-podal identification becomes the indentification under the involution $\vec{x} \to -\frac{\vec{x}}{\vec{x}^2}$. We may further conformal transform to a cylinder $R_t \times S_{d-1}$, in which the involution becomes a "PT" transformation $(t,\vec{\Omega}) \to (-t,-\vec{\Omega})$. On the cylnder, one may introduce the crosscap state $|C\rangle$ that satisfies $M_{ab}|C\rangle = (P_a + K_a)|C\rangle =0$ at $t=0$ to implement the identification.

Solving the CFT on the real projective space is equivalent to  computing one-point functions of all the scalar primary operators $\langle O_i (\vec{x}) \rangle = \frac{A_i}{(1+\vec{x}^2)^{\Delta_i}}$ so that we can compute all the correlation functions by repeated use of operator product expansions (OPE) given a flat space CFT data. The same one-point functions appear in the expression of the physical (Cardy-like) crosscap state:
\begin{align}
|C \rangle\rangle = \sum_i A_i \Gamma(\Delta_i-d/2+1) \left(\frac{\sqrt{p^2}}{2}\right)^{\frac{d}{2}-\Delta_i}J_{\Delta_i-\frac{d}{2}}(\sqrt{p^2})|O_i\rangle ,
\end{align}
where $p$ is the momentum operator.

To test our philosophy of solving CFTs on curved space-time from the given CFT data on the flat space-time, let us try the bootstrap analysis in which we study the consistency of the two-point functions on the real projective space \cite{Nakayama:2016cim}. 
The scalar two-point function takes the form
\begin{align*}
 \langle O_{1}(\vec{x}_1) O_{2}(\vec{x}_2) \rangle = \frac{(1+\vec{x}^2_1)^{\frac{-\Delta_1 + \Delta_2}{2}}   (1+\vec{x}^2_2)^{\frac{-\Delta_2+\Delta_1}{2}}}{ (\vec{x}_1 - \vec{x}_2)^{2 (\frac{\Delta_1+\Delta_2}{2})}} G_{12}(\eta) \ ,
\end{align*}
where $
\eta =  \frac{(\vec{x}_1 - \vec{x}_2)^2}{(1+\vec{x}_1^2)(1+\vec{x}_2^2)} $ is so-called crosscap cross ratio. By noticing the identification $O_i(\vec{x}) \sim O_i(-\frac{\vec{x}}{\vec{x}^2})$, we demand the crossing symmetry
\begin{align*}
\left(\frac{1-\eta}{\eta^2}\right)^{\frac{\Delta_1+\Delta_2}{6}} G_{12}(\eta) = \left(\frac{\eta}{(1-\eta)^2}\right)^{\frac{\Delta_1+\Delta_2}{6}} G_{12}(1-\eta) \ .
\end{align*}
One may solve the crossing equation exactly in $d=2$ minimal models by using the Virasoro conformal block decomposition of $G(\eta)$.

In $d>2$ dimensions, we use the global conformal block decomposition
\begin{align*}
G_{12}(\eta) = \sum_i C_{12i} A_i \eta^{\Delta_i/2}  {}_2F_1 \left(\frac{\Delta_1 - \Delta_2 + \Delta_i}{2},\frac{\Delta_2-\Delta_1+\Delta_i}{2}; \Delta_i + 1-\frac{d}{2};\eta\right) \ , 
\end{align*}
but here we may have to take the sum over infinitely many primary operators (except for certain favorable cases like free field theories). In the numerical analysis, we truncate the sum.

Let us focus on the Ising-like theories and study the two-point functions of spin operators with the OPE $[\sigma] \times [\sigma] = 1 + [\epsilon] + [\epsilon'] + \cdots$. In $d=4-\epsilon$ dimensions, we find the exact (i.e. to all order in $\eta$ expansion) expression for the crossing symmetric two-point function
\begin{align*}
G(\eta) &= 1 + (1-a\epsilon)\eta^{1-\frac{\epsilon}{2}+a\epsilon}{}_2F_1(1-\frac{\epsilon}{2} + a \epsilon,1-\frac{\epsilon}{2} +  a \epsilon,2-\epsilon + 2 a \epsilon + 1-\frac{4-\epsilon}{2};\eta) \cr
 &+ \frac{a \epsilon}{2} \eta^{2-\epsilon} {}_2F_1 (2-\epsilon,2-\epsilon,4-2\epsilon +1-\frac{4-\epsilon}{2};\eta) +O(\epsilon^2)
\end{align*}
 up to $O(\epsilon^2)$ corrections, where $a$ is arbitrary but is related to $\gamma_{\phi^2} = 2a\epsilon$. 

Beyond this order, we need an infinite number of primary operators to satisfy the crossing symmetry at $\eta=0$ or $\eta=1$, but the convergence is exponentially fast at $\eta = 1/2$. In $d=2$ dimensions, the truncated bootstrap gives the numerical prediction

\begin{tabular}{|c||c|c|c|c|c|c|}
\hline
 &Exact & 2 & 3 & 4 & 5  \\ \hline \hline
$C_{\sigma \sigma \epsilon} A_\epsilon$ &$0.20711$&$0.20407$&$0.20757$&$0.20693$&$0.20710$\\ \hline
$C_{\sigma \sigma \epsilon'} A_{\epsilon'}$ &$0.01563$&$0.01702$&$0.01539$&$0.01572$&$0.01563$ \\ \hline
\end{tabular}

which is in good agreement with the exact value $C_{\sigma \sigma \epsilon}= \frac{\sqrt{2}-1}{2} = 0.20711$. 

In $d=3$ dimensions, we predict

\begin{tabular}{|c||c|c|c|c|c|c|}
\hline
 &(2,0) & (4,0)$_{\mathrm{A}}$ & (4,0)$_{\mathrm{B}}$ & (6,0)$_{\mathrm{B}}$ & (4,0)$_{\mathrm{S}}$  \\ \hline \hline
$C_{\sigma \sigma \epsilon} A_\epsilon$ &0.690&0.7015&0.7022&0.70197&0.6908\\ \hline
$C_{\sigma \sigma  \epsilon'} A_{\epsilon'}$ &0.054&0.0475&0.0470&0.04714&0.0549 \\ \hline
\end{tabular}

The similar analysis can be done in the critical Lee-Yang model, whose $\epsilon$ expansion is discussed in \cite{Hasegawa:2016piv}. In the future, it would be interesting to study holographic models such as $N=4$ super Yang-Mills theory or Liouville theory and compare them with the holographic predictions.

%%%%%%%%%%%%%%%%%%%%%%%%%%%%%%%%%%%%%%%%%%%%%%%%%SCHWEIGERT
\section[Conformal field theories with and without boundaries: from a holographic picture to logarithmic CFT]{Conformal field theories with and without boundaries: from a holographic picture to logarithmic CFT \protect\footnote{This section was authored by Christoph Schweigert from Fachbereich Mathematik, Universit{\"a}t Hamburg Bereich Algebra und Zahlentheorie, Bundesstra{\ss}e 55, 20146 Hamburg, Germany.}}\label{sec:schweigert}

Correlators in rational conformal field theories can be obtained by the TFT 
construction. This construction, which is summarized e.g.\ in \cite{Schweigert:2006af},
amounts to the following. Assume we are given a chiral conformal field theory for
which the conformal blocks and their monodromy are encoded in a (semisimple) modular 
tensor category $\cal C$. Such a category allows for the construction of a three-dimensional
topological field theory $\mathrm{tft}_{\mathcal C}$ of Reshetihkin-Turaev type.
By the principle of holomorphic factorization, a correlator on an 
(oriented) surface $X$ for these chiral data $\mathcal C$ is an element 
$\mathrm{Cor}_X\,{\in}\, \mathrm{tft}_{\mathcal C}(\widehat X)$, where $\widehat X$ 
is the oriented double of $X$. A consistent set of correlators is the datum
of such a vector $\mathrm{Cor}_X\,{\in}\, \mathrm{tft}_{\mathcal C}(\widehat X)$
for every surface $X$, such that the vector for $X$ is invariant under the action of the
mapping class group of $X$ on the vector space $\mathrm{tft}_{\mathcal C}(\widehat X)$,
and that the vectors for different $X$ are compatible with the factorization
(or sewing) of surfaces.

Such a collection of correlators can be constructed for any special symmetric 
Frobenius algebra in $\mathcal C$ in terms of the topological field theory 
$\mathrm{tft}_{\mathcal C}$ as follows: Given a surface $X$, one constructs a 
three-manifold $M_X$ with boundary $\partial M_X \,{=}\, \widehat X$ and with
an embedded ribbon graph such that $\mathrm{Cor}_X \,{=}\, 
\mathrm{tft}_{\mathcal C}(\emptyset\,{\stackrel{M_X}\longrightarrow} \widehat X)(1) 
\,{\in}\, \mathrm{tft}_{\mathcal C}(\widehat X)$. This constitutes a mathematically 
precise holographic approach to RCFT correlators which uses a three-dimensional
topological field theory in the bulk. In this rigorous framework one can prove 
that consistent systems of correlators on all surfaces are in bijection with (classes
of) special symmetric Frobenius algebras in $\mathcal C$, and OPE coefficients as
well as coefficients of partition functions can be expressed as
invariants of links in three-manifolds.

The Frobenius algebra enters the construction of the correlator for a surface $X$
via a trivalent graph, dual to a triangulation of $X$. This dual triangulation can be
seen as the remnant of a surface defect in the three-dimensional
topological field theory; see \cite{Kapustin:2010if} and
\cite[Sect.\,6]{Fuchs:2012dt} for the relation to special
symmetric Frobenius algebras.

More recent developments transcend the realm of rational CFT, dealing with 
logarithmic conformal field theories, whose chiral data are described by a 
non-semisimple finite modular tensor category $\mathcal C$. (Examples of such 
logarithmic conformal field theories are the minimal models of type 
${\mathcal W}_{1,p}$.) For such theories the physical idea of ``summing over 
all intermediate states'' can be implemented by the categorical notion of a 
coend \cite{Lyubashenko:1994tm}. Combining this idea with the one of a 
Lego-Teichm\"uller game in the sense of \cite{BakalovKirillov}, in which a 
modular functor is built from three-point conformal blocks on the sphere via sewing, one 
arrives at a 
mathematically precise description of the monodromy data of the
conformal blocks of such a chiral logarithmic conformal field theory. 

A construction of correlators of a logarithmic conformal field theory is achieved by 
implementing these ideas for the category ${\mathcal C}\boxtimes {\mathcal C}^{\mathrm{rev}}$
which accounts for the presence of both
left- and right-movers. (Here ${\mathcal C}^{\mathrm{rev}}$ is the modular category 
in which over-braidings are replaced by under-braidings and vice versa.) This way
it has been shown \cite{Fuchs:2016wjr} that consistent systems of correlators of 
bulk fields are in bijection with modular Frobenius algebras in 
${\mathcal C}\boxtimes {\mathcal C}^{\mathrm{rev}}$. 
These are commutative symmetric Frobenius algebras satisfying one additional
property that ensures consistency of higher genus correlators. The physical
interpretation of the Frobenius algebra is as the algebra of bulk fields.

In the case that $\mathcal C$ is the category of finite-dimensional modules over
a finite-dimensional factorizable ribbon Hopf algebra $H$ it is known \cite{Fuchs:2011mg}
that the coend object
$$ 
  \int^{c\in \cal C}\! \omega(c)\boxtimes c^\vee
  ~\in {\mathcal C}\boxtimes {\mathcal C}^{\mathrm{rev}}
$$
has the structure of a modular Frobenius algebra, for any ribbon automorphism $\omega$ 
of $H$. This can be interpreted as a generalization of modular invariant
partition functions of automorphism type to logarithmic conformal field theories. 

Specifically, for $\omega$ the identity, the result can be regarded as a generalization
of the Cardy case to logarithmic conformal field theories.  In this case the 
coefficients of the torus partition function are \cite {Fuchs:2012ya} given by the entries 
of the Cartan matrix of the category $\mathcal C$, i.e.\ by the dimensions 
$$
  c_{ij} = \dim\mathrm{Hom}_{\mathcal C} (P_i,P_j) 
$$
of Hom-spaces involving the indecomposable projective objects in $\cal C$.

%%%%%%%%%%%%%%%%%%%%%%%%%%%%%%%%%%%%%%%%%%%%%%%%%FRIEDAN
\section[A new kind of quantum field theory
of (n-1)-dimensional defects in 2n dimensions]{A new kind of quantum field theory
of (n-1)-dimensional defects in 2n dimensions\protect\footnote{This section was authored by Daniel Friedan from the New High Energy Theory Center at Rutgers University, Piscataway, New Jersey, USA and the Natural Science Institute at the University of Iceland, Reykjav\'ik, Iceland.}}\label{sec:friedan}

This is a summary of the main points of \cite{Friedan:2016mvo}.
A short summary is \cite{Friedan:EQFT}.
% References can be found there.
% A less condensed version of this note is \cite{Friedan:EQFT}.
% , based
% on a talk at the Royal Society Workshop,
% Boundary and Defect Conformal Field Theory, Chicheley Hall, UK, 
% September 8, 2017 \cite{Friedan2017:ChicheleyHall}.
More expositions appear at \cite{Friedan:webpage}.

%\newsec{1}
Let $M$ be euclidean space-time: an oriented conformal 
$2n$-manifold, compact, without boundary.
When $n=1$, $M$ is a Riemann surface.
The basic examples are $M=S^{2n} = \Reals^{2n}\cup\{\infty\}$.
The Hodge $*$-operator acting on $n$-forms
is conformally invariant, with $*^{2}=(-1)^{n}$.
% \eq
% (*\omega)_{\nu_{1}\cdots\nu_{n}}(x)
% =
% \omega_{\mu_{1}\cdots\mu_{n}}(x)
% \,\frac1{n!} \,
% \epsilon^{\mu_{1}\cdots\mu_{n}}{}_{\nu_{1}\cdots\nu_{n}}(x)
% \qquad
% *^{2}=(-1)^{n}
% \en
Nothing else is used of the conformal structure on $M$.

%\newsec{2}
The physical objects are represented mathematically
as the integral $(n{-}1)$-currents in $M$,
as constructed in Geometric Measure Theory \cite{FedererFleming}.
A $k$-{\it current} $\xi$ in $M$ is a distribution on the smooth $k$-forms,
$\omega\mapsto \int_{\xi}\omega\in\Reals$.
% \eq
% \omega\in\Omega_{k}(M)\mapsto 
% \int_{\xi}\omega
% =
% \int_{M} 
% \frac1{k!} \omega_{\mu_{1}\ldots\mu_{k}}(x)
% \,\xi^{\mu_{1}\ldots\mu_{k}}(x) d^{2n}x 
% \en
The boundary operator $\partial$ is dual to the exterior 
derivative, $\int_{\partial\xi}\omega = \int_{\xi}d\omega$,
$\partial^{2}=0$.
% \eq
% \int_{\partial\xi}\omega = \int_{\xi}d\omega
% \qquad
% \partial^{2}=0
% \en
A singular $k$-chain  is an
integer linear combination
$\sigma=\sum_{i}m_{i}\sigma_{i}$
of $k$-simplices
$\sigma_{i}\colon\Delta^{k}\rightarrow M$
in $M$.
% \eq
% \sigma\colon\Delta^{k}\rightarrow M
% \qquad
% \int_{[\sigma]} \omega = \int_{\Delta^{k}} \sigma^{*}\omega
% \en
The {\it singular $k$-currents} $\cD^{\sing}_{k}(M)$
are the currents $\int_{[\sigma]} \omega = \sum_{i}m_{i}\int_{\Delta^{k}} 
\sigma_{i}^{*}\omega$
that represent the singular $k$-chains.
Examples are the $k$-submanifolds.
The current represents the physical object in $M$ independent of
its expression as a combination of simplices.

The physical difference between singular $k$-currents is
measured by
the {\it flat metric} $\norm{\xi_{1}-\xi_{2}}$,
$
\norm{\xi} = \inf\{\text{vol}_{k}(\xi-\partial \xi')
+\text{vol}_{k+1}(\xi')
\colon
\xi'\in \cD^{\sing}_{k+1}(M)
\}
$.
The space  of {\it integral} 
\noindent
$k$-currents $\cD^{\integral}_{k}(M)$ is the metric completion:
$
\cD^{\sing}_{k}(M) \subset \cD^{\integral}_{k}(M) \subset
\cD^{\distr}_{k}(M)
$,
a complete metric space and an 
abelian group.
The boundary of an integral current is an integral current.

Recall the 2d gaussian model: the free 1-form cft in 2d.
$j(x)$ is a 1-form
on a Riemann surface
satisfying
$dj = 0$, 
$d(*j)=0$.
The integrals of $j$ and $*j$
are 0-forms  $\phi$, $\phi^{*}$
which take values in dual circles,
$d\phi = j$,
$d\phi^{*} = {*}j$,
$\phi(x) \in \Reals/2\pi R\Integers$,
$\phi^{*}(x) \in \Reals/2\pi R^{*}\Integers$,
$R R^{*} =1$.
They are determined up to $U(1){\times} U(1)$ global symmetries
$\phi(x)\rightarrow \phi(x)+a$,
$\phi^{*}(x)\rightarrow \phi^{*}(x)+a^{*}$.
The vertex operator $V_{p,p^{*}}(x)= e^{ip\phi(x) + 
ip^{*}\phi^{*}(x)}$
describes a point defect
of charges $p,p^{*}\in \frac1R \Integers \times\frac1{R^{*}} 
\Integers$.
It transforms by
$V_{p,p^{*}} \rightarrow V_{p,p^{*}}\, e^{ipa + 
ip^{*}a^{*}}$.

Recall the free $n$-form cft in $2n$ dimensions.
$F(x)$ is an $n$-form on the $2n$-manifold $M$ satisfying
$dF = 0$, $d(*F) = 0$.
The integrals of $F$ and $*F$
are $(n{-}1)$-forms  $A$, $A^{*}$,
$dA = F$,
$dA^{*} = *F$
which take values in dual circles
$\int_{\xi} A \in \Reals/2\pi R\Integers,\;\;
\int_{\xi} A^{*} \in \Reals/2\pi R^{*}\Integers\;
\forall\xi \in \cD^{\sing}_{n-1}(M)$,
$R R^{*} =1$.
$A$, $A^{*}$ are
determined up to $U(1){\times} U(1)$ local gauge symmetries 
given by $(n{-}2)$-forms $f$, $f^{*}$,
$A\rightarrow A+df$,
$A^{*}\rightarrow A^{*}+df^{*}$.
$(n{-}1)$-dimensional defects are 
described by fields on $\cD^{\sing}_{n-1}(M)$,
$V_{p,p^{*}}(\xi) = e^{ip\phi(\xi) +ip^{*}\phi^{*}(\xi)}$,
$\phi(\xi) = \int_{\xi} A$,
$\phi^{*}(\xi) = \int_{\xi} A^{*}$,
$p,p^{*}\in \frac1R \Integers \times\frac1{R^{*}} \Integers$
transforming by
$V_{p,p^{*}}(\xi)\rightarrow V_{p,p^{*}}(\xi)
\;e^{ipa(\partial\xi)+ip^{*}a^{*}(\partial\xi)}$,
$a(\partial\xi)  = \int_{\partial\xi} f $,
$a^{*}(\partial\xi) = \int_{\partial\xi} f^{*}$.
Fix an $(n{-}2)$-boundary $\partial\xi_{0}$ and consider
the abelian subgroup
$\cD^{\sing}_{n-1}(M)_{\Integers\partial\xi_{0}}
=\left\{\xi\colon
\:\partial\xi \in \Integers\partial\xi_{0}
\right\}
\;\subset
\cD^{\sing}_{n-1}(M)
$.
On $\cD^{\sing}_{n-1}(M)_{\Integers\partial\xi_{0}}$
the gauge 
symmetries act as a global $U(1){\times} U(1)$ generated by the
numbers $a(\partial\xi_{0})$ and $a^{*}(\partial\xi_{0})$.

%\newsec{5}
Calculus is needed on 
$\cD^{\sing}_{n-1}(M)_{\Integers\partial\xi_{0}}$
to continue the analogy with the 2d gaussian model.
Go to the metric completion,
writing it $Q=\cD^{\integral}_{n-1}(M)_{\Integers\partial\xi_{0}}$.
Geometric Measure Theory provides
the spaces $\cD^{\integral}_{j}(Q)$ of integral $j$-currents in 
any such complete metric space \cite{AmbrosioKirchheim}.
Define the $j$-forms on $Q$ as the real duals of the currents
and the exterior derivative as the dual of the boundary operator,
$\Omega_{j}(Q) = \Hom(\cD^{\integral}_{j}(Q),\Reals)$,
$d\omega(\eta) = \omega(\partial\eta)$.
The infinitesimal
$j$-simplices  generate $\cD^{\integral}_{j}(Q)$,
so the tangent bundle $TQ$ can be defined as the set of infinitesimal 
1-simplices in $Q$.  The 1-forms then become the sections of the dual 
cotangent bundle $T^{*}Q$.
The
equivalences of simplices $\Delta^{j}\times \Delta^{n-1}\cong \Delta^{j+n-1}$
give natural maps
$\Pi_{j,n-1}\colon\,\cD^{\integral}_{j}(Q)\rightarrow 
\cD^{\integral}_{j+n-1}(M)$,
$\partial \Pi_{j,n-1} = \Pi_{j-1,n-1} \partial$.
The map $\Pi_{1,n-1}$
identifies each tangent space $ T_{\xi}Q$ 
with a certain subspace $\cV_{n}\subset\cD^{\distr}_{n}(M)$.
That $*\cV_{n} = \cV_{n}$
is a crucial technical point
whose demonstration uses
the flat metric completion.
Then $*$ acts on each tangent space $ T_{\xi}Q$.

The forms $F$, $*F$, $A$, $A^{*}$ pull back to $Q$,
$j=\Pi_{1,n-1}^{*}F$, 
${*}j = \Pi_{1,n-1}^{*}(*F)$,
$\phi = \Pi_{0,n-1}^{*} A$,
$\phi^{*} = \Pi_{0,n-1}^{*} A^{*}$,
with $d\phi = j$,  $d\phi^{*} = {*}j$.
So there is the classical 2d gaussian model on each of the spaces $Q$,
except that $*^{2}=1$ for $n$ even,
while $*^{2}=-1$ in 2d.
Define
$J=\epsilon_{n}*$,
$\epsilon_{n}^{2} =(-1)^{n-1}$,
so $J^{2}=-1$.
$J$ is imaginary when $n$ is even, so the currents
have to be complexified
in order that $J$ act on the tangent spaces $T_{\xi}Q$,
$Q= \cD^{\integral}_{n-1}(M)_{\Integers\partial\xi_{0}}
\oplus i \partial \cD^{\integral}_{n}(M)$.
Now, for all $n$, on each of the 
spaces $Q$ there is a 2d gaussian model
written in terms of 
the chiral fields
$d\phi_{\pm}= j_{\pm}$,
$j_{\pm}=\frac12(1\pm i^{-1}J)j$.

%\newsec{6}
Quantization of a free field theory is
expressed by the Schwinger-Dyson equation on the 2-point functions.
In the 2d gaussian model, the chiral fields are (anti-)holomorphic.
The S-D equation on $\langle \bar\phi_{\pm}\,j_{\pm}\rangle$
is the Cauchy-Riemann equation
$\frac{\partial}{\partial \bar z} \;\frac{1}{z-z'} = \pi
\delta^{2}(z-z')$
which is the foundation for complex analysis on Riemann 
surfaces.
The 2d gaussian model would have led to complex analysis on Riemann surfaces
had that not existed already.
For the free $n$-form in $2n$ dimensions,
the S-D equation 
has an expression containing no explicit mention of $n$,
$\langle
\int_{\bar \xi_{1}} \bar A_{\alpha}
\int_{ \xi_{2}} d F_{\beta}
\rangle
=
2\pi i c_{\alpha\beta}
\IM{\bar\xi_{1}}{\xi_{2}}
$,
$c_{\alpha\beta}= -c_{\beta\alpha}$,
$c_{+-}=1$.
The lhs is the 2-pt function
$
\langle
\bar A_{\pm}(x)
\,d F_{\pm}(x')
\rangle
$
smeared against the
$(n{-}1)$-current $\bar\xi_{1}$
and the $(n{+}1)$-current $\xi_{2}$.
The rhs is a slight modification of the intersection number,
which is
nonzero only if $k_{1}+k_{2}=2n$,
$I_{M}(\xi_{1},\xi_{2})
=
\int_{M}
\,
\frac1{k_{1}!\;k_{2}!}
\xi_{1}^{\mu_{1}\cdots}(x)
\,
\xi_{2}^{\nu_{1}\cdots\mu_{k_{1}}}(x)
{\epsilon_{\mu_{1}\cdots \mu_{k_{1}}\nu_{1}\cdots\nu_{k_{2}}}}(x)
\;d^{2n}x
$.
The modification 
$\IM{\bar\xi_{1}}{\xi_{2}} = 
\epsilon_{n,k_{2}-n} I_{M}(\bar\xi_{1}, \xi_{2})
$,
$\epsilon_{n,m} = (-1)^{n m+m(m+1)/2} \,\epsilon_{n}^{-1}$
is such that $\IM{\bar\xi_{1}}{\xi_{2}}$
has properties independent of $n$:
$\IM{\bar\xi_{1}}{\xi_{2}}$ is skew-hermitian;
$\IM{\overbar{\partial\xi_{1}}}{\xi_{2}} 
= -\IM{\bar\xi_{1}}{\partial\xi_{2}}$;
$\IM{\bar\xi_{1}}{J\xi_{2}}$ on $n$-currents is hermitian and 
positive definite.

Pulled back to $Q$, the S-D equation of the free $n$-form cft is
$
\langle
\int_{\bar\eta_{1}} \bar\phi_{\alpha}
\int_{ \eta_{2}} d j_{\beta}
\rangle
=
2\pi i c_{\alpha\beta}
I_{Q}\langle\bar \eta_{1},\eta_{2}\rangle
$.
The rhs is $\IM{\bar\xi_{1}}{\xi_{2}}$
pulled back to a skew-hermitian form on currents in $Q$,
$
I_{Q}\langle\bar \eta_{1},\eta_{2}\rangle =  \IM{\Pi_{j_{1},n-1}\bar\eta_{1}}{\Pi_{j_{2},n-1}\eta_{2}}
$
which
is nonzero only if $(j_{1}+n-1)+(j_{2}+n-1) = 2n$,
which is $j_{1}+j_{2}=2$,
just like the intersection number of currents in a 2-manifold.
The S-D equation on $Q$ is formally analogous to the S-D equation
of the 2d gaussian model on a Riemann surface,
which is the Cauchy-Riemann equation.

%\newsec{7}
The free $n$-form cft on $M$ has now become the
2d gaussian model on each of the metric spaces $Q$.  
Moreover, each  $Q$ has the structure
needed to write the Cauchy-Riemann equation.
This is taken to be the defining structure of a {\it quasi Riemann
surface}.
The $Q=\cD^{\integral}_{\Integers\partial\xi_{0}}$
are  the quasi Riemann surfaces.
They are the fibers of a bundle of quasi Riemann surfaces
$\cQ(M) \rightarrow \PBM$,
$\PBM =\{\text{maximal infinite cyclic subgroups }\Integers\partial\xi_{0}
\subset \partial\cD^{\integral}_{n-1}(M)\}$.
On each fiber $Q$
there is a 2d gaussian model
with its global $U(1){\times} U(1)$ symmetry group, collectively
comprising a local gauge symmetry over $\PBM$.

%\newsec{8}
All of the constructions of 2d cft are based on
the Cauchy-Riemann equation and on the 2d gaussian model.
So there is the prospect of carrying out those constructions on each 
of the fibers $Q$
to obtain, for 
every 2d cft, a new cft of defects in $M$.
The 2d cft on each fiber $Q$ will be ambiguous up to its global 2d symmetry 
group.
The collection of global symmetry groups on the fibers
will forms a local gauge symmetry group over $\PBM$.

%\newsec{9}
There are many basic problems to be worked on:
opportunities to leverage 2d qft to develop a new technology of qft
in $2n$ dimensions.
Some are the following.

Complex analysis on quasi Riemann surfaces needs to be 
developed in analogy with ordinary Riemann surfaces.

Conjecturally, every quasi Riemann surface $Q$ is 
isomorphic to $\cD^{\integral}_{0}(\Sigma)$ for $\Sigma$ the 2d conformal 
space with the same Jacobian as $Q$,
the Jacobian being the complex torus made from the homology in the 
middle dimension.

The conjectured isomorphism
would allow constructing a 2d cft on each $Q$ by lifting an 
ordinary 2d cft from $\Sigma$ to $\cD^{\integral}_{0}(\Sigma)$,
which is a purely 2d problem,
then to each $Q=\cD^{\integral}(M)_{\Integers\partial\xi_{0}}$ via 
one of the conjectured isomorphisms.
As an example of the first step,
the 2d gaussian model is to be lifted
by extending the renormalization of the vertex operators $V_{p,\bar p}(\xi)$
from singular 0-currents $\xi$ to integral 0-currents.

If the conjecture is true, there will be some universal 
objects to study.  The automorphism group $\Aut(Q)$ will encode 
information about all the conformal groups of the conformal manifolds $M$
and about the global symmetry groups of all 2d cfts.
There will be a universal homogeneous bundle of quasi Riemann surfaces 
with structure group $\Aut(Q)$
in which all the bundles $\cQ(M)\rightarrow\PBM$ are embedded.

There should be a large collection of structure preserving maps from 
the complex disk $\Disk_{1}$ into $Q$. Meromorphic 
functions on $Q$ will pull back to ordinary meromorphic functions on $\Disk_{1}$.  
The local structure of a cft on $Q$ will be 
expressed as a collection of ordinary 2d cfts on each of these {\it 
local quasi holomorphic 
curves}, each with its radial quantization, pair of Virasoro 
algebras, and operator product expansion.
Explicit
constructions of local quasi holomorphic curves are needed, say for $M=S^{2n}$.

The local gauge symmetry in the bundle of quasi Riemann surfaces
needs a space-time interpretation.
What, for example,  is the space-time interpretation
of the local $SU(2){\times} SU(2)$ symmetry over $\PBM$ corresponding to the global 
$SU(2){\times} SU(2)$ at the self-dual 
point $R=1$ of the 2d gaussian model?

It should be possible to imitate on the quasi Riemann surfaces the 
usual constructions of 2d cft such as 
orbifolding and perturbation theory.

Tiny defects look like points in $M$, so fields $\Phi(\xi)$ 
when restricted to the small $\xi$ in $Q$
will give ordinary local quantum fields on $M$.
Will these form new local qfts in $2n$-dimensions?

%%%%%%%%%%%%%%%%%%%%%%%%%%%%%%%%%%%%%%%%%%%%%%%%%

\vskip6pt

\enlargethispage{20pt}

%\ethics{Insert ethics text here.}

\dataccess{This article has no additional data.}

\aucontribute{Section 1 was written by M.B., A.O'B., and B.R. The authorship of the individual sections 2 through 17 is listed in the footnote to each section title.} 

\competing{There are no competing interests to report.}

\funding{M.B. and A.O'B. are partially supported by Royal Society University Research Fellowships. M.B.'s work was  supported by the Royal Society under the grant, \lq\lq New Constraints and Phenomena in Quantum Field Theory."}. A.O'B. and B.R. were also supported by STFC consolidated grant ST/L000296/1. Funding sources reported by the authors of sections 2 through 17 are listed along with their names and affiliations in the footnote associated with each section title.}

\ack{All of the authors would like to acknowledge the Royal Society for providing the opportunity to participate in this workshop. AB would like to thank Fernando Alday for collaboration on \cite{Alday:2017xua} and related topics.  PED would like to thank Vincent Caudrellier, Ed Corrigan,  Giuseppe Mussardo, Tomasz Romanczukiewicz and Yasha Shnir for helpful discussions as well as Matt Buican and Andy O'Bannon for the invitation to speak at such an enjoyable workshop. JE would like to thank Andy O'Bannon, Carlos Hoyos, Mario Flory, Max Newrzella, Jackson Wu, and Ioannis Papadimitriou for fruitful collaboration and the participants of the Workshop for stimulating discussions. CK would like to hank the organizers of the workshop "Boundary and Defect Conformal Field Theory: Open Problems and Applications" for the invitation to speak there and for creating an inspiring atmosphere and thank all of her collaborators for illuminating discussions. SR would like to acknowledge discussions with Gil Young Cho, Andreas Ludwig, Ken Shiozaki, Bo Han, Apoorv Tiwari, Chang-Tse Hsieh, Xueda Wen, Shunji Matsuura, Shoucheng Zhang, Masamichi Miyaji, Tadashi Takayanagi, and Takahiro Morimoto.}

\end{document}